\documentclass[%
 amsmath,amssymb,
 aps,
 prd,
 twocolumn,
 eqsecnum,
 floatfix,
 superscriptaddress,
 showpacs,
]{revtex4-1}

\usepackage{graphicx}
\usepackage{dcolumn}
\usepackage{bm}
\usepackage{placeins} 
\usepackage{gensymb} 
\usepackage{color} 
\usepackage{hyperref}

\usepackage{siunitx}
\usepackage{epstopdf}
\usepackage{multirow}

\begin{document}


\title{Measurements of the atmospheric neutrino flux by Super-Kamiokande: \protect\\
       energy spectra, geomagnetic effects, and solar modulation}

\newcommand{\AFFicrr}{\affiliation{Kamioka Observatory, Institute for Cosmic Ray Research, University of Tokyo, Kamioka, Gifu 506-1205, Japan}}
\newcommand{\AFFkashiwa}{\affiliation{Research Center for Cosmic Neutrinos, Institute for Cosmic Ray Research, University of Tokyo, Kashiwa, Chiba 277-8582, Japan}}
\newcommand{\AFFipmu}{\affiliation{Kavli Institute for the Physics and
Mathematics of the Universe (WPI), The University of Tokyo Institutes for Advanced Study,
University of Tokyo, Kashiwa, Chiba 277-8582, Japan }}
\newcommand{\AFFmad}{\affiliation{Department of Theoretical Physics, University Autonoma Madrid, 28049 Madrid, Spain}}
\newcommand{\AFFubc}{\affiliation{Department of Physics and Astronomy, University of British Columbia, Vancouver, BC, V6T1Z4, Canada}}
\newcommand{\AFFbu}{\affiliation{Department of Physics, Boston University, Boston, MA 02215, USA}}
\newcommand{\AFFbnl}{\affiliation{Physics Department, Brookhaven National Laboratory, Upton, NY 11973, USA}}
\newcommand{\AFFuci}{\affiliation{Department of Physics and Astronomy, University of California, Irvine, Irvine, CA 92697-4575, USA }}
\newcommand{\AFFcsu}{\affiliation{Department of Physics, California State University, Dominguez Hills, Carson, CA 90747, USA}}
\newcommand{\AFFcnm}{\affiliation{Department of Physics, Chonnam National University, Kwangju 500-757, Korea}}
\newcommand{\AFFduke}{\affiliation{Department of Physics, Duke University, Durham NC 27708, USA}}
\newcommand{\AFFfukuoka}{\affiliation{Junior College, Fukuoka Institute of Technology, Fukuoka, Fukuoka 811-0295, Japan}}
\newcommand{\AFFgifu}{\affiliation{Department of Physics, Gifu University, Gifu, Gifu 501-1193, Japan}}
\newcommand{\AFFgist}{\affiliation{GIST College, Gwangju Institute of Science and Technology, Gwangju 500-712, Korea}}
\newcommand{\AFFuh}{\affiliation{Department of Physics and Astronomy, University of Hawaii, Honolulu, HI 96822, USA}}
\newcommand{\AFFkek}{\affiliation{High Energy Accelerator Research Organization (KEK), Tsukuba, Ibaraki 305-0801, Japan }}
\newcommand{\AFFkobe}{\affiliation{Department of Physics, Kobe University, Kobe, Hyogo 657-8501, Japan}}
\newcommand{\AFFkyoto}{\affiliation{Department of Physics, Kyoto University, Kyoto, Kyoto 606-8502, Japan}}
\newcommand{\AFFmiyagi}{\affiliation{Department of Physics, Miyagi University of Education, Sendai, Miyagi 980-0845, Japan}}
\newcommand{\AFFnagoya}{\affiliation{Solar Terrestrial Environment Laboratory, Nagoya University, Nagoya, Aichi 464-8602, Japan}}
\newcommand{\AFFpol}{\affiliation{National Centre For Nuclear Research, 00-681 Warsaw, Poland}}
\newcommand{\AFFsuny}{\affiliation{Department of Physics and Astronomy, State University of New York at Stony Brook, NY 11794-3800, USA}}
\newcommand{\AFFokayama}{\affiliation{Department of Physics, Okayama University, Okayama, Okayama 700-8530, Japan }}
\newcommand{\AFFosaka}{\affiliation{Department of Physics, Osaka University, Toyonaka, Osaka 560-0043, Japan}}
\newcommand{\AFFregina}{\affiliation{Department of Physics, University of Regina, 3737 Wascana Parkway, Regina, SK, S4SOA2, Canada}}
\newcommand{\AFFseoul}{\affiliation{Department of Physics, Seoul National University, Seoul 151-742, Korea}}
\newcommand{\AFFshizuokasc}{\affiliation{Department of Informatics in
Social Welfare, Shizuoka University of Welfare, Yaizu, Shizuoka, 425-8611, Japan}}
\newcommand{\AFFskk}{\affiliation{Department of Physics, Sungkyunkwan University, Suwon 440-746, Korea}}
\newcommand{\AFFtokyo}{\affiliation{The University of Tokyo, Bunkyo, Tokyo 113-0033, Japan }}
\newcommand{\AFFtodai}{\affiliation{Department of Physics, University of Tokyo, Bunkyo, Tokyo 113-0033, Japan }}
\newcommand{\AFFtoronto}{\affiliation{Department of Physics, University of Toronto, 60 St., Toronto, Ontario, M5S1A7, Canada }}
\newcommand{\AFFtriumf}{\affiliation{TRIUMF, 4004 Wesbrook Mall, Vancouver, BC, V6T2A3, Canada }}
\newcommand{\AFFtokai}{\affiliation{Department of Physics, Tokai University, Hiratsuka, Kanagawa 259-1292, Japan}}
\newcommand{\AFFtsinghua}{\affiliation{Department of Engineering Physics, Tsinghua University, Beijing, 100084, China}}
\newcommand{\AFFuw}{\affiliation{Department of Physics, University of Washington, Seattle, WA 98195-1560, USA}}

\AFFicrr
\AFFkashiwa
\AFFmad
\AFFbu
\AFFubc
\AFFbnl
\AFFuci
\AFFcsu
\AFFcnm
\AFFduke
\AFFfukuoka
\AFFgifu
\AFFgist
\AFFuh
\AFFkek
\AFFkobe
\AFFkyoto
\AFFmiyagi
\AFFnagoya
\AFFpol
\AFFsuny
\AFFokayama
\AFFosaka
\AFFregina
\AFFseoul
\AFFshizuokasc
\AFFskk
\AFFtokai
\AFFtokyo
\AFFipmu
\AFFtoronto
\AFFtriumf
\AFFtsinghua
\AFFuw

\author{E.~Richard}
\AFFkashiwa 
\author{K.~Okumura}
\AFFkashiwa
\AFFipmu 

\author{K.~Abe}
\AFFicrr
\AFFipmu
\author{Y.~Haga}
\AFFicrr
\author{Y.~Hayato}
\AFFicrr
\AFFipmu
\author{M.~Ikeda}
\AFFicrr
\author{K.~Iyogi}
\AFFicrr 
\author{J.~Kameda}
\author{Y.~Kishimoto}
\author{M.~Miura} 
\author{S.~Moriyama} 
\author{M.~Nakahata}
\AFFicrr
\AFFipmu 
% ...

\collaboration{The Super-Kamiokande Collaboration}
\noaffiliation


\date{\today}

\begin{abstract}

\newpage

A comprehensive study of the atmospheric neutrino flux 
in the energy region from sub-GeV up to several TeV 
using the Super-Kamiokande water Cherenkov detector
is presented in this paper.
The energy and azimuthal spectra, and variation over time, of the
atmospheric $\nu_e$$+$$\bar{\nu}_e$ and $\nu_{\mu}$$+$$\bar{\nu}_\mu$ 
fluxes are measured.
The energy spectra are obtained using an iterative unfolding method
by combining various event topologies with differing energy responses.
The azimuthal spectra depending on energy and zenith angle,
and their modulation by geomagnetic effects,
are also studied. A predicted east-west asymmetry
is observed in both the $\nu_e$ and $\nu_{\mu}$ samples
at 8.0~$\sigma$ and 6.0~$\sigma$ significance, respectively, and an indication
that the asymmetry dipole angle changes depending on the zenith angle
was seen at the 2.2~$\sigma$ level.
The measured energy and azimuthal spectra are consistent with the current flux models 
within the estimated systematic uncertainties.
A study of the long-term correlation between the
atmospheric neutrino flux and the solar magnetic activity cycle is performed, 
and a weak preference for a correlation was seen at the 1.1~$\sigma$ level,
using SK I-IV data spanning a 20~year period.
For several particularly strong solar activity periods, corresponding to Forbush decrease events,
no theoretical prediction is available but a deviation below the typical neutrino event rate
is seen at the 2.4~$\sigma$ level.
The seasonal modulation of the neutrino flux is also examined,
but the change in flux at the SK site is predicted to be negligible,
and as expected no evidence for a seasonal correlation is seen.

\end{abstract}




\maketitle


%
%

\section{\label{sec:intro}Introduction}

Atmospheric neutrinos are one of the main experimentally available neutrino sources,
observed in a wide energy region from 100~MeV up to the PeV scale~\cite{Gaisser:2002jj}. 
They are generated after the interaction of cosmic rays with the air molecules in Earth's atmosphere,
from the decay of secondary particles such as $\pi$ and $K$ mesons.
The form of the energy spectrum is well approximated by a power-law,
although it is relatively suppressed below the GeV scale due to the rigidity cutoff effect
on the primary cosmic rays, caused by Earth's magnetic field.
The flavor ratio of $\nu_\mu$+$\bar{\nu}_\mu$ to $\nu_e$+$\bar{\nu}_e$ below the GeV scale
is approximately two, considering the dominance of the $\pi^\pm$ decay chains,
but increases towards higher energies.
In neutrino detectors, atmospheric neutrinos are observed coming from all directions,
as the Earth is mostly transparent for neutrinos below the PeV scale;
the flight length from the production point to the detector thus varies from
$O(10)$ to $O(10^4)$~km, depending on the zenith angle of the arrival direction.
The neutrino flux in the horizontal direction is generally higher than that in the vertical direction,
due to the longer path of the parent particles in the atmosphere;
however in the sub-GeV region there is an east-west asymmetry
due to the azimuthal dependence in the rigidity cutoff energy of the primary cosmic rays.
Towards the higher energies, fewer neutrinos are produced as the $\pi$ and $K$ decay lengths
become longer than their path lengths in the atmosphere, and the
parent particles reach the ground before decaying.
Above around 100~TeV, so called ``prompt'' neutrinos
coming from the fast decay of charmed mesons
are expected to dominate, due to their much shorter decay length.

Since the first detection of the atmospheric neutrino
in underground experiments in the 1960s~\cite{kolar-exp,Reines},
further measurements brought the discovery of
neutrino oscillation
in 1998~\cite{sk-evidence}.
The continuing series of independent neutrino oscillation measurements in solar~\cite{SK-solar,SNO}, 
atmospheric~\cite{SK1-fullpaper,Wendell}, reactor~\cite{KamLAND,DAYA-BAY,RENO,DOUBLE-CHOOZ},
and accelerator~\cite{K2K,MINOS,T2K-numu,T2K-nue} sourced neutrinos are 
consistent with three neutrinos mixing as described by
the 3$\times$3 PMNS matrix~\cite{Maki:1962mu,Pontecorvo:1968fh}, which is generally accepted
as the standard framework of neutrino oscillation
while the CP-violation phase and the mass ordering between the second and 
third mass states are not yet known.

The study of the atmospheric neutrino is generally based on
predictions of the expected flux, coming from Monte Carlo simulations.
In this paper we will discuss three such simulations performed by the
HKKM~\cite{honda06,honda2011}, Bartol~\cite{bartol},
and FLUKA~\cite{fluka} research groups.
The HKKM model defines a primary cosmic ray spectrum
based on BESS~\cite{BESS2000,BESS2004} and 
AMS~\cite{AMS2000,AMS2000-2} measurements. For interactions in the atmosphere,
the DPMJET-III~\cite{Roesler:2000he} hadronic interaction model is used,
with some customized tuning for better agreement with the cosmic ray muon data. 
The JAM nuclear interaction model~\cite{Niita20061080}, 
which has better agreement with the $\pi$ production measurements
by the HARP experiment~\cite{HARP2008-2}, 
is also introduced below 32~GeV in the more recent HKKM11 model~\cite{honda2011},
resulting in a relative increase of the neutrino flux 
below 1~GeV when compared to HKKM07~\cite{honda06}. 
The Bartol model adopts 
a primary proton spectrum that is relatively lower
below 50~GeV, and the high energy nucleon flux 
has a flatter energy dependence~\cite{Agrawal:1995gk}.
TARGET version 2.1~\cite{TARGET} is used for
the decay and interaction generator.
The FLUKA flux model is so named as it uses the FLUKA
Monte Carlo simulation code~\cite{FlukaMC},
a widely used hadronic and electro-magnetic interaction model.
In all of these simulations, three-dimensional particle tracking
is performed for the primary and secondary particles.

The estimated uncertainties on the atmospheric neutrino flux predictions
are currently 
between 5$\sim$25\% in the 100~MeV$\rightarrow$100~GeV range.
As the neutrino energy increases above 10~GeV,
the uncertainties in the $\pi$ and $K$ fluxes
become the dominant error sources~\cite{Honda:2011vda,Barr:2006it}.
The neutrino flux predictions
are consistent within $\sim$20\% below 32~GeV~\cite{honda2011},
with larger differences in the higher 
energy region~\cite{honda06}. 

The current generation of neutrino detection experiments 
have increased their statistics and reduced systematic errors,
such that direct measurements can now be made with uncertainties
comparable to those on the flux models. In this paper, we discuss
several types of direct measurements made by the Super-Kamiokande detector.
First we discuss briefly these measurements in a conceptual sense,
especially taking into account the fact that the Super-Kamiokande atmospheric data 
was previously used to make an original measurement of the neutrino oscillation 
parameters.

In a general sense,
the measured event rate $N$ of an atmospheric neutrino detector is 
expressed by the convolution of atmospheric neutrino flux $\Phi$, 
neutrino oscillation probability $O$, neutrino cross section $\sigma$, 
and detector efficiency $\epsilon$:
\begin{equation}
N = \Phi \otimes O \otimes \sigma \otimes \epsilon \label{eqn:evrate}.
\end{equation}
To measure the oscillation probability $O$, 
the quantities $\Phi$, $\sigma$, and $\epsilon$ must be 
determined by other independent measurements in advance. 
As described above $\Phi$ is calculated by 
Monte Carlo simulations, $\sigma$ is modeled 
based on the results of neutrino beam experiments,
and $\epsilon$ is determined 
based on detector calibrations and simulations.
Combining these inputs and the measured event rate $N$,
the oscillation probability $O$ and thus the oscillation parameters
have been previously measured by Super-Kamiokande,
such as the mass difference $|\Delta m^2_{32}|$.
Conversely, it is possible to measure the atmospheric neutrino flux 
$\Phi$ if the neutrino oscillation probability $O$ is given.
The measurements of the oscillation parameters by the Super-Kamiokande
atmospheric data have since been confirmed by several indendent experiments,
and a consistent and accurate three-flavor oscillation parameterization
has emerged, by combining data from each of the independent experiments.
Thus, by taking the PDG values of the oscillation parameters~\cite{PDG2014}
as the input, the neutrino flux $\Phi$ can be measured at Super-Kamiokande.

We separate our measurements of the atmospheric neutrino flux
into three main categories, as follows.

The first measurement
is the energy spectra of the $\nu_e$ and $\nu_\mu$ fluxes.
Over the past decade, the simulations have improved their statistics and calculation methods
and now provide accurate predictions across a wide energy range
from sub-GeV to 10~TeV~\cite{honda2011,fluka,bartol}.
These predictions are important for studies of the atmospheric neutrino itself,
and also as a background model for rare event searches
such as proton decay~\cite{PDK2009,PDK2014} or dark matter~\cite{Billard:2013qya}. 

An initial measurement of the energy spectra was made by the Frejus experiment in 1995~\cite{frejus},
before the existence of neutrino oscillation was known.
More recently measurements were made above TeV energies by the cubic-kilometer size
detector IceCube~\cite{amanda2-ff,amanda2-unfold,icecube-flux2011,icecube-numu-ff,icecube-fluxnue,Aartsen:2015xup}.
At these energies the understanding of the atmospheric flux is important with
respect to the searches of astronomical neutrinos, 
a flux of which was recently discovered
by the IceCube collaboration~\cite{Aartsen:2013bka,Aartsen:2013jdh,Aartsen:2014gkd}.

A precise measurement of atmospheric neutrino flux below 100~GeV
has not yet been published, and
the Super-Kamiokande detector is able to make several significant improvements
with respect to the Frejus measurement in this region.
The larger detector size increases event statistics by around two orders of magnitude,
and extends the measurement down to 100 MeV and up to around 10~TeV,
which overlaps with the low energy end of the
cubic-kilometer detector measurements.
The high resolution of the Cherenkov ring imaging technique used in Super-Kamiokande
leads to excellent identification of the neutrino flavor and background elimination.
Combined with the progress in understanding neutrino cross-sections
(see e.g. Section~49 of \cite{PDG2014} and references therein) and oscillations
by several independent experiments, an accurate measurement of the flux
based on the number of observed events in the detector is possible.

The second measurement
is the azimuthal distribution of the atmospheric neutrino flux.
The geomagnetic field deflects incoming primary cosmic rays,
depending on their momentum and nuclear composition;
the geomagnetic rigidity is defined as $pc/Ze$, for momentum
$p$, atomic number $Z$ and the elementary charge $e$,
and is often given in units of GV.
For a given arrival location and direction on the Earth's surface,
only cosmic rays with rigidity above a certain threshold
will have been able to traverse the geomagnetic field to that point,
excluding trajectories that had intersected the Earth;
this threshold is known as the geomagnetic rigidity cutoff.
The structure of the geomagnetic field causes variations in this cutoff,
the primary effect of which is an east-west oriented anisotropy in
cosmic rays at the Earth's surface, originally detected in the 1930s~\cite{Rossi}
and used to infer that cosmic rays are generally positively charged.

A similar east-west anisotropy is expected in the atmospheric neutrino flux,
and in a previous Super-Kamiokande measurement~\cite{skeastwest}
was discovered in the $\nu_e$ sample ($5\sigma$) but seen
only with low significance in the $\nu_{\mu}$ sample ($2\sigma$).
Since that measurement the flux simulations have also progressed
from simple one-dimensional to three-dimensional calculations,
and began to use complex geomagnetic field models instead of dipole approximations.
The HKKM and Bartol simulations have further included
bending of secondary particles due to the geomagnetic field in the atmosphere.
These changes have led to significant modifications in the predictions of the
azimuthal anisotropies~\cite{honda2004,Battistoni2000315,Lipari2000153}.
Measuring in detail the angular distributions,
in addition to the previously well-studied zenith distributions,
can thus be used as a further test of the flux simulations
and their implementations of geomagnetic effects,
and also to confirm the discovery of such azimuthal anisotropies.

The third measurement
is the modulation of the neutrino flux over time.
The solar cycle is an oscillatory change in the solar activity,
such as the level of plasma emissions,
with an average period of approximately 11 years.
The cosmic ray flux at Earth is well known to be
anti-correlated
with the solar activity~\cite{solmod_cosmic_rays}.
This is essentially because the plasma flux (or ``solar wind'') from the sun can scatter cosmic rays
entering the solar system, and therefore during periods of high solar activity
the cosmic ray flux is relatively reduced. Consequently, the atmospheric neutrino flux is predicted to
also be anti-correlated
with the solar cycle, although this has not previously been measured.

Historically, the solar activity was measured by its correlation with the appearance of sunspots.
However, since 1948 the use of neutron monitors (NMs) provides a method to accurately and continuously track
the neutron flux at the Earth's surface resulting from cosmic ray impacts~\cite{nm}, and
the NM counts are generally believed to be well-correlated with the primary cosmic ray activity.
In this paper we test for the anti-correlation of the atmospheric neutrino flux with the solar cycle,
by searching for a correlation between the neutrino flux at SK and the neutron detection
rates at various NMs operated by other institutes. This method is able to test on
short timescales of $O(\SI{1}{\hour})$, as both neutrino and neutron observations take
place on Earth, and the propagation of the solar wind within the solar system
(with speed of the order \SI{100}{km.s^{-1}}) need not be considered.
We assume that the effect of the solar wind is uniform in the neighborhood of the Earth,
such that we may expect a good correlation between the neutrino flux
at the SK site and the neutron flux monitored at the NMs in various locations around the Earth.

Yearly changes in the atmospheric neutrino flux
are also expected, due to seasonal temperature variations.
In the summer months the atmospheric density is increased at higher altitudes,
and relatively more neutrinos are created by secondary particles decaying in-flight.
Such changes are predicted~\cite{honda2015} to be strongest at the polar regions,
with a normalization change of a few percent around the GeV to TeV scale,
but to become negligible moving towards to the equator
where the seasonal variation in air density is less.
While the variation of the atmospheric neutrino flux at SK is thus expected to be minimal,
we also test for such a correlation in this paper.

This paper continues by explaining
the detector and atmospheric neutrino dataset in Section \ref{sec:detector},
the energy spectrum measurement in Section \ref{sec:flux},
the azimuthal spectrum measurement in Section \ref{sec:eastwest},
the solar modulation measurement in Section \ref{sec:solarmodulation}, 
and concludes with a summary in Section \ref{sec:summary}.

%
%
%
\section{\label{sec:detector}Detector and Atmospheric Neutrino Dataset}

\subsection{\label{sec:det_detector}Super-Kamiokande Detector}
%

%
%
Super-Kamiokande (SK) is a water Cherenkov detector located 
in the Mozumi mine of Gifu prefecture, Japan~\cite{skdetector},
at geographic coordinates 
36$\degree$25$^\prime$32.6$^{\prime\prime}$N 137$\degree$18$^\prime$37.1$^{\prime\prime}$E and altitude 370~m
in the WGS-84 system~\cite{gps}.
About 11,000 20-inch photomultipliers (PMTs) are mounted on the wall of the 
detector facing inwards, and observe the Cherenkov light emitted by charged particles
produced by neutrino interactions in the ultra-pure water.
An optically-separated region on the outer side also contains
about 1,800 PMTs, which act as a veto for incoming cosmic ray events.
The detector has excellent particle identification (PID) capability
by using the Cherenkov ring's pattern and opening angle, separating showering-type events
from track-type events, which are denoted as $e$-like or $\mu$-like events respectively.

%
%
SK has four experimental phases so far.
These are designated as SK-I (1996--2001), SK-II (2002--2005), SK-III (2006--2008), and SK-IV (2008--).
The major detector changes that distinguish these periods are as follows.
The SK-I period was ended when an implosion accident destroyed about half of the PMTs;
the remaining PMTs were rearranged for even coverage during SK-II. For SK-III, full PMT coverage
was restored, and with SK-IV came an improved front-end electronics system~\cite{Nishino:2009zu}.
SK-IV is an ongoing experiment, and the data sets used in this paper include data until September 2014
for the energy spectrum and azimuthal analyses described in Section~\ref{sec:flux} and \ref{sec:eastwest}, 
and until April 2015 for the solar modulation analysis in Section~\ref{sec:solarmodulation}.

\subsection{\label{sec:det_types}Atmospheric Data Types}
%

The atmospheric neutrino data events are separated into three main samples
with different event topologies.
In the Fully-Contained (FC) sample, the reconstructed neutrino interaction vertex
is inside the 22.5~kton fiducial volume,
which is the inner region with a boundary 2~m inside the inner wall,
and all visible secondary particles are contained inside the inner detector. 
In the Partially-Contained (PC) sample, the vertex is also inside this fiducial volume,
but outgoing particles are allowed to exit the inner detector.
PC events typically have longer charged particle tracks,
which are therefore mostly muons induced by $\nu_\mu$ charged-current (CC) interactions. 
In the UPward-going-MUon event sample (UPMU), neutrinos interacting with 
the surrounding rock create muons which enter the detector from below
(down-going muons are ignored, as these are overwhelmingly
produced by cosmic rays).
The UPMU sample is also a predominantly $\nu_\mu$-induced sample.
The efficiency in selecting true neutrino events
is estimated for fully-contained events as 
$>$97\%, and for partially contained events as $>$80\%    
(improving to $>$95\% for SK-III and IV)~\cite{euan_phd}.  
Non-neutrino background contamination from cosmic ray muons or
light-flashing PMTs is less than a few percent for all samples.
%

In this analysis these three main event samples are further divided as follows, according to their
detailed properties identified by the event reconstruction algorithm,
such as PID, number of reconstructed Cherenkov rings, reconstructed energy,
and presence or absence of electrons from delayed muon decays.
These sub-sample definitions are similar to the standard ones used in the SK atmospheric neutrino
oscillation analysis~\cite{Wendell},
although compared to that analysis some sub-samples are combined or discarded in this paper.
The detector cannot distinguish neutrinos from antineutrinos by measuring the sign of the produced lepton,
as no magnetic field is applied in the detector,
therefore when we refer to $\nu_\mu$ or $\nu_e$ samples this
includes also $\bar{\nu}_\mu$ and $\bar{\nu}_e$ respectively
(some separation techniques in fact exist~\cite{Abe:2011ph,maggie_phd,tristan_phd}
that take advantage of various kinematic differences, but are not used in this paper). 

Fully-contained events are divided into $e$-like and $\mu$-like samples
according to the PID of the most energetic Cherenkov ring,
corresponding generally to $\nu_e$ and $\nu_\mu$ samples respectively.
They are further divided according to whether the number of the identified Cherenkov rings is one 
(single-ring) or more (multi-ring). 
Events are also classified into ``sub-GeV'' or ``multi-GeV''
samples at a threshold of 1.33~GeV in total visible energy, which is the sum of the
visible energy among all fitted Cherenkov rings.
Single-ring events are of higher purity in neutrino flavor,
while multi-ring events tend to have higher neutrino energy. 
Additional selection criteria to reduce the background of neutral-current (NC)
and wrong-flavor CC events are applied;
for the sub-GeV single-ring $e$-like events, NC produced $\pi^0$ backgrounds are reduced by 
a $\pi^0$ finder algorithm~\cite{polfit} which identifies events using a likelihood method, based around the
reconstructed invariant mass assuming a $\pi^0 \to 2\gamma$ decay event
(where the second $\gamma$ ring was originally missed due to low energy or ring overlap).
Sub-GeV multi-ring $e$-like events are also not used in these analyses
in order to further avoid NC $\pi^0$ background.
Another likelihood-based cut is applied for multi-GeV multi-ring $e$-like events,
which is the same method as described in \cite{Wendell} but extended to include SK-IV,
to reduce background contamination from hadronic events dominated by $\gamma$ rings, produced by 
$\pi^0$ created in CC $\nu_\mu$ and NC induced events.

Partially-contained events are divided into two samples, ``PC stopping'' and ``PC through-going'', 
based on whether the muon particle stops in the outer detector, or leaves the detector completely.
The separation is made according to the charge sum in the outer detector
around the muon exiting point, considering the detector geometry~\cite{oscillationevidence}. 

Upward-going-muon events are divided into three samples: ``UPMU stopping'', ``UPMU non-showering'', 
and ``UPMU showering''. 
UPMU stopping events have only an entering hit cluster in the outer detector,
while the other two samples have both entering and exiting hit clusters. 
UPMU non-showering and showering events
are separated by the estimated deposited energy per unit muon range,
with the aim to separate muon events with radiative emission, such as bremsstrahlung, which
dominates when the muon's energy is greater than around 1~TeV~\cite{Desai:2007ra}. 
\subsection{\label{sec:det_sets}Data and Monte Carlo sets}
%

%
%
\begin{table*}[htbp]
\begin{center}
\scalebox{0.9}{
\begin{tabular}{l c r r r r r r r r c r r r r r r }
\hline
\hline
\textbf{Sub-Sample} &&  \multicolumn{2}{c}{\textbf{SK-I}}   & \multicolumn{2}{c}{\textbf{SK-II}} & \multicolumn{2}{c}{\textbf{SK-III}} & \multicolumn{2}{c}{\textbf{SK-IV}} && \multicolumn{2}{c}{\textbf{Total}} &\multicolumn{4}{c}{}  \\
\hline
&& \multicolumn{15}{c}{\textbf{Livetime (days)}} \\
FC and PC  && \multicolumn{2}{c}{1489}   & \multicolumn{2}{c}{799} & \multicolumn{2}{c}{518} & \multicolumn{2}{c}{1993} && \multicolumn{2}{c}{4799} &\multicolumn{4}{c}{} \\
UPMU   && \multicolumn{2}{c}{1646}   & \multicolumn{2}{c}{828} & \multicolumn{2}{c}{636} & \multicolumn{2}{c}{1993} && \multicolumn{2}{c}{5103}&\multicolumn{4}{c}{}  \\
\hline
&& \multicolumn{11}{c}{\textbf{Number of Events}}  & \multicolumn{4}{c}{\textbf{Interaction [\%]}}  \\
&& \multicolumn{11}{c}{}                           & $\nu_e$CC & $\nu_\mu$CC & $\nu_\tau$CC & NC  \\
FC $e$-like  && \multicolumn{13}{c}{}  \\
~~~sub-GeV single-ring   && 3288 & (3104.7) & 1745 & (1632.8) & 1209 & (1100.7) & 4251 & (4072.8) && 10493 & (9911.0) & 94.1 &  1.5 &  $<$0.1 & 4.4  \\
~~~multi-GeV single-ring && 856 & (842.8)   &  396 & (443.7)  &  274 & (299.5) & 1060 & (1080.0) &&  2586 & (2666.0) & 86.3 &  3.2 & 1.7 &  8.8 \\
~~~multi-GeV multi-ring  && 449 & (470.1)   &  267 & (252.1)  & 140  & (161.9) &  634 & (654.9)  &&  1490 & (1539.0) & 73.0 &  7.6 &  3.3 & 16.1 \\
FC $\mu$-like && \multicolumn{13}{c}{}  \\
~~~sub-GeV single-ring   && 3184 & (3235.6) & 1684 & (1731.8) & 1139 & (1152.0) &  4379 & (4394.7) && 10386 & (10514.0) & 0.9  & 94.2 & $<$0.1 & 4.9 \\
~~~multi-GeV single-ring && 712  & (795.4)  & 400  & (423.9)  &  238 & (273.9)  & 989   & (1051.5) && 2339 & (2544.7)   &  0.4 &  99.1 & 0.3 & 0.2 \\
~~~multi-GeV multi-ring  && 603  & (656.5)  &  337 & (343.8)  &  228 & (237.9)  &  863  & (927.8) && 2031 & (2166.0)  &  3.4 &  90.5 &  0.6 & 5.5 \\
PC && \multicolumn{13}{c}{}  \\
~~~stop                  && 143  & (145.3)  &   77 &  (73.2)  &  54  & (53.3)   &  237  & (229.0) && 511 & ( 500.8) &  12.7 &  81.7 &  0.9 & 4.6 \\
~~~thru                  && 759  & (783.8)  &  350  & (383.0) &  290 & (308.8)  &  1093 & (1146.7) && 2492 & (2622.3) &  0.8 &  98.2 & 0.4 &  0.5 \\
UPMU  && \multicolumn{13}{c}{}  \\
~~~stop                  && 432.0 & (433.7)   & 206.4 & (215.7)  &  193.7 & (168.3)  &  492.7 & (504.1) && 1324.8 &(1321.8) & 1.0 &  97.7 & 1.0 & 0.3 \\
~~~non-showering         && 1564.4 &  (1352.4) &   726.3 & (697.5) &   612.9 & (504.1) &  1960.7 & (1690.3) && 4864.3 &(4244.4) & 0.2 &  99.4 & 0.3 & 0.1 \\
~~~showering             && 271.7 & ( 291.6)  &  110.1 & (107.0)  &  110.0 & (126.0)  &  350.1 & (274.4) &&  841.9 & (799.0)  &  0.1 &  99.8 &  $<$0.1 & 0.1 \\
\hline
\hline
\end{tabular}
}
\end{center}
\caption{ \small
    Livetimes and numbers of events for each SK period and each sub-sample.
    Numbers in brackets are the MC expectation based on the HKKM11~\cite{honda2011} flux model, 
    which are calculated separately for each SK period and then scaled to the data livetime.
    The oscillation and solar activity weights are included.
    UPMU data livetimes are longer due to less strict conditions for good run selection,
    and UPMU event numbers are fractional due to the subtraction of horizontally-going 
    cosmic muon backgrounds.
    The fractions of the neutrino interaction modes for each sub-sample, 
    estimated from the MC datasets averaging over all SK periods, are shown in 
    the last four columns.
}
\label{tab:sub_sample}
\end{table*}

%
%
\begin{figure}[htbp]
\begin{center}
\begin{minipage}{\columnwidth}
\includegraphics[width=0.95\columnwidth]{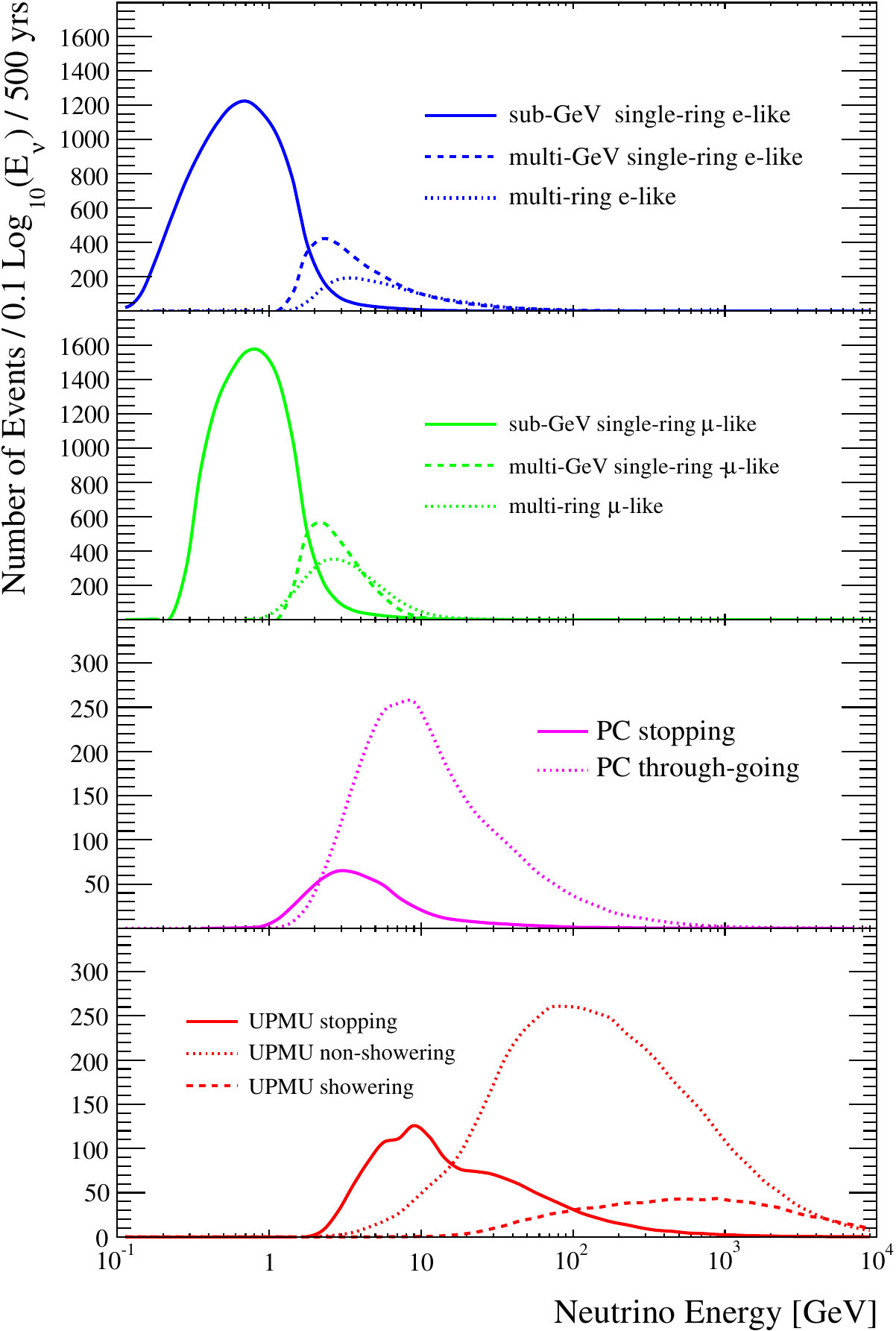}
\end{minipage}
\end{center}
\caption{ \small
(color online) Monte Carlo neutrino true energy distributions for each sub-sample,
corresponding to the total expected number of events across the SK I-IV periods
accounting for neutrino oscillation.
From top, the FC $e$-like, FC $\mu$-like, PC, and UPMU samples are shown. 
}
\label{fig:nu_energy_sample}
\end{figure}

%
%
Table~\ref{tab:sub_sample} summarizes the detector livetimes and numbers of events
in the data and Monte Carlo (MC) sets used by this analysis,
where the MC is generated corresponding to 500~years of livetime for each SK period.
Neutrino interactions are generated from the HKKM07~\cite{honda06} flux and
NEUT~\cite{NEUT} neutrino interaction models, and then passed to our detector simulation.
The recent updates to NEUT and the detector simulation are described in \cite{sk4_sterile}.
The data reduction and event reconstruction processes,
which are the standard ones used for the SK atmospheric neutrino analysis~\cite{Wendell}, 
are applied to both the data and MC samples.

Monte Carlo correction weights from the HKKM07 model to the HKKM11~\cite{honda2011} model
are applied on an event-by-event basis, considering the differences
in the energy, azimuthal, and zenith angle distributions.
In some cases in Section~\ref{sec:flux}, weights will instead be applied
to reweight to other flux models, according to their predicted energy spectra.
Weights due to neutrino oscillation are also applied per event,
using the three-flavor parameterization in \cite{PDG2014}, 
under the assumptions of normal hierarchy ($\Delta m_{32}^2$$>$0) and CP symmetry ($\delta_{\rm CP}$=0).
Finally, weights due to changes in the average degree of solar activity
were estimated based on cosmic NM data~\cite{neutron_data} and applied depending on the SK period, 
except for the solar modulation analysis in Section~\ref{sec:solarmodulation},
where a more accurate per-event solar correction is described and used.

The fractions of the interaction mode for each sub-sample (i.e. $\nu_e$CC, $\nu_\mu$CC, or NC)
are estimated from the MC datasets, and are also shown in Table~\ref{tab:sub_sample}.
We see that all $e$-like and $\mu$-like sub-samples have a high purity of $\nu_e$ and 
$\nu_\mu$ CC interactions, respectively.
The systematic uncertainties on these purity estimations
were calculated during this study, and found to be relatively small; 
less than 4 and 1.5 percentage points
for all $e$-like and $\mu$-like sub-samples, respectively. 
Figure~\ref{fig:nu_energy_sample} shows the MC true neutrino energy distributions
of the sub-samples,
showing that each sample has a different energy sensitivity.
The FC $\mu$-like sample extends above 10~GeV,
while the $e$-like sample extends up to 100~GeV,
indicating that $\mu$-like events in the fiducial volume beyond 10~GeV
tend to exit the inner detector and be classified as PC events.
The FC $e$-like sample has a lower neutrino energy threshold than the $\mu$-like sample
(100~MeV as opposed to 200~MeV), due to the difference in Cherenkov threshold
between the electrons/positrons and the muons which are produced in the CC interactions in water.
The PC and UPMU samples approximately range from 1~GeV to 10~TeV, and
classification into the respective sub-samples is also seen to cover different energy regions.

%
%
%
\section{\label{sec:flux}Energy Spectrum Analysis}

\subsection{\label{sec:flux_data}Event Classification}

In order to measure the flux of neutrinos as a function of their energy,
we select all events shown in Table~\ref{tab:sub_sample}.
This gives us a generally high purity selection of data induced by either $\nu_e$ or $\nu_\mu$
charged-current (CC) interactions, across a wide energy range.
%
Noting that PC and UPMU samples generally contain $\nu_\mu$ interactions, and that
the FC sample may be well separated by its $\nu_e$ and $\nu_\mu$ components,
we define two data samples for the energy spectrum measurement: 
a $\nu_e$ sample containing FC $e$-like events,
and a $\nu_\mu$ sample from FC $\mu$-like, PC, and UPMU events.
For the $\nu_e$ ($\nu_\mu$) sample, the signal event is defined as a $\nu_e$ ($\nu_\mu$) 
CC interaction, and the backgrounds are CC interactions of other flavor neutrinos and 
NC interactions of all flavor neutrinos.

\begin{figure*}[htbp]
\begin{center}
\includegraphics[width=0.6\textwidth]{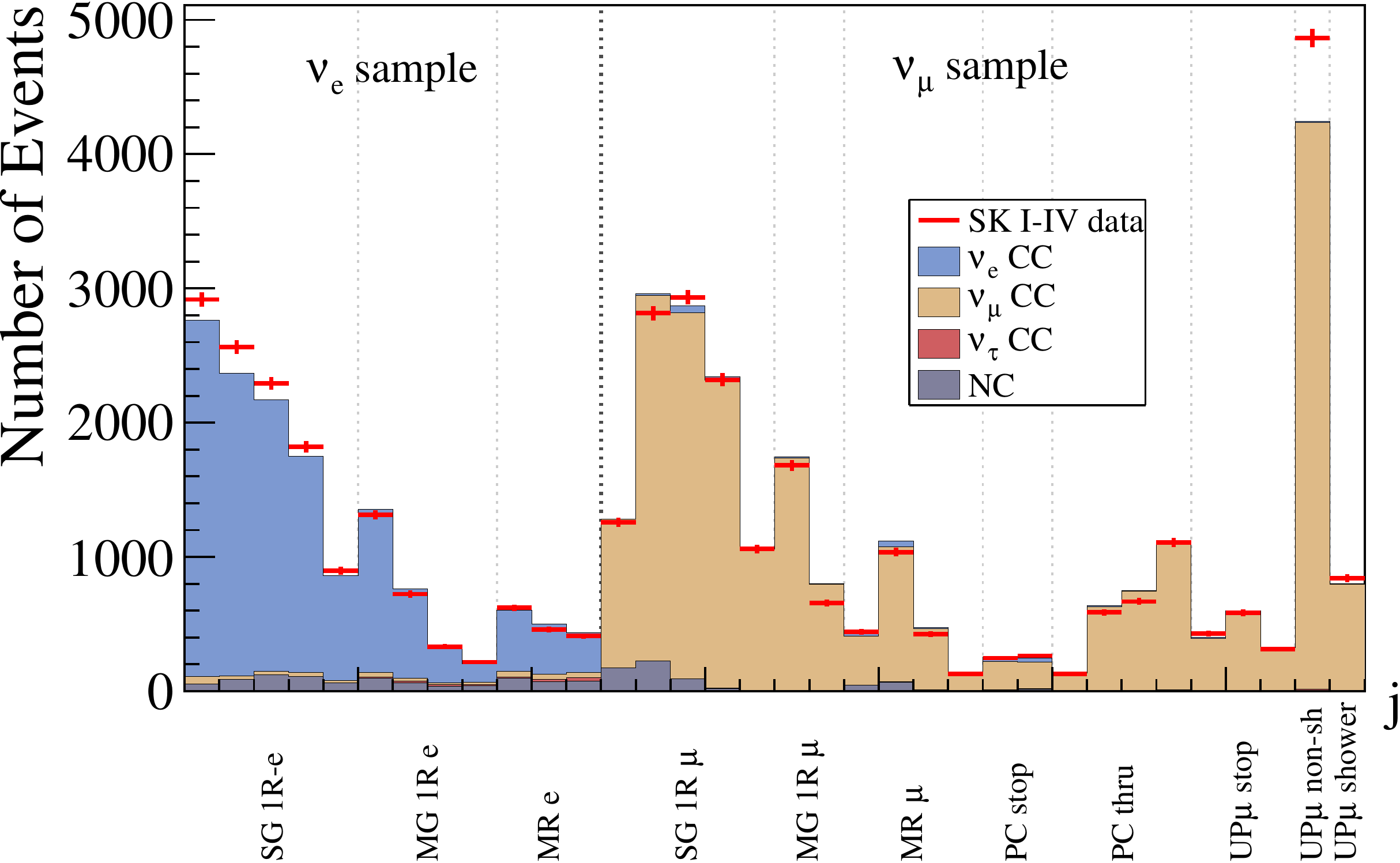}
\end{center}
\caption{ \small
(color online) Number of events for each reconstructed energy bin $j$ for each sample.
Colored regions represent $\nu_e$ CC,  $\nu_\mu$ CC, $\nu_\tau$ CC and NC contributions
from the MC,
and red crosses represent the data values,
for the entire SK I-IV period.
Monte Carlo distributions are normalized by the data livetimes.
Only statistical uncertainties are shown in the error bars of the data points.
}
\label{fig:Mj}
\end{figure*}

In each sub-sample, the linear sum of the reconstructed momentum of each visible particle
(or Cherenkov-ring) is used as a estimator of the neutrino energy, and denoted as 
the reconstructed energy $E_{\rm rec}$.
The reconstructed momentum of each particle is itself calculated
using the observed charge pattern of the associated ring and the particle's PID.
The energy binning is defined as shown in Table~\ref{tab:event-binning}, in $\log_{10}E_{\rm rec}$.
These bin widths are determined considering the energy resolution of each sub-sample, which is
typically $\sim$0.2 in $\Delta \log_{10}(E_{\nu}/\rm{GeV})$ for FC,
but 0.2$\sim$0.3 for PC, $\sim$0.4 for UPMU stopping, and $\sim$0.7 for other UPMU samples,
since the energy deposited outside the detector cannot be observed.
Thus, no energy binning is performed for UPMU showering or non-showering events.
We refer hereafter to the number of events in each data bin as $M_j$ for bin index $j$.
Figure~\ref{fig:Mj} shows the values of $M_j$ for SK I-IV data.

\begin{table*}
\begin{center}
\scalebox{0.9}{
\begin{tabular}{l l c c c}
\hline
\hline
Sub-sample  &  & NBins &  $j$ &   Bin edges in $\log_{10}(E_{\rm rec}/\rm{GeV})$ \\ 
\hline
FC $e$-like  &  & & \\
~~sub-GeV single-ring & (SG 1R-e)    & 5 & 1 - 5    &  -1.0, -0.6, -0.4, -0.2, 0.0, 0.2 \\
~~multi-GeV single-ring & (MG 1R-e)  & 4 & 6 - 9    & 0.0, 0.4, 0.7, 1.0, 3.0 \\
~~multi-GeV multi-ring & (MR e)  & 3 & 10 - 12  & 0.0, 0.4, 0.7, 3.0  \\
FC $\mu$-like &  &  & \\
~~sub-GeV single-ring & (SG 1R-$\mu$)   & 5 & 13 - 17  & -0.8, -0.6, -0.4, -0.2, 0.0, 0.2 \\
~~multi-GeV single-ring & (MG 1R-$\mu$) & 2 & 18, 19  & 0.0, 0.4, 2.0 \\
~~multi-ring         & (MR $\mu$)   & 4 & 20 - 23  & -1.0, 0.0, 0.4, 0.7, 2.0 \\
PC &  &  & \\
~~stopping  & (PC stop)                & 2 & 24, 25  & -1.0, 0.4, 2.0 \\
~~through-going & (PC thru)                  & 4 & 26 - 29  & -1.0, 0.0, 0.4, 0.7, 3.0 \\
UPMU &  &  & \\
~~stopping  & (UP$\mu$ stop)                & 3 & 30 - 32  & 0.0, 0.4, 0.7, 3.0 \\
~~non-showering & (UP$\mu$ non-sh)         & 1 & 33 & -- \\
~~showering & (UP$\mu$ shower)             & 1 & 34 & -- \\
\hline
\hline
\end{tabular}
}
\caption{ \small 
Binning definitions for the data bins $M_j$ (by the reconstructed energy $E_{\rm rec}$)
for the 11 sub-samples. There are 34 bins in total, among which 12 bins for $\nu_e$ and 22 bins for $\nu_\mu$ sample are assigned.
}
\label{tab:event-binning}
\end{center}
\end{table*}

%
\subsection{\label{sec:flux_unfolding}Flux Unfolding}
To measure the energy spectrum, we employ an unfolding method.
This is a class of method in which a true spectrum is deconvolved
from an experimentally measured one, based on a knowledge of the experimental reconstruction.
This differs from forward-fitting methods,
where the experimentally measured spectrum is not deconvolved, but compared directly
to model predictions that have been passed through a simulated experimental reconstruction;
for example, in the case of measuring the energy spectrum, a simple model could be defined
such as a power-law spectrum with normalization and spectral index parameters.
The reconstructed energy spectra depending these parameters would then be predicted,
and the values of the parameters measured by finding the
best-fit to the experimentally measured spectra.
The benefit of unfolding methods, on the other hand,
is that they allow direct comparisons of the unfolded spectra between experiments,
without restricting the measurement to a particular choice of parameterization.

In our case the reconstructed energy spectrum $M_j$
will be unfolded into a true neutrino energy spectrum, of correct flavor CC
interactions only, which we denote $N^{CC}_i$
(the binning of this unfolded spectra,
and the conversion to the actual neutrino flux values,
will be explained later in this section).
In a general sense, the relationship between the true and reconstructed spectra is
expressed by the detector response matrix $A_{ji}$ as
\begin{equation}
M_j = \sum_i A_{ji} N^{CC}_i \label{eqn:response_matrix}
\end{equation}
where $A_{ji}$ can be estimated by the detector MC, and accounts for the
inability to reconstruct perfectly the true neutrino energy.
We can write the inverse relationship using the unfolding matrix $U_{ij}$ as
\begin{equation}
N^{CC}_i = \sum_j U_{ij} M_j. \label{eqn:unfolding_matrix}
\end{equation}
However, taking $U_{ij}$ as $A^{-1}_{ji}$ is in principle a poor approach,
as the response matrix will not be estimated perfectly (and may not even be invertible).
Among the several more advanced algorithms available,
we adopt an iterative Bayesian method~\cite{D'Agostini1995,D'Agostini2010},
using the ``RooUnfold'' library~\cite{Roounfold} for the practical implementation.
The method is known as Bayesian due to its use of Bayes' theorem
in the construction of the unfolding matrix at each iterative step.
We chose this library as it is known to be reliable and easy to implement.

\begin{figure}[htbp]
\begin{center}
\includegraphics[width=0.48\textwidth]{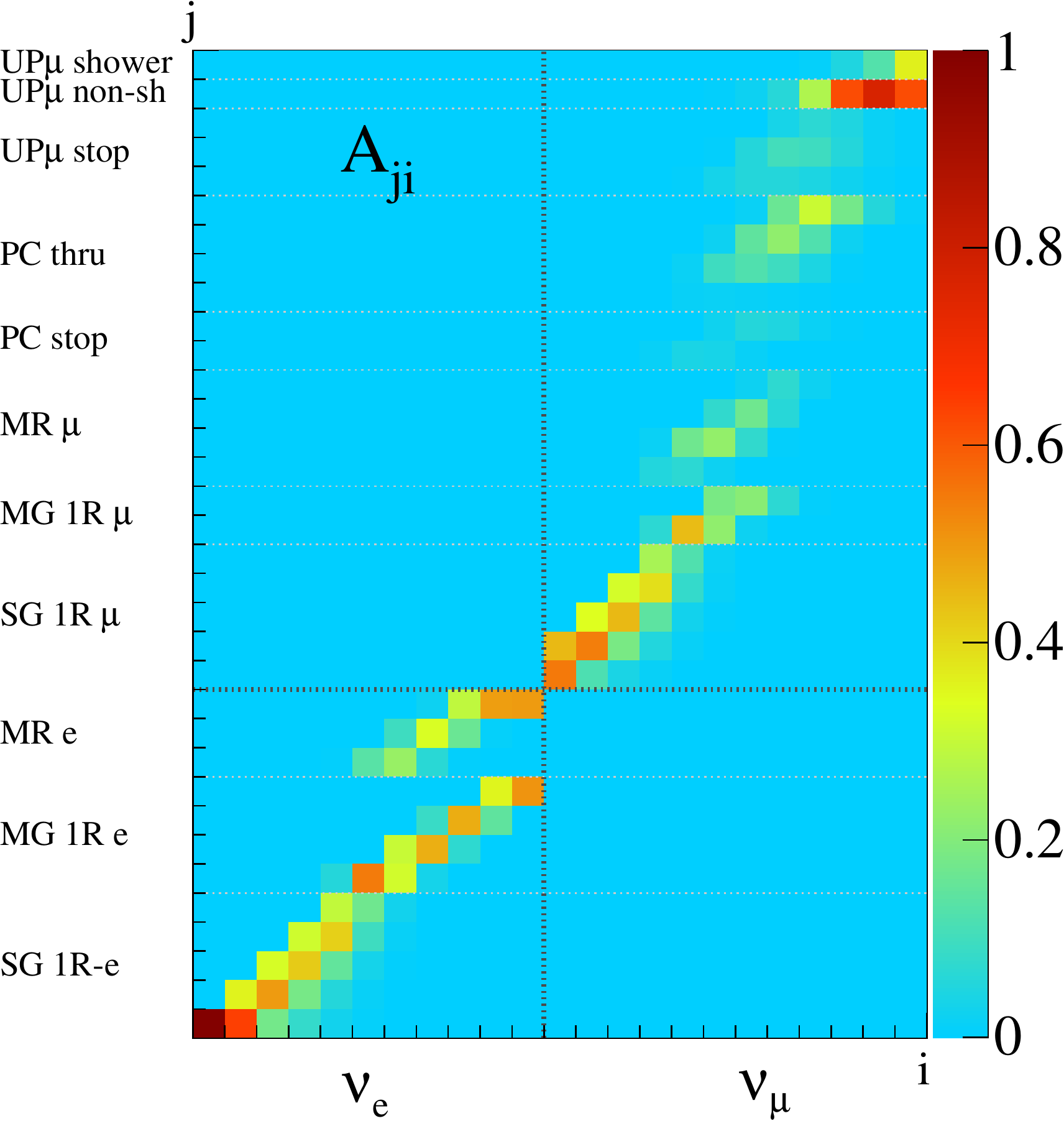}
\end{center}
\caption{ \small
(color online) Response matrix $A_{ji}$ for the SK-IV period, constructed using MC events. 
The vertical axis represents the binning by sub-sample and reconstructed energy as defined in Table~\ref{tab:event-binning}, and 
the horizontal axis represents the $\nu_e$ and $\nu_\mu$ sample binnings defined later in Table~\ref{tab:flux_table_nue_numu}.
The content of the matrix is normalized so that the sum of each column becomes unity. 
}
\label{fig:flux_response}
\end{figure}

The specific details of the iterative unfolding procedure are as follows~\cite{D'Agostini1995}.
The first estimation of the detector response matrix $A_{ji}$
is made using the SK MC dataset,
using correct flavor CC events reconstructed into the final samples.
The contributions of background events (wrong flavour or NC events) to each reconstructed
bin $j$ are also recorded, and will be accounted for in the final unfolded spectrum.
We also apply the normalization condition $\sum_j A_{ji}$=1,
which means that at this stage detector inefficiencies are not accounted for in the matrix.
Figure~\ref{fig:flux_response} shows the estimated response matrix for the SK-IV period,
not including the background events.

To construct the unfolding matrix, we first define the notation $P(j|i)$
as the probability for an event in true energy bin $i$
to be detected in the reconstructed energy bin $j$,
where these probabilities are exactly equivalent to the values of the response matrix $A_{ji}$.
We also define a set of prior probabilities $P_0(i)$
for a single event to fall in true energy bin $i$ as
\begin{equation}
P_0(i) = \frac{ N^{CC}_{MC,i}}{\sum_k N^{CC}_{MC,k}}, \label{eq:bayes1}
\end{equation}
where $N^{CC}_{MC,i}$ is the default events spectrum taken from the MC,
and the denominator ensures the probabilities sum to unity.
We can then state Bayes' theorem
\begin{equation}
P(i|j) = \frac{P(j|i)P_0(i)}{P_0(j)} \label{eq:bayes2}
\end{equation}
where
\begin{equation}
P_0(j) = \sum_i P(j|i)P_0(i). \label{eq:bayes3}
\end{equation}
Now that we have the estimated inverse probabilities $P(i|j)$,
we can use our data $M_j$ to make a \emph{first estimation}
of the true number of events as
\begin{equation}
\hat{N}^{CC}_i = \sum_j P(i|j)M_j. \label{eq:bayes4}
\end{equation}
Considering that this estimation takes inputs
from both MC and real data, it generally lies
somewhere between the default MC spectrum and the true values.
Thus, it is useful to proceed in an iterative way,
by using the normalized $\hat{N}^{CC}_i$
as new set of prior probabilities $P_0(i)$ to generate new $P(i|j)$,
and in turn iteratively update $\hat{N}^{CC}_i$.
The final iteration is denoted $N^{CC}_i$, and
the operation of the above procedure on the data
essentially takes the place of the unfolding matrix $U_{ij}$ in Equation~(\ref{eqn:unfolding_matrix}).

This unfolding method is seen to
depend on the number of iterations; a low number of iterations may
be too close to the statistically smooth prior values and not fully reflect
the information input by the data, while a high number of iterations will lead
statistical fluctuations in the data to distort the shape of the unfolded spectrum.
In general any prior will thus cause some small bias in the
unfolded spectrum which cannot be perfectly corrected for, but importantly can be accurately
estimated and included as a systematic error; this will be shown in Section~\ref{sec:flux_errors}.
In practice, according to tests with our MC data,
the iterative Bayesian method is strongly data-driven and
produces good results compared to other unfolding methods
after only a few iterations. 
We set the number of iterations
based on a study using several sets of toy MC data generated from the HKKM spectra,
by modifying the normalization and spectral index, 
and testing their reproducibility depending on the number of iterations.
Five iterations was chosen since 
the unfolded spectra were generally well reproduced and stable.



We next discuss our choice of binning for $N^{CC}_i$.
The energy range of the unfolded atmospheric neutrino flux spectrum is defined
for the $\nu_e$ flux as [-0.8,2.0] in units of $\log_{10}(E_{\nu}/\rm{GeV})$,
and divided into 11 bins of variable size;
for the $\nu_\mu$ flux, the range is [-0.6,4.0], which is divided into 12 bins.
The binning definition is shown later, together with the measurement results, in
Table~\ref{tab:flux_table_nue_numu}.

These energy ranges and bin widths are determined considering the neutrino true energy coverage and 
resolution of the flux data samples. 
Smaller bins of width 0.2 in $\log_{10}(E_\nu/\rm{GeV})$
are adopted below 10~GeV due to the relatively finer
energy resolution of the FC sample, 
and wider bins are adopted for higher energies due to 
the deterioration of the energy resolution in the PC and UPMU samples. 
The global flux binning containing all $\nu_e$ and $\nu_\mu$ bins is denoted as $i$$=$1...23.

We obtain the measured flux values $\Phi_i$ 
observed at the detector position from the predicted $\nu_\alpha + \bar{\nu}_\alpha$ flux values $\Phi^{\nu_\alpha}_{MC}(\bar{E}_i)$
(where $\alpha=e$,$\mu$), 
the expected number of CC interactions $N^{CC}_{MC,i}$,
and the number of CC events obtained by unfolding the data ${N}^{CC}_i$.
Considering the difference in detector configuration by each operational period,
we perform an unfolding separately for each period denoted by the subscript $SK$, so
\begin{equation}
 \Phi_i = \Phi^{\nu_\alpha}_{MC}(\bar{E}_i) \cdot \frac{ \sum^{SK} {N}^{CC}_{i,SK} } { \sum^{SK} N^{CC}_{MC,i,SK} } 
 \label{eqn:unfolding_by_skperiod}
\end{equation}
where $\sum^{SK}$ is a sum over the four SK periods.
The predicted flux values
$\Phi^{\nu_\alpha}_{MC}(\bar{E}_i)$ are themselves calculated as follows.
First, the predicted unoscillated $\nu_\alpha$ and $\bar{\nu}_\alpha$ differential fluxes
$\phi^{\nu_\alpha,\bar{\nu}_\alpha}_{MC}(E_\nu,\theta_z,\phi_A)$
as a function of neutrino energy $E_\nu$, and
zenith and azimuthal angles of arrival direction $\theta_z$ and $\phi_A$,
are calculated by interpolating the tabulated HKKM11 flux in energy and direction as described in \cite{honda2011}.
The predicted oscillated flux at the detector position $\Phi^{\nu_\alpha}_{MC}(E_{\nu})$
is then calculated by integrating the differential flux 
over a 4$\pi$ solid angle as
\begin{widetext}
\begin{equation}
\Phi^{\nu_\alpha}_{MC}(E_\nu) = \int_{4\pi} d\Omega \sum_{\beta}^{e,\mu} \left\{ \phi^{\nu_\beta}_{MC}(E_\nu,\theta_z,\phi_A) O^{\nu_\beta \to \nu_\alpha}(E_\nu,\theta_z,\boldsymbol{\theta}_O) +  \phi^{\bar{\nu}_\beta}_{MC}(E_\nu,\theta_z,\phi_A) O^{\bar{\nu}_\beta \to \bar{\nu}_\alpha}(E_\nu,\theta_z,\boldsymbol{\theta}_O) \right\} \label{eqn:flux_integ_1} \\
\end{equation}
\end{widetext}
where $O^{\nu_\beta \to \nu_\alpha}(E_\nu,\theta_z,\boldsymbol{\theta}_O)$
is the $\nu_\beta$ $\to$ $\nu_\alpha$ oscillation probability 
from the production point to the detector, 
calculated with the standard three flavor oscillation model with parameters $\boldsymbol{\theta}_O$ 
as shown in Table~\ref{tab:osc-param}
and including the matter effect when propagating inside the Earth.
Finally, the predicted flux values at the mean energy of the $i$-th energy bin $\bar{E}_i$ 
are calculated according to \cite{Lafferty:1994cj}, using 
and the lower and upper edges of the $i$-th energy bin $E_{i}$ and $E_{i+1}$, as
\begin{equation}
\Phi^{\nu_\alpha}_{MC}(\bar{E}_i) = \frac{1}{E_{i+1}-E_{i}} \int_{E_{i}}^{E_{i+1}} \Phi^{\nu_\alpha}_{MC}(E_\nu) dE_\nu, \label{eqn:mean_energy}
\end{equation}
which are the values appearing in Equation~\ref{eqn:unfolding_by_skperiod}.

\begin{figure}[htbp]
\begin{center}
\begin{minipage}{\columnwidth}
\includegraphics[width=\columnwidth]{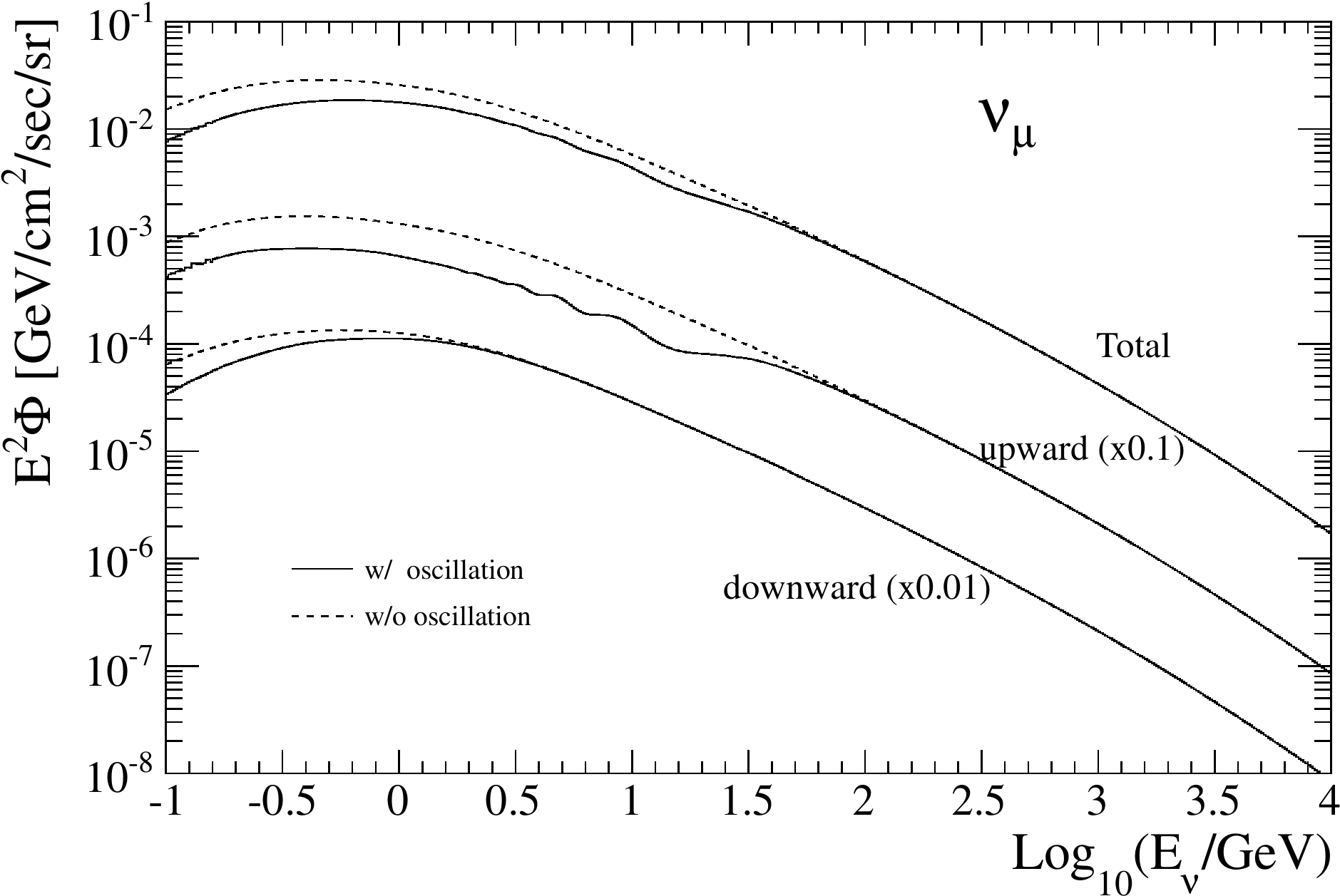}
\end{minipage}
\end{center}
\caption{ \small
Calculated $\nu_\mu$ flux in Equation~(\ref{eqn:flux_integ_1}) at the SK site,
based on the HKKM11 flux model with and without applying neutrino oscillation. 
The calculations of only upward-going ($\cos\theta_z$$<$0) and 
downward-going ($\cos\theta_z$$>$0) are also shown separately with 
their normalizations scaled by $\times$0.1 and $\times$0.01. 
}
\label{fig:hkkm11_numu_updown}
\end{figure}

For comparison, the predicted neutrino fluxes without oscillation are also calculated by replacing 
$O^{\nu_\beta \to \nu_\alpha}(E_\nu,\theta_z,\boldsymbol{\theta}_O)=0 (1)$ for $\alpha\neq\beta$  
($\alpha=\beta$) in Equation~(\ref{eqn:flux_integ_1}).
Figure~\ref{fig:hkkm11_numu_updown} shows the comparison between the 
predicted $\nu_\mu$ flux with and without oscillation,
also showing the upward-going and downward-going flux components separately.
It can be seen that the oscillation effect is different between the upward and downward fluxes,
since the neutrino path length is not a linear function of the zenith angle and changes 
rapidly near the horizontal direction.

%
%
%
%


\begin{table}
\begin{center}
\begin{tabular}{lcc}
\hline \hline
Parameter   &  Nominal &    Uncertainty \\ 
\hline
$\Delta m^2_{21}$ &  7.50$\times$10$^{-5}$  & (7.30-7.70)$\times$10$^{-5}$ \\
$|\Delta m^2_{32}|$ &  2.32$\times$10$^{-3}$  &  (2.20-2.44)$\times$10$^{-5}$ \\ 
$\sin^2\theta_{12}$ & 0.31 &   0.296-0.333 \\
$\sin^2\theta_{23}$ & 0.50 &   0.388-0.612  \\
$\sin^2\theta_{13}$ & 0.024 &  0.0217-0.0263  \\
$\delta_{CP}$       & 0   &  0-2$\pi$ \\
sign of $\Delta m^2_{32}$    &  $>$0 &  Both $>$0 and $<$0 \\ 
\hline
\hline
\end{tabular}
\caption{ \small 
Nominal value and 1~$\sigma$ uncertainty (except for CP-violation phase and mass hierarchy) 
of the oscillation parameters based on \cite{PDG2014}.
The unit of mass square difference is eV$^2$.
The probability density distributions of the error are assumed to be Gaussian distributions
while the uniform probability is considered for $\delta_{CP}$ and mass hierarchy.
The correlations among these parameters are not taken into account.
}
\label{tab:osc-param}
\end{center}
\end{table}

%
%
\subsection{\label{sec:flux_errors}Systematic Uncertainties}

The systematic uncertainties on the number of events in each data bin $M_j$,
which we define to include the uncertainties on the MC prediction, are estimated as follows.
Using a systematic error database maintained by the SK collaboration, 
the effects of errors relating to the event selection and reconstruction in the detector (116 errors),
and those relating to neutrino interaction (17 errors) can be calculated.
For the detector related errors, 
the same 29 systematic uncertainty sources described in \cite{Wendell} are considered and estimated for 
each SK period, giving 116 total errors.
Their effects are evaluated as systematic error coefficients $f_{jk}$,
representing the fractional shift 
in each data bin $M_j$ resulting from a 1~$\sigma$ shift of the $k$-th systematic error source.
The modified expectation of the number of events in each bin, summing over each error, $\tilde{M}_j$ , 
is thus
\begin{eqnarray}
\tilde{M}_j(\mathbf{g}) = M_{MC,j} \times \left( 1 + \sum_k^{N_{\rm{sys}}} f_{jk} g_k \right) 
\end{eqnarray}
where $M_{MC,j}$ are the nominal MC expectations without any systematic effect,
$N_{\rm{sys}}$ is the number of systematic error sources, 
and $\mathbf{g}$=($g_1$,$\cdots$,$g_{N_{\rm{sys}}}$ ) represents the applied strength of each systematic in units of $\sigma$.
The effects of the oscillation parameters are treated separately, being directly calculated
by shifting their values in Equation~(\ref{eqn:flux_integ_1}). 
The uncertainties of the oscillation parameters are shown in Table~\ref{tab:osc-param}.

%
%
For propagation of both the systematic and statistical uncertainties
from $M_j$ to the flux values $N^{CC}_i$ we employ a toy MC method,
since while RooUnfold provides an accurate theoretical calculation
of the statistical errors~\cite{Roounfold},
simultaneous treatment of the systematic uncertainties within the software is not implemented.
In this method, toy data sets are generated by randomly fluctuating MC data
according to their systematic and statistical error PDFs, as follows:
first systematically-shifted data $\tilde{M}_j(\mathbf{g})$ are generated
using a random set of $N_{\rm{sys}}$ systematic strengths $\mathbf{g}$,
where the probability density function (PDF) for each $g_k$ parameter is a Gaussian distribution, 
and that there is no correlation between the $g_k$ parameters. 
The final toy data $M^{\rm{toy}}_j$ is then generated by a Poisson distribution
with the mean values of $\tilde{M}_j(\mathbf{g})$.
Two thousand sets of toy data are generated and
analyzed with the same analysis method as described in the previous section.
The variations among these toy flux measurements are taken as the uncertainties,
including both statistical and systematic uncertainties,
and the covariances of the unfolded flux among the energy bins are also recorded into 
the covariance matrix, which will be used for the $\chi^2$ calculation in Section~\ref{sec:flux_results}.
The method is also repeated for statistical errors only,
to allow separate calculation of the statistical and systematic components.

\begin{figure*}[htbp]
\begin{center}
\begin{minipage}{0.9\columnwidth}
\includegraphics[width=0.9\columnwidth]{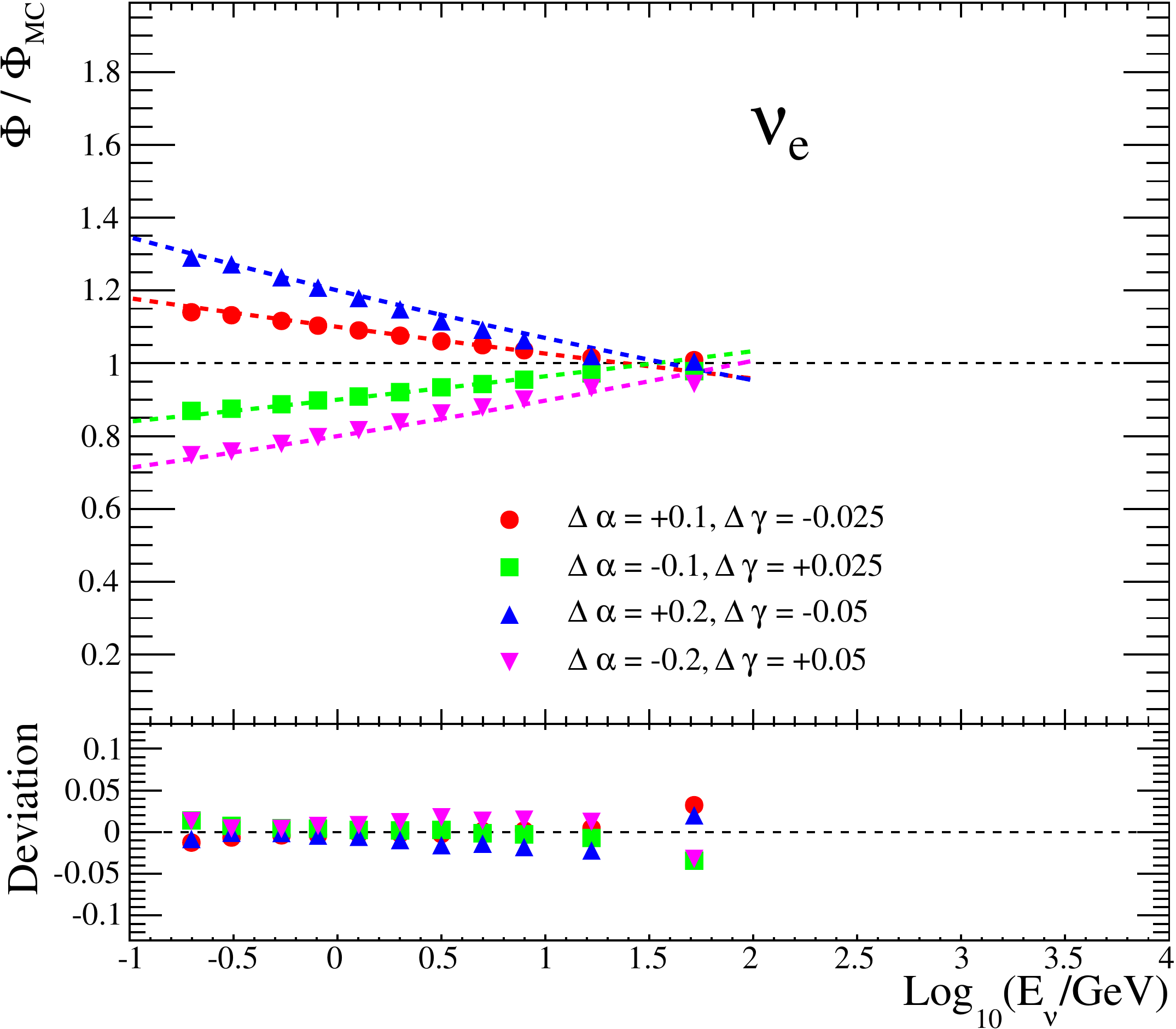}
\end{minipage}
\begin{minipage}{0.9\columnwidth}
\includegraphics[width=0.9\columnwidth]{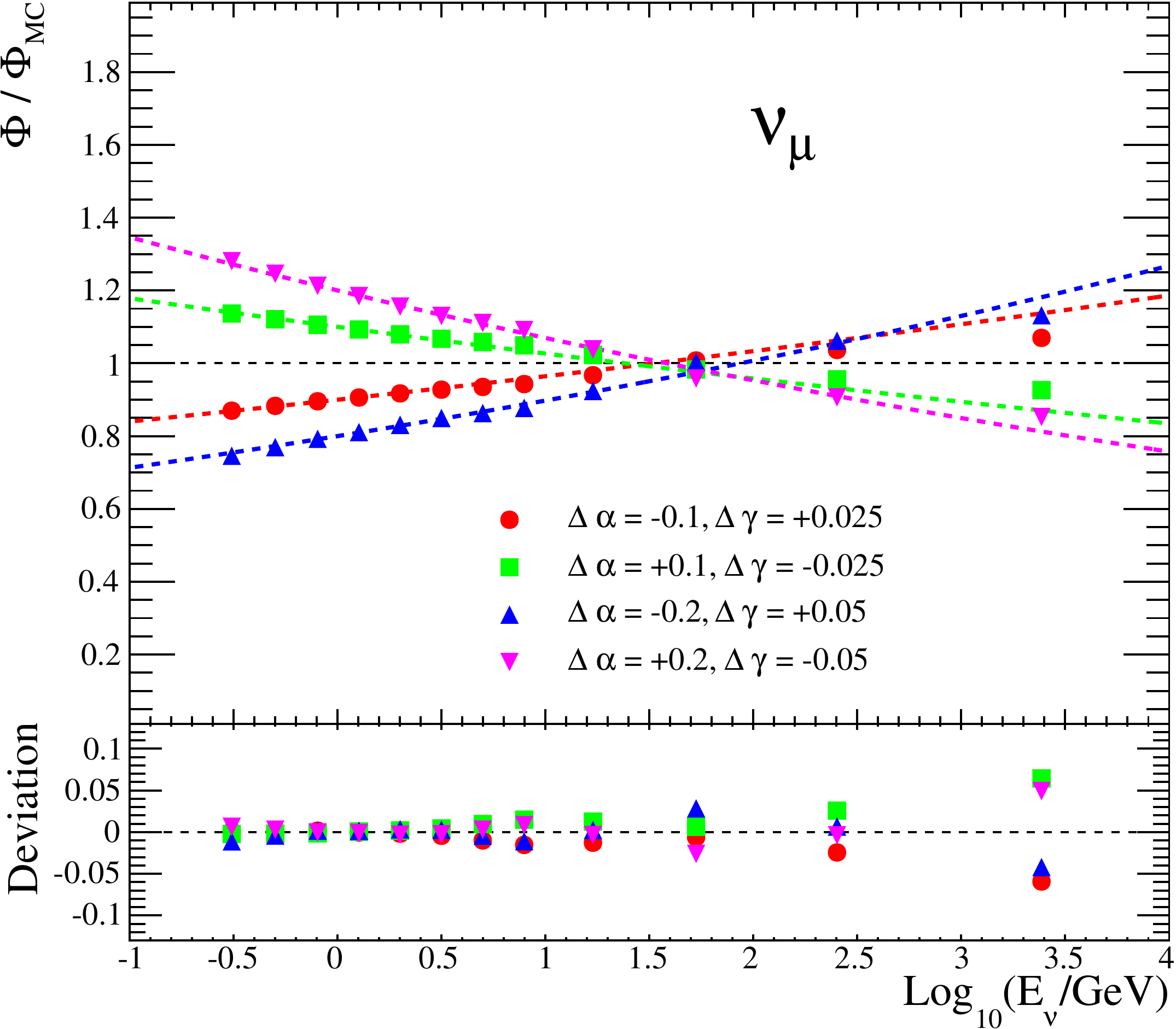}
\end{minipage}
\end{center}
\caption{ \small
(color online) Pseudo-data input and the unfolded energy spectra (upper) and their ratio (lower) for $\nu_e$ (left) and $\nu_\mu$ (right), respectively.
Vertical axis is  the ratio to HKKM11~\cite{honda2011} ($\Phi_{MC}$) in the upper figure.
Four sets of pseudo-data inputs, where the deviations of flux normalization 
and spectral index ($\Delta \alpha$, $\Delta \gamma$) = ($\pm$10\%,$\pm$0.025), ($\pm$20\%,$\pm$0.05), 
are tested.
}
\label{fig:flux_bias_study}
\end{figure*}

%
%
A final contribution to the uncertainty in the values $N^{CC}_i$ comes from the regularization error,
i.e. the bias that may be caused by inaccuracy in the initial estimation of the response matrix
and Bayesian prior.
As mentioned in section~\ref{sec:flux_unfolding}, such a bias should be small, but it may be noticeable
in unfolded bins with low statistics. This bias cannot be corrected for exactly, but we may
estimate an associated error by unfolding pseudo-data sets with an energy spectrum
reasonably far from the MC prediction, and observing the resulting difference.
Such pseudo-data sets were produced from the MC data
by re-weighting events to define a modified energy spectrum $\Phi^{\prime}_{MC,i}$ as
\begin{equation}
\Phi^{\prime}_{MC,i} =  (1+\Delta \alpha)\left(\frac{\bar{E}_i}{1~\rm{GeV}}\right)^{\Delta \gamma}\Phi_{MC,i}
    \label{eqn:modify_spectrum}
\end{equation}
where $\Delta \alpha$ and $\Delta \gamma$ represent the deviation of
the normalization and spectral index from each model,
and are modified in the ranges of $\pm$0.05 and $\pm$20\% respectively,
and $\Phi_{MC,i}$ is shorthand for the predicted flux values $\Phi_{MC}(\bar{E}_i)$.
In Fig.~\ref{fig:flux_bias_study} the comparisons of
each input pseudo-data set and its unfolded flux spectra are shown,
along with the fractional deviation between the input and unfolded output in each case.
The difference becomes largest at the highest energy bin for both $\nu_e$ and $\nu_\mu$ spectra,
and the error sizes are $\pm$6\% and $\pm$8\% at most in the case of a $\pm$0.05 and $\pm$20\% change. 
These differences are taken as the estimated uncertainties due to regularization in the unfolded
energy spectrum, and included in the systematic uncertainties. 

\begin{figure}[htbp]
\begin{center}
\begin{minipage}{\columnwidth}
\includegraphics[width=\columnwidth]{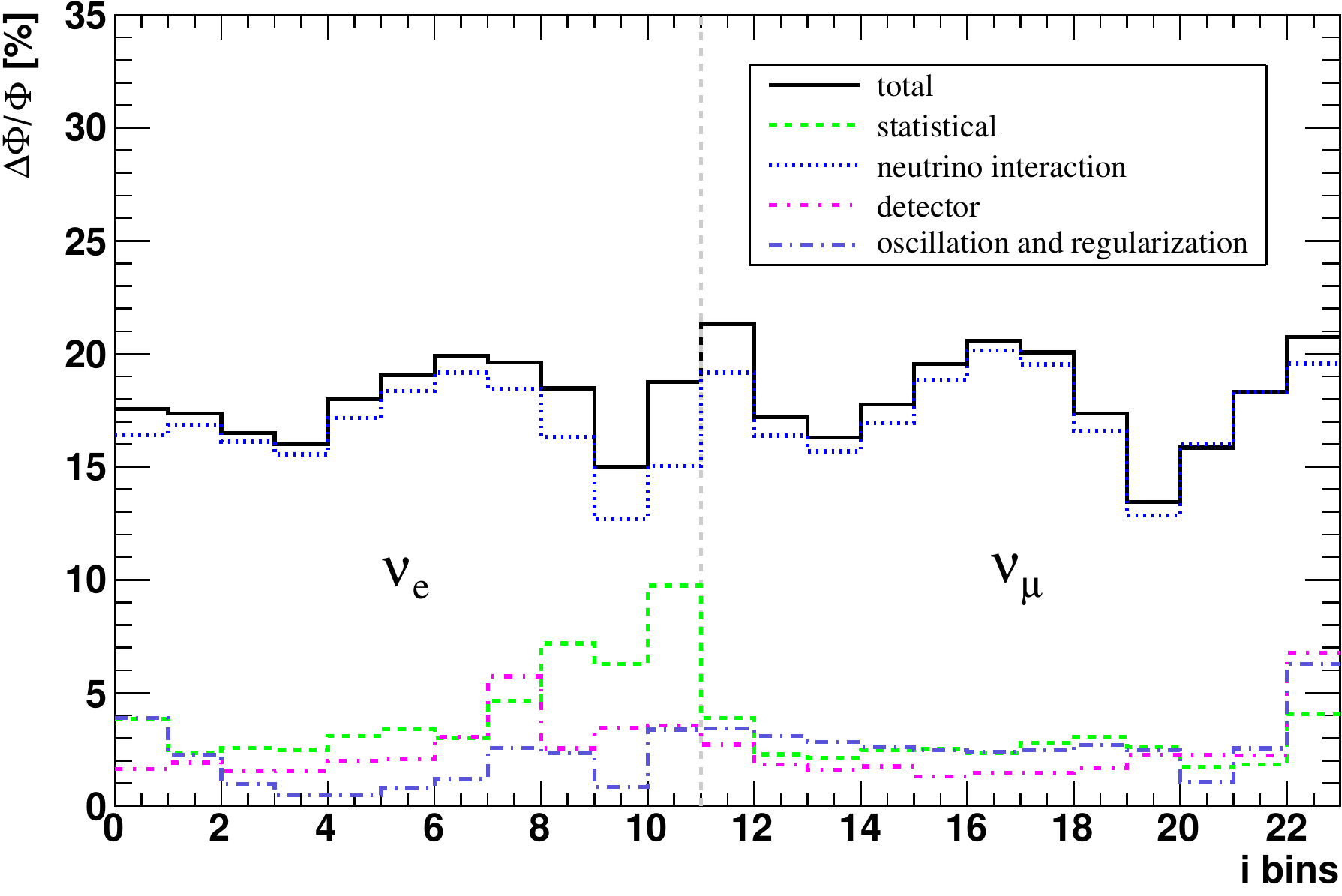}
\end{minipage}
\end{center}
\caption{ \small
(color online) Breakdown of the uncertainties in measured flux estimated for each error group,
showing neutrino cross section, detector response, and neutrino oscillation $+$ regularization. 
}
\label{fig:syst_breakdown}
\end{figure}

%
%
Figure~\ref{fig:syst_breakdown} shows the breakdown of the estimated uncertainties
for each error source group. The absolute errors of four groups
(statistical, neutrino interaction, detector response, and neutrino oscillation $+$ regularization),
are individually calculated and compared.
Currently the neutrino interaction errors are the dominant source of systematic uncertainty.

%
%

\subsection{\label{sec:flux_results}Results and Discussions}

\begin{figure*}[htbp]
\begin{center}
\includegraphics[width=0.6\textwidth]{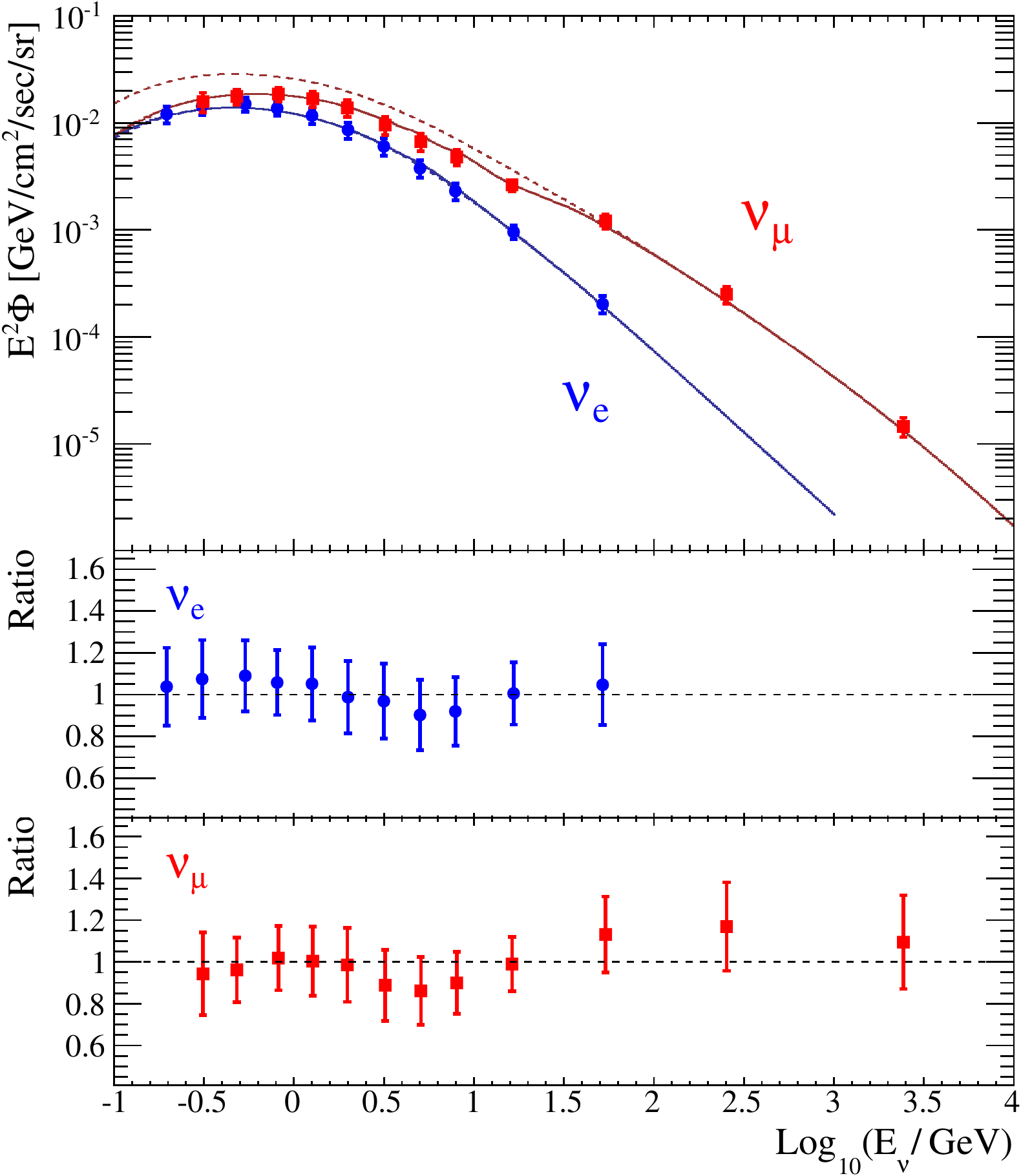}
\end{center}
\caption{ \small
(color online) Measured flux spectra, using all SK I-IV data, for $\nu_e$ (blue points) and 
$\nu_\mu$ (red points) as a function of the logarithm of the neutrino energy. 
Error bars correspond to the uncertainties, including all statistical and 
systematic errors.
The $\nu_e$ (blue line) and $\nu_\mu$ (red line) flux curves show the HKKM11~\cite{honda2011} model
with (solid) and without (dashed) neutrino oscillation.
In the lower part of the figure, the data / HKKM11 ratio is shown.
}
\label{fig:flux}
\end{figure*}

Figure~\ref{fig:flux} shows the obtained $\nu_e$ and $\nu_\mu$ energy spectra using all SK~I-IV data.
The energy binning, mean energy, measured flux, and error are also described in 
Table~\ref{tab:flux_table_nue_numu} for $\nu_e$ and $\nu_\mu$.
The measured energy spectrum agrees with the oscillated HKKM11 flux within the estimated uncertainties,
which as mentioned above are dominated by neutrino interaction uncertainties.
The unoscillated flux is also plotted, such that
the deficit of $\nu_\mu$ flux due to neutrino oscillation becomes apparent below 100~GeV.

\begin{table}
\begin{center}
\scalebox{0.9}{
\begin{tabular}{r c c c c c}
\hline
\hline
$i$ & {\small $\log_{10}(E/\rm{GeV})$} & ~~{\small $\log_{10}(\bar{E}_i/\rm{GeV})$}  &&   $\bar{E}_i^2\Phi^{\nu}_{i}$ &  $\Delta \Phi^{\nu}_i/\Phi^{\nu}_i$  \\ 
& & &&   {\small [GeV/cm$^{2}$/sec/sr]}  & {\small [\%]}  \\
\hline
\multicolumn{1}{c}{$\nu_e$} & & & & & \\ 
~~1 & -0.8 -- -0.6 &  -0.71 &&               1.21$\times$10$^{-2}$  & $\pm$18  \\
~~2 & -0.6 -- -0.4 &  -0.51 &&               1.46$\times$10$^{-2}$  & $\pm$17  \\
~~3 & -0.4 -- -0.2 &  -0.27 &&               1.50$\times$10$^{-2}$  & $\pm$16  \\
~~4 & -0.2 -- 0.0  &  -0.09 &&               1.37$\times$10$^{-2}$  & $\pm$15  \\
~~5 &  0.0 -- 0.2  &   0.10 &&               1.16$\times$10$^{-2}$  & $\pm$17  \\
~~6 &  0.2 -- 0.4  &   0.30 &&               8.55$\times$10$^{-3}$  & $\pm$17  \\
~~7 &  0.4 -- 0.6  &   0.50 &&               6.09$\times$10$^{-3}$  & $\pm$18  \\
~~8 &  0.6 -- 0.8  &   0.70 &&               3.73$\times$10$^{-3}$  & $\pm$19  \\
~~9 &  0.8 -- 1.0  &   0.90 &&               2.32$\times$10$^{-3}$  & $\pm$18  \\
~~10 &  1.0 -- 1.5  &   1.22 &&               9.42$\times$10$^{-4}$  & $\pm$15  \\
~~11 & 1.5 -- 2.0  &   1.72 &&               2.03$\times$10$^{-4}$  & $\pm$18  \\
\hline 
\multicolumn{1}{c}{$\nu_\mu$}  & & & & & \\ 
~~12 & -0.6 -- -0.4 &  -0.51 &&               1.58$\times$10$^{-2}$  & $\pm$21  \\
~~13 & -0.4 -- -0.2 &  -0.32 &&               1.77$\times$10$^{-2}$  & $\pm$16  \\
~~14 & -0.2 --  0.0 &  -0.09 &&               1.86$\times$10$^{-2}$  & $\pm$15  \\
~~15 &  0.0 --  0.2 &   0.10 &&               1.68$\times$10$^{-2}$  & $\pm$16  \\
~~16 &  0.2 --  0.4 &   0.30 &&               1.38$\times$10$^{-2}$  & $\pm$18  \\
~~17 &  0.4 --  0.6 &   0.51 &&               9.59$\times$10$^{-3}$  & $\pm$19  \\
~~18 &  0.6 --  0.8 &   0.71 &&               6.68$\times$10$^{-3}$  & $\pm$19  \\
~~19 &  0.8 --  1.0 &   0.90 &&               4.79$\times$10$^{-3}$  & $\pm$17  \\
~~20 &  1.0 --  1.5 &   1.21 &&               2.62$\times$10$^{-3}$  & $\pm$13  \\
~~21 &  1.5 --  2.0 &   1.73 &&               1.20$\times$10$^{-3}$  & $\pm$16  \\
~~22 &  2.0 --  3.0 &   2.40 &&               2.49$\times$10$^{-4}$  & $\pm$18  \\
~~23 &  3.0 --  4.0 &   3.39 &&               1.46$\times$10$^{-5}$  & $\pm$21   \\
\hline \hline
\end{tabular}
}
\caption{ \small 
Electron and muon neutrino fluxes measured by SK I-IV data. 
Error written in percentage including both statistical and systematic uncertainties.
}
\label{tab:flux_table_nue_numu}
\end{center}
\end{table}

\begin{figure*}[htbp]
\begin{center}
\includegraphics[width=0.6\textwidth]{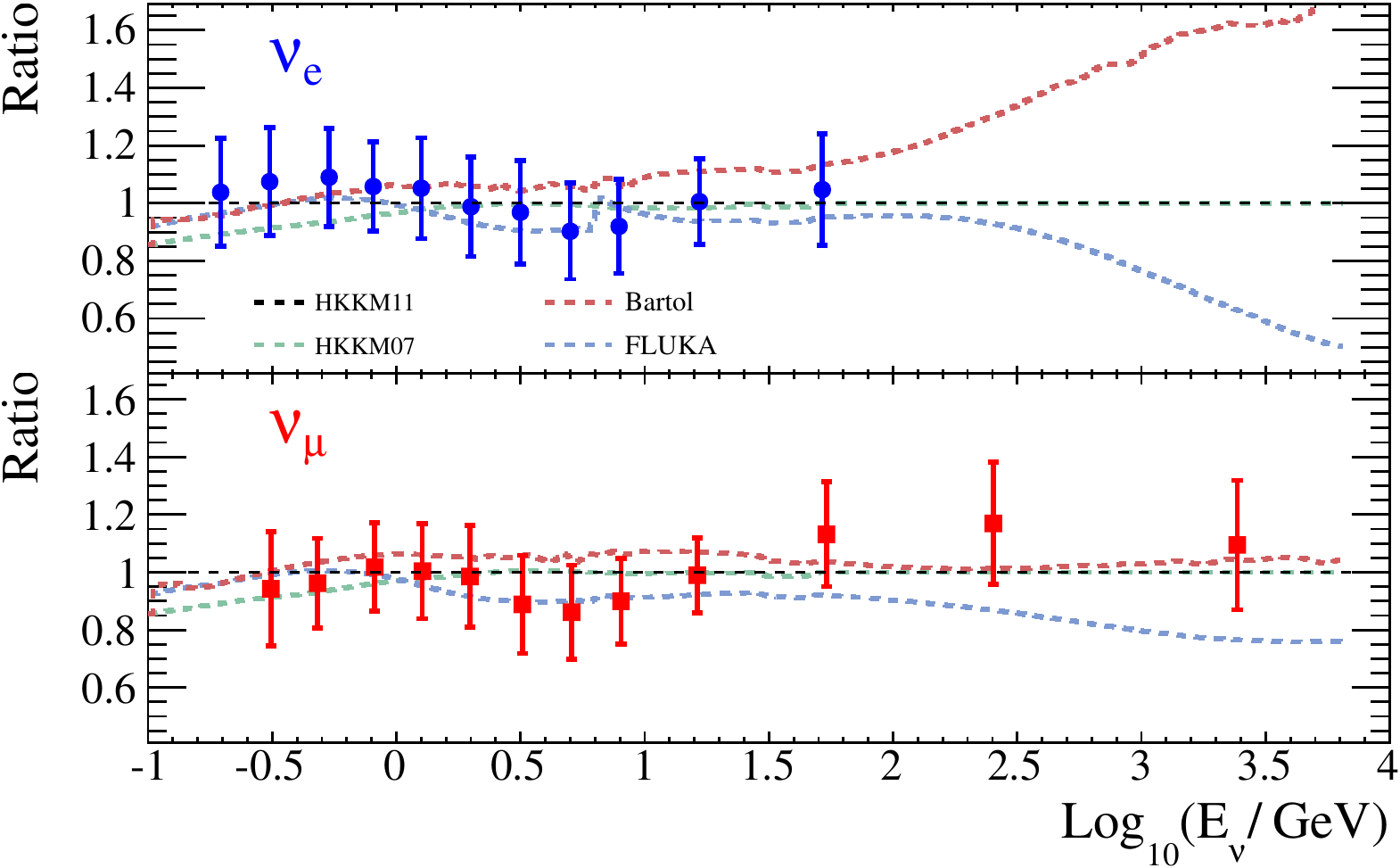}
\end{center}
\caption{ \small
(color online) Measured energy spectra compared to the flux model predictions (dotted lines) from the
HKKM11~\cite{honda2011} (black), HKKM07~\cite{honda06} (light green), 
Bartol~\cite{bartol}, and FLUKA~\cite{fluka} models.
The vertical axis represents the flux ratio from the data (or each model) to HKKM11.
Error bars include both statistical and systematic uncertainties.
}
\label{fig:flux_ratio_with_models}
\end{figure*}

\begin{figure}[htbp]
\begin{center}
\includegraphics[width=0.73\columnwidth]{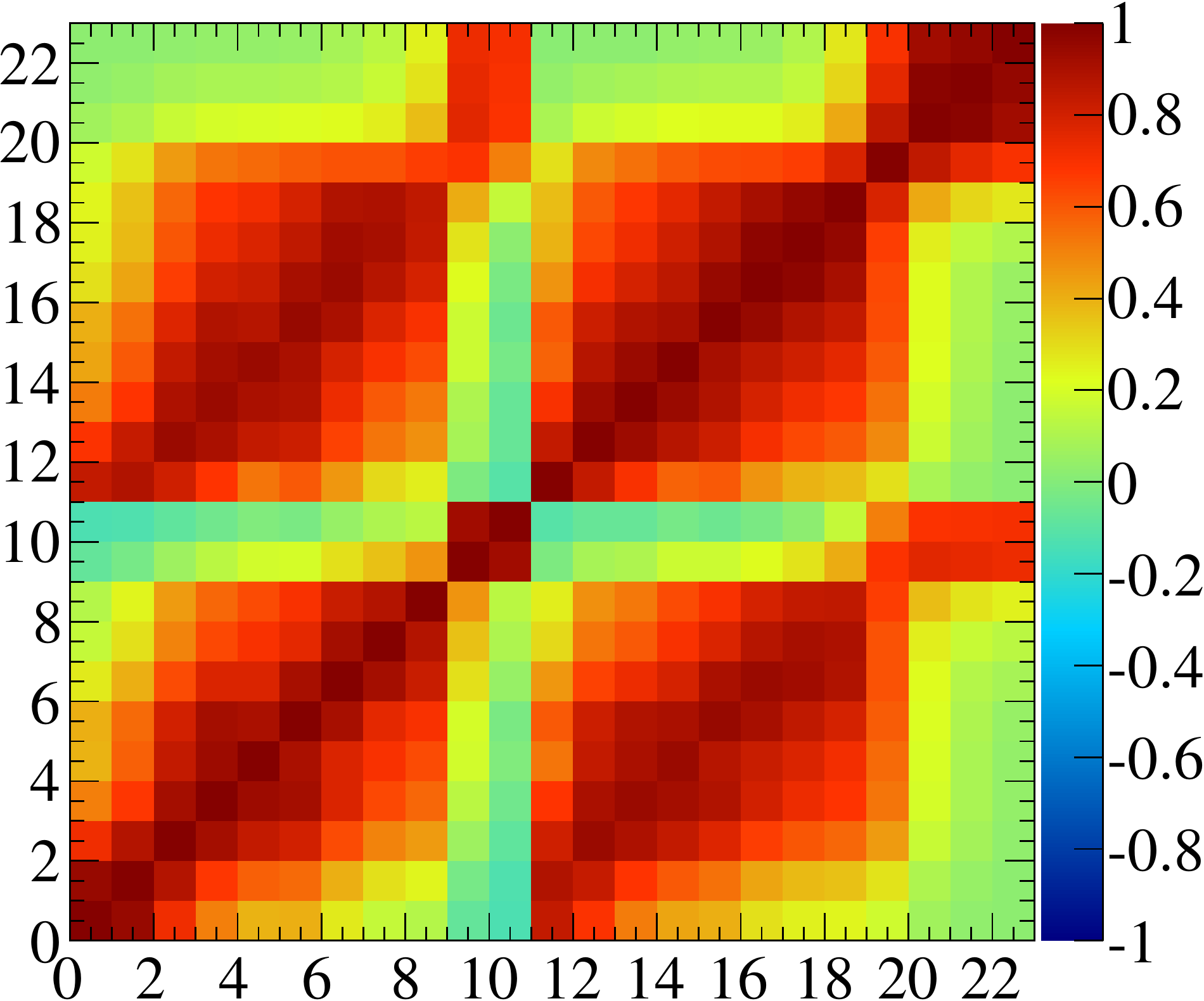}
\end{center}
\caption{ \small
(color online) Error correlation matrix, $\hat{C}_{ij}$, of the observed spectrum bins.
Used for the $\chi^2$ calculation.
The matrix contains both $\nu_e$ and $\nu_\mu$, with
$\nu_e$ ranging from the 1st to the 11th bin, and 
$\nu_\mu$ from the 12th to the 23rd bin.
Strong correlations are seen between $\nu_e$ and $\nu_\mu$
as many systematics apply to both flavors in similar ways.
}
\label{fig:flux_cormat}
\end{figure}

%
%
The observed fluxes are compared to several flux models, including
HKKM11~\cite{honda2011}, HKKM07~\cite{honda06}, FLUKA~\cite{fluka}, and Bartol~\cite{bartol}, 
in Fig.~\ref{fig:flux_ratio_with_models}.
In order to perform a quantitative comparison, $\chi^2$ values are calculated between 
the observed energy spectrum and the flux model predictions
while taking into account the error correlations between the energy bins:
\begin{equation}
\chi^2 = \sum_i^{N} \sum_j^{N} \left(\Phi_i - \Phi_{MC,i})^T C_{ij}^{-1} (\Phi_j - \Phi_{MC,j}\right) \label{eqn:chi2} 
\end{equation}
where $N$ is the number of the data bins,
$\Phi_i$ and $\Phi_{MC,i}$ are the observed flux and
the expectation of the chosen flux model at the $i$-th energy bin respectively,
and $C_{ij}$ is the covariance matrix of the observed spectrum $\Phi_i$ (which is calculated by 
observing the correlation of the error fluctuations for $N^{CC}_i$ within the set of toy MCs
described in Section~\ref{sec:flux_errors}).
Figure~\ref{fig:flux_cormat} shows the correlation matrix, where each element 
$\hat{C}_{ij}$ is obtained by normalizing the diagonal term of the covariance matrix 
$C_{ij}$ to one, i.e. 
$\hat{C}_{ij} = C_{ij} / (\Delta \Phi_i \Delta \Phi_j)$,
where $\Delta \Phi_i$ is the error of the observed flux at $i$-th bin shown in 
Table~\ref{tab:flux_table_nue_numu}.
The value of each element of the correlation matrix is given in Appendix~\ref{sec:corr_matrix}.
Both statistical and systematic uncertainties are taken into account in the $\chi^2$ calculation.

The calculation is performed for these three cases; both $\nu_e$ and $\nu_\mu$ (degrees-of-freedom $N=23$),
$\nu_e$ only ($N=11$), and $\nu_\mu$ only ($N=12$), and
the obtained $\chi^2$s for each flux model are shown in Table~\ref{tab:chi2_model}.
The test including $\nu_e$ and $\nu_\mu$ together takes the systematic correlations between
the two flavours into account.
In this test the $\chi^2$ values are not strongly inconsistent
for any of the flux models, though
HKKM11 has the smallest values among these models;
the $p$-values are $p$=0.51 for HKKM11, $p$=0.32 for FLUKA and $p$=0.13 for Bartol.
As HKKM11 is preferred above HKKM07, we see that
the updates to the hadron simulation~\cite{honda2011} in the HKKM11 model
(which cause changes in the energy region below 1~GeV)
seem to bring better agreement.
For the individual tests, both $\nu_e$ and $\nu_\mu$ agree well.

\begin{figure}[htbp]
\begin{center}
\includegraphics[width=0.82\columnwidth]{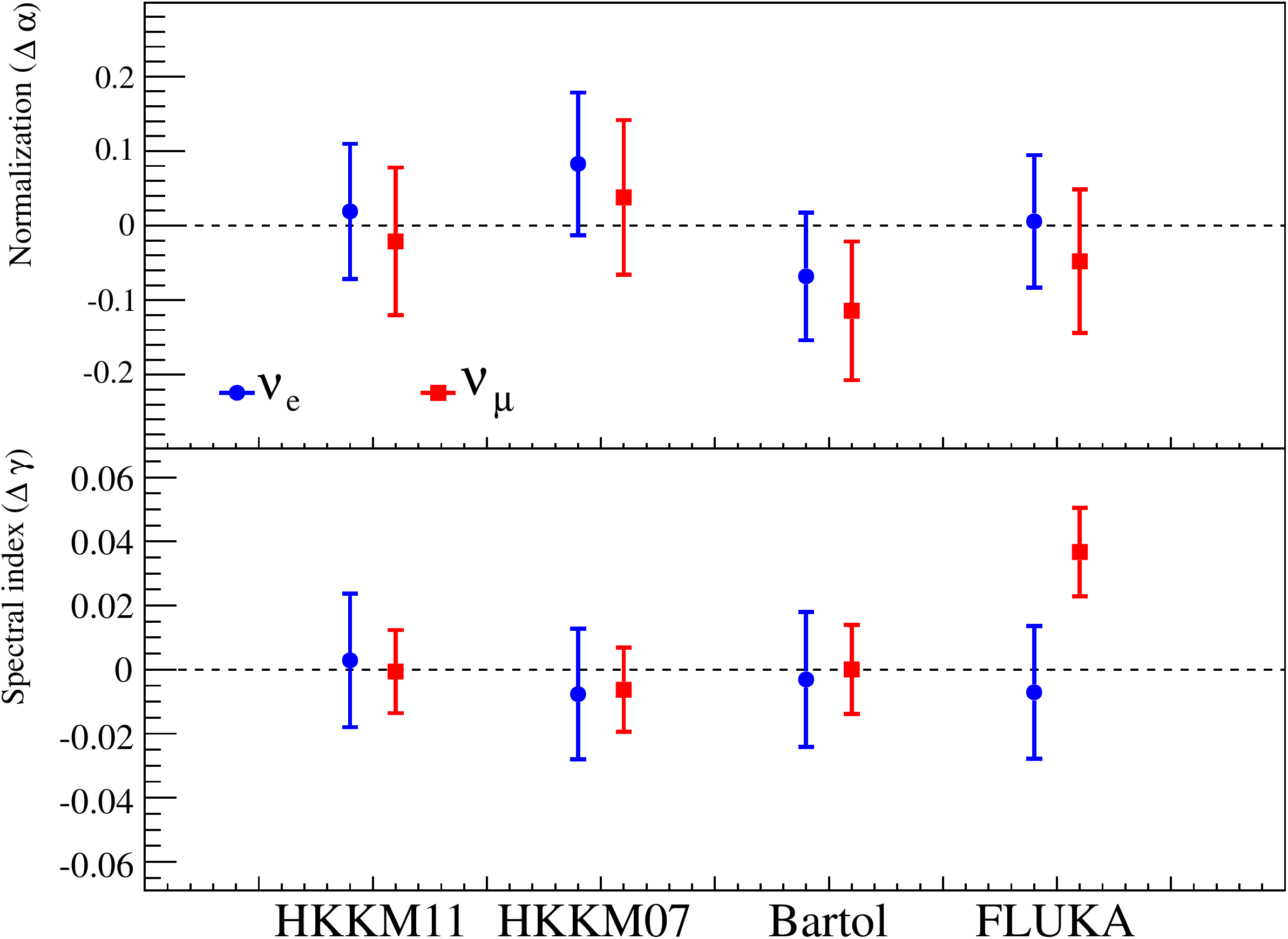}
\end{center}
\caption{ \small
(color online) Best-fit and 1~$\sigma$ error of $\Delta \alpha$ and $\Delta \gamma$ parameters 
for each flux model by $\chi^2$ calculation, 
representing the deviation in normalization and spectral index from the flux prediction, respectively.
}
\label{fig:flux_specfit}
\end{figure}

We also evaluate each model against the data by fitting an energy spectrum with variable parameters, 
$\Delta\alpha$ and $\Delta\gamma$, which represent the deviations
in normalization and spectral index from each model respectively,
in the same way as was defined in Equation~(\ref{eqn:modify_spectrum}).
The $\chi^2$ statistic is defined by Equation~(\ref{eqn:chi2}) with $\Phi_{MC,i}$ replaced with $\Phi^{\prime}_{MC,i}$,
and the best-fit values and 1~$\sigma$ errors of $\Delta \alpha$ and $\Delta \gamma$ parameters
are obtained by minimizing the $\chi^2$.
Figure~\ref{fig:flux_specfit} shows the result for each flux model case.
The normalization and spectral index agrees within the 1~$\sigma$ error for every model,
except from the fitted spectral index of FLUKA $\nu_\mu$ which deviates by 2.7~$\sigma$. 

%
%
\begin{table}
\begin{center}
\begin{tabular}{l c c c c}
\hline
\hline
         & &  \multicolumn{3}{c}{$\chi^2$} \\ 
~~~Flux model & &  ~~~$\nu_e$ and $\nu_\mu$~~~  & ~~~$\nu_e$ only~~~ & ~~~$\nu_\mu$ only~~~~~ \\ 
\hline
~~~HKKM11~\cite{honda2011} &&  22.2 & 5.3  & 12.2 \\   
~~~HKKM07~\cite{honda06} &&  22.5  & 6.8 & 12.1  \\
~~~Bartol~\cite{bartol} &&     30.7 & 6.6 & 17.0  \\
~~~FLUKA~\cite{fluka}  &&     25.6 & 5.4  & 15.2  \\
~~~(~~DOF &&  23  & 11  &  12~~)\\ 
\hline
\hline
\end{tabular}
\caption{ \small 
$\chi^2$ values calculated by testing the measured flux against each flux model prediction
according to Equation~(\ref{eqn:chi2}).
The number of degrees-of-freedom (DOF) in each test is also shown.
}
\label{tab:chi2_model}
\end{center}
\end{table}

%
%

%
%

\begin{figure}[htbp]
\begin{center}
\begin{minipage}{\columnwidth}
\includegraphics[width=0.9\columnwidth]{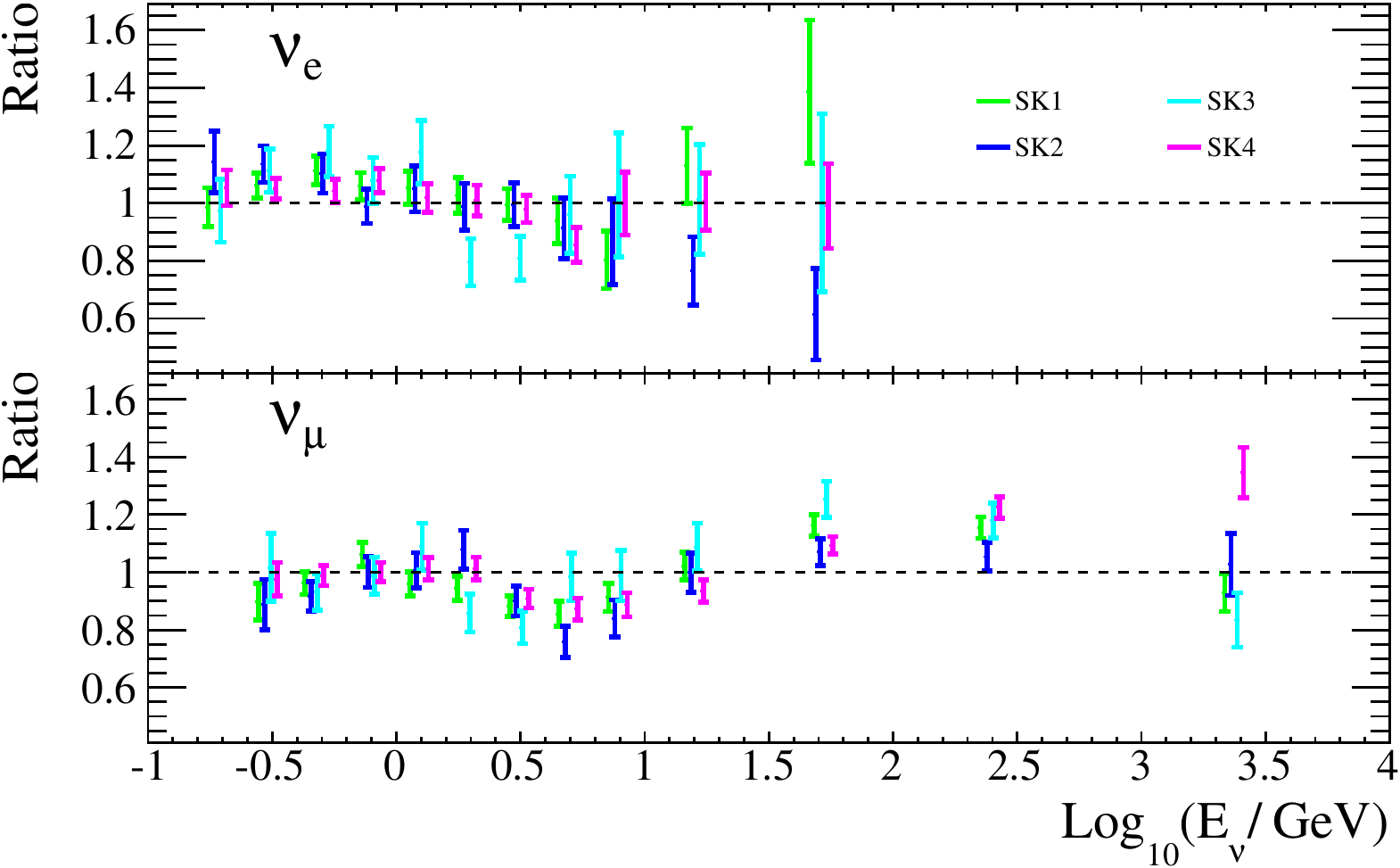}
\end{minipage}
\end{center}
\caption{ \small
(color online) Ratio of the measured flux to the HKKM11 model for each SK period,
SK-I (green), SK-II (blue), SK-III (cyan), SK-IV (magenta).
Error bars show the statistical uncertainty only. 
}
\label{fig:flux_ratio_sk1234}
\end{figure}

%
%
We notice that 
a similar wavy structure is seen for both the $\nu_e$ and $\nu_\mu$ unfolded fluxes.
In order to verify the consistency of this structure across each SK period,
the unfolded energy spectrum of each period 
is compared in Fig.~\ref{fig:flux_ratio_sk1234}.
Though there are larger statistical fluctuations (especially in the high energy regions in $\nu_e$
due to the smaller number of observed events),
the spectral shape seems consistent across all SK periods within the statistical errors.
This shape is however consistent within the range of the estimated systematic effects,
in particular the neutrino interaction uncertainties.

\begin{figure}[htbp]
\begin{center}
\begin{minipage}{\columnwidth}
\includegraphics[width=0.98\columnwidth]{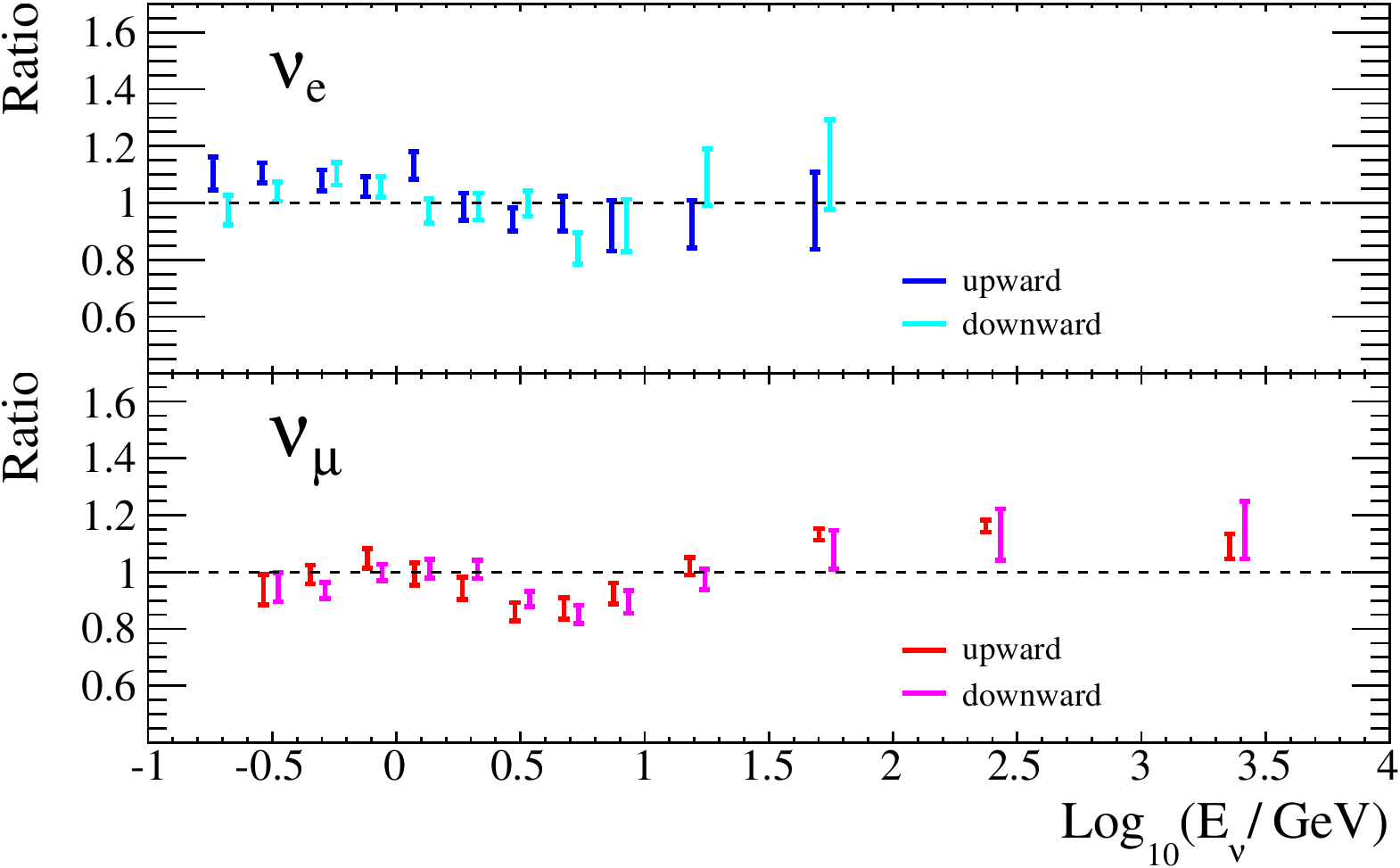}
\end{minipage}
\end{center}
\caption{ \small
(color online) Ratio of the measured flux to the HKKM11 model, separated into upward-going (blue)
and downward-going (red) data. Error bars show the statistical uncertainty only. 
}
\label{fig:flux_ratio_updn}
\end{figure}

%
%
The $\nu_e$ and $\nu_\mu$ fluxes are also separated into upward-going and downward-going datasets,
using their reconstructed direction, and measured in Fig.~\ref{fig:flux_ratio_updn}.
This is in order to check any possible bias in the flux calculation due to the differences of the 
datasets; for example, neutrino oscillation has a stronger effect
in the upward-going data above GeV energies, and also the UPMU data is an upward-going sample only. 
As seen in the figure, no obvious difference exists
between the fluxes measured using these two datasets.

\begin{figure}[htbp]
\begin{center}
\begin{minipage}{\columnwidth}
\includegraphics[width=0.98\columnwidth]{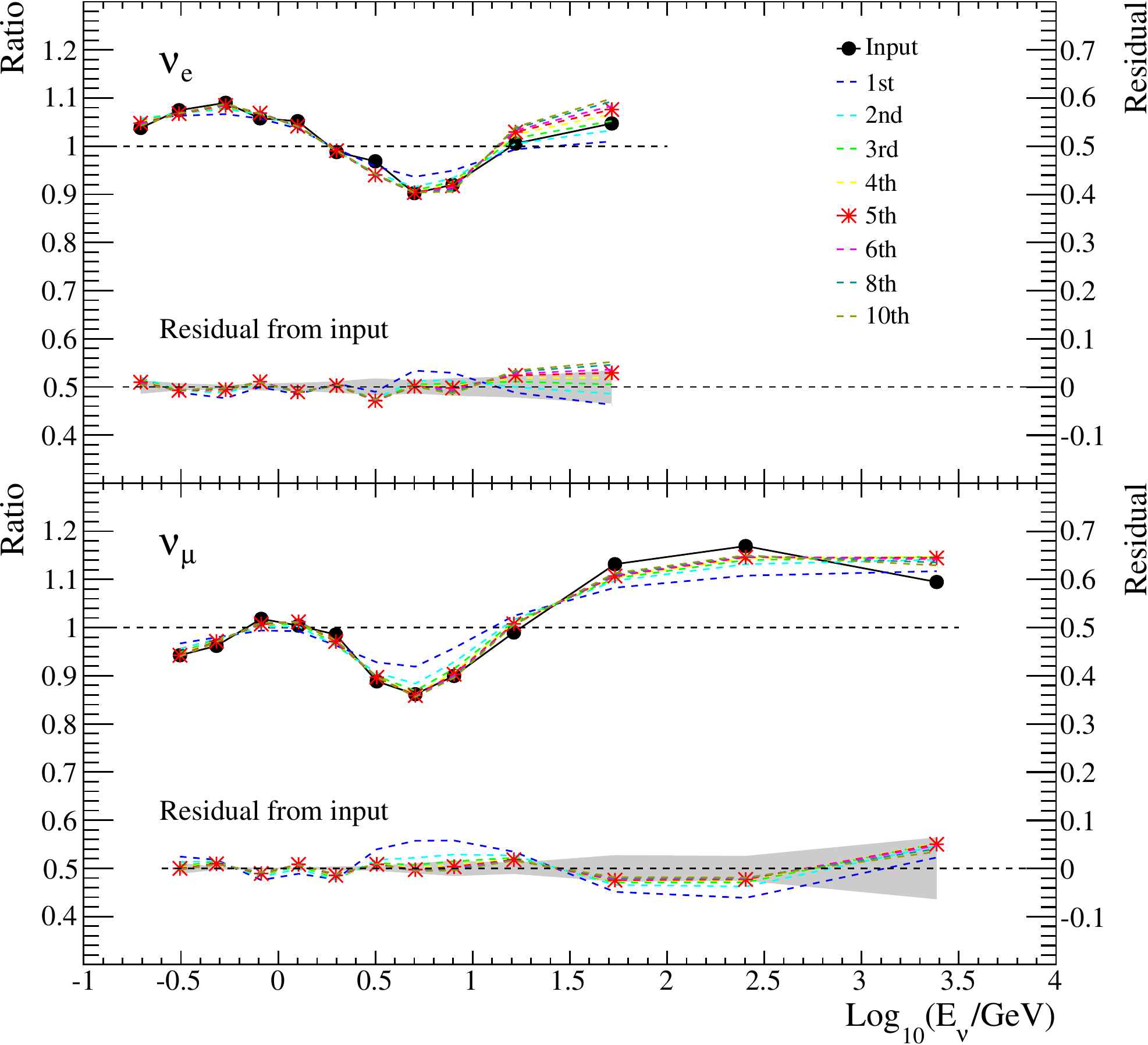}
\end{minipage}
\end{center}
\caption{ \small
(color online)
Validation of the unfolding procedure, showing
the unfolded spectra at each iteration
(dashed lines with different colors) compared to the truth input
(solid line with dots), which in this case is set identically
to our spectrum unfolded from the data.
The number of iterations was set to five in this paper
(red dashed line with asterisks).
The vertical axis on the left side shows the ratio of the unfolded flux
to the HKKM11 flux, 
and on the right side the residuals compared to the input spectrum.
The shaded area shows the size of the previously-estimated
regularization error.
}
\label{fig:flux_iteration_effect}
\end{figure}

%
%
Although we thoroughly validated the accuracy of the unfolding
procedure using the HKKM-like pseudo-data
and included the estimated regularization bias as a systematic uncertainty, 
there has still been some concern about the ability of the unfolding procedure to accurately reproduce
more complicated spectral shapes, such as the wavy shape that was eventually obtained,
and in particular whether or not such a shape would be more strongly affected by the number of iterations.
We therefore perform a further post hoc check using the same validation method as before,
but using our actual unfolded spectra from the data as the pseudo-data truth input.
Fig.~\ref{fig:flux_iteration_effect} shows the unfolded
spectrum as a function of number of iterations, from one up to ten,
on top of the previously-estimated regularization error.
It is seen that at around five iterations, which we had adopted based on the pseudo-data test,
the unfolded spectra are stable.
Furthermore, the spectra are reproduced approximately within the estimated regularization uncertainties.
Therefore, we conclude that the shapes of our unfolded spectra are not
due to an unexpected additional bias from the unfolding procedure.

%
%

\begin{figure*}[htbp]
\begin{center}
\includegraphics[width=0.8\textwidth]{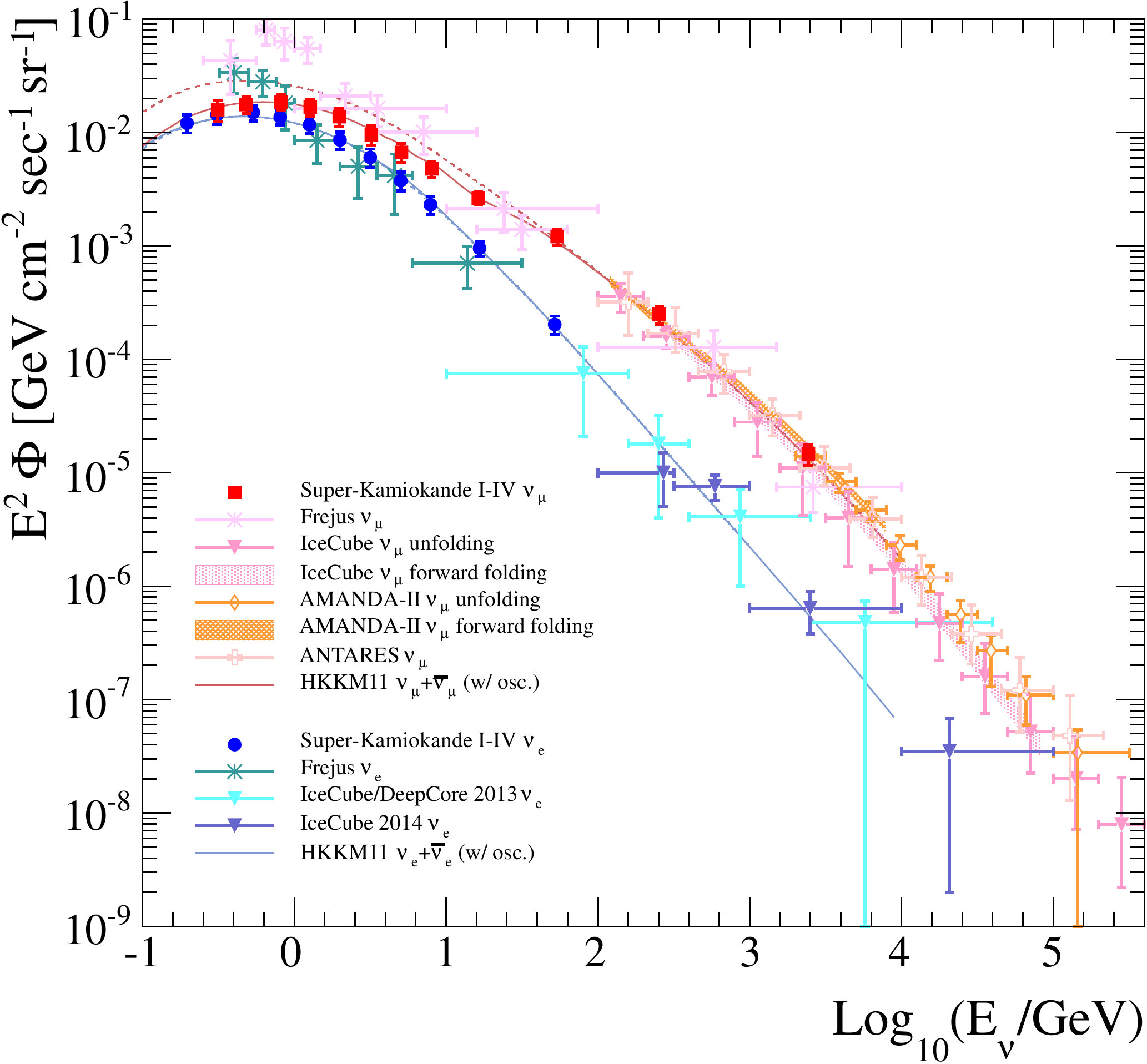}
\end{center}
\caption{ \small
(color online) The measured energy spectra of the atmospheric $\nu_e$ and $\nu_\mu$ fluxes by SK, shown with 
measurements by other experiments, 
Frejus~\cite{frejus}, AMANDA-II~\cite{amanda2-ff,amanda2-unfold}, 
IceCube~\cite{icecube-flux2011,icecube-numu-ff,icecube-fluxnue,Aartsen:2015xup}, and ANTARES~\cite{Adrian-Martinez:2013bqq}.
The phrase "forward folding" used by IceCube and AMANDA-II is synonymous with forward-fitting.
The HKKM11 flux model predictions for the Kamioka site are also shown
in solid (with oscillation) and dashed (without oscillation) lines.
The error bars on the SK measurement include all statistical and systematic uncertainties.
}
\label{fig:flux_summary}
\end{figure*}

In Fig.~\ref{fig:flux_summary} this flux measurement is shown with the results from other experiments.
Our measured data provide significantly improved precision below 100~GeV. 
The minimum of the observed energy range is extended below 1~GeV, 
and at higher energies overlaps with $\nu_\mu$ measurements by AMANDA-II, IceCube , and ANTARES, 
which should allow a better constraint on the flux normalization
at the energies beyond 100~GeV. 
The measured fluxes are consistent with these measurements 
in the overlap energy region greater than 100~GeV.
There is some discrepancy with the Frejus measurement below 1~GeV,
however due to the difference in geomagnetic cutoff at the Frejus
site, the low energy flux is expected to be a few tens of percent
higher at that location~\cite{honda_private}.

%
%

%
%
%
%
\clearpage
\section{\label{sec:eastwest}Azimuthal Spectrum Analysis}
\subsection{\label{sec:ew_data}Data Sample}
The azimuthal analysis uses only the fully-contained (FC)
sample described in Section~\ref{sec:det_types},
and selects sub-GeV and multi-GeV $e$-like and $\mu$-like events with a single reconstructed Cherenkov ring.
For the sub-GeV $e$-like events, the NC $\pi^0$-like events identified
by the $\pi^0$ finder algorithm are removed.
These samples have comparitively high statistics among the SK data samples,
as seen in Table~\ref{tab:sub_sample} in Section~\ref{sec:det_sets},
and cover a wide energy range from 100~MeV to $\sim$10~GeV,
which includes the range where the east-west asymmetry is expected to be visible.
The single-ring events have the highest accuracy for vertex and directional reconstruction,
and the neutrino flavor accuracy in this sample is estimated as
93.6\% for $\nu_e$ and 98.5\% for $\nu_\mu$, CC and NC inclusive.

The azimuthal angle $\phi$ is defined by a particle's forward-going direction,
clockwise from true south at the SK detector position
(i.e. $\phi=0\degree$ for south-going, $\phi=90\degree$ for west-going,
$\phi=180\degree$ for north-going, and $\phi=270\degree$ for east-going).
The zenith angle $\theta$ is given in terms of $\cos\theta$,
where $-1$ represents an upward-going particle.
Each neutrino event is binned using the reconstructed properties of the produced lepton,
including azimuthal angle $\phi_{\rm rec}$ (binned in 12 evenly-sized bins),
zenith angle $\theta_{\rm rec}$ (binned in 5 evenly-sized bins in $\cos\theta_{\rm rec}$),
and energy $E_{\rm rec}$.
The reconstructed energy $E_{\rm rec}$ was described in Section~\ref{sec:flux_data},
and is binned as 4 energy bins starting at 0.1 (0.2 for $\mu$-like), 0.4, 1.33, and 3.0~GeV,
where the last bin is unbounded from above.
This binning was chosen following the MC prediction, in order to show the zenith and energy regions
in which the predicted east-west effects have various different strengths, while
keeping statistics high in each bin.

All data from the SK periods I-IV are summed together,
for a total of 13,079 $e$-like and 12,725 $\mu$-like events.
The MC events, as described in Section~\ref{sec:det_sets}, are binned in the same way
using their reconstructed variables,
with weights for the data livetimes, solar activity, and neutrino oscillation.
In this section we consider, unless otherwise stated, the MC generated according to the HKKM11 flux model,
which is the only model that publishes comprehensive azimuthal distributions.
The HKKM11 model uses the IGRF-10 geomagnetic field model~\cite{IGRF}.

\subsection{\label{sec:ew_analysis}Analysis Method}
Unlike the energy spectrum analysis, a forward-fitting method is used
instead of an unfolding procedure. This is because
the azimuthal distributions are unique to the SK site and will not be directly compared
with the results of other experiments, and as such
we can compare our data distributions with the MC predictions,
without needing to reconstruct the true flux distributions.
The azimuthal distributions are plotted first using events
in each energy bin while summing over all zenith angle bins, 
then for each zenith bin while summing over all energy bins.
To quantify the east-west dipole asymmetry in each plot, we define the parameter
\begin{equation}
A = \frac{n_{\rm east}-n_{\rm west}}{n_{\rm east}+n_{\rm west}}
\label{eq:a_parameter}
\end{equation}
where $n_{\rm west}$ represents the total number of events
with azimuth angle between 0--180 degrees, i.e. west going events,
and $n_{\rm east}$ represents east going events with azimuth angle of 180--360 degrees. 

To calculate the significance of a nonzero east-west effect, 
we test a reduced sample of events in the energy range from 0.4 to 3.0~GeV, and 
with $\lvert\cos\theta_{\rm rec}\rvert<0.6$. These criteria are optimized by MC to select
the events giving the largest predicted value of $A / \Delta A$,
which is defined as the significance in units of $\sigma$. 

\begin{figure}[htbp]
\begin{center}
\includegraphics[width=0.48\textwidth]{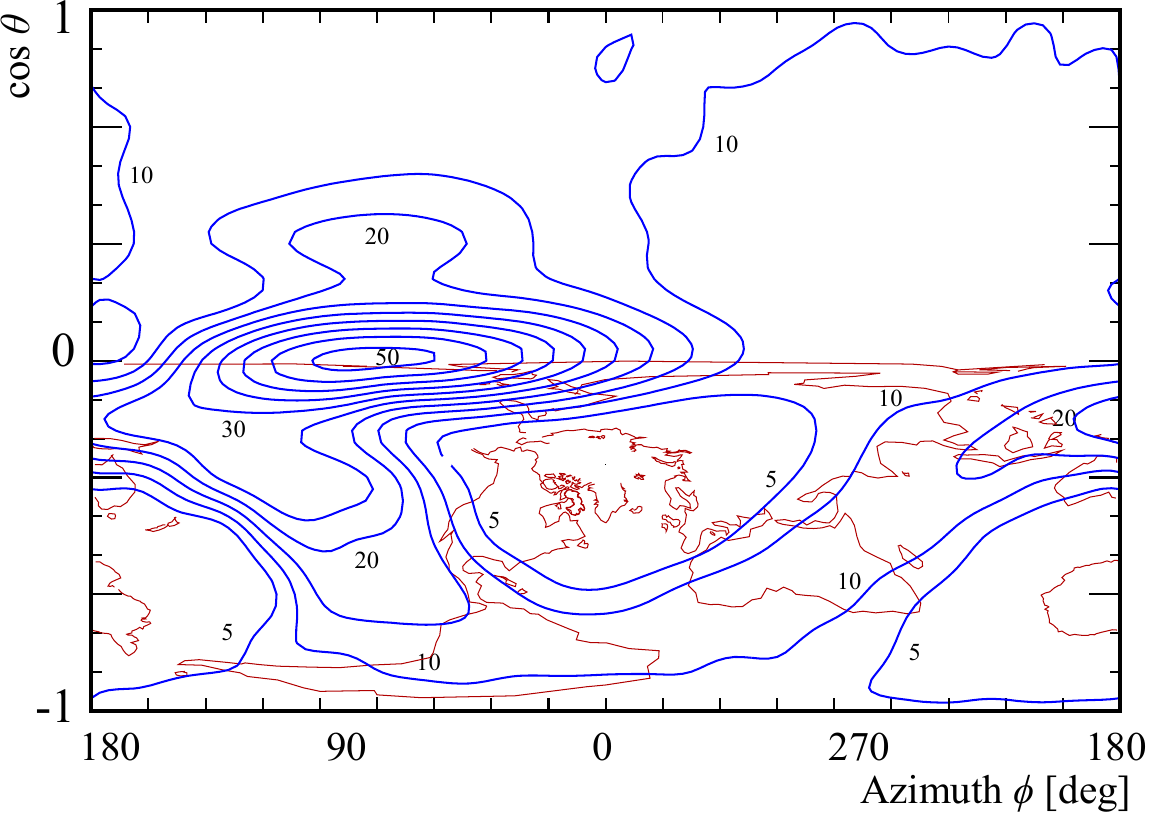}
\end{center}
\caption{ \small
(color online) Contour map of the rigidity cutoff, seen from the perspective of the SK detector.
The red lines show the projection of the Earth's landmasses,
and the blue contours and labels represent the rigidity cutoff in GV.
Reproduced from~\cite{honda2002} with permission.
While the axes are flipped compared with other plots in this paper,
the labeling of the azimuth $\phi$ and zenith $\theta$ angles uses the same convention.
}
\label{fig:contour_map}
\end{figure}

By examining the rigidity cutoff at SK as a function of both $\phi$ and $\theta$,
shown in Fig.~\ref{fig:contour_map}, we can see that while the simple dipole effect is clearly visible
near the horizontal, in general the structure of the rigidity contours is more complicated.
This is due to the difference between geomagnetic and geophysical north,
and irregular contributions from the non-dipole components of the geomagnetic field~\cite{geomag}.
In order to test if this predicted structure is visible at SK within the limited
statistics available, we define the parameter $B$ by fitting the function
\begin{equation}
k_1\sin(\phi + B) + k_2
\end{equation}
to the azimuthal distribution of the data
in each zenith bin, where $k_1$, $k_2$ are free parameters.
This parameter essentially measures any phase shift in the azimuthal
alignment of the asymmetry dipole, depending on zenith angle.

\begin{figure}[htbp]
\begin{center}
\includegraphics[width=0.48\textwidth]{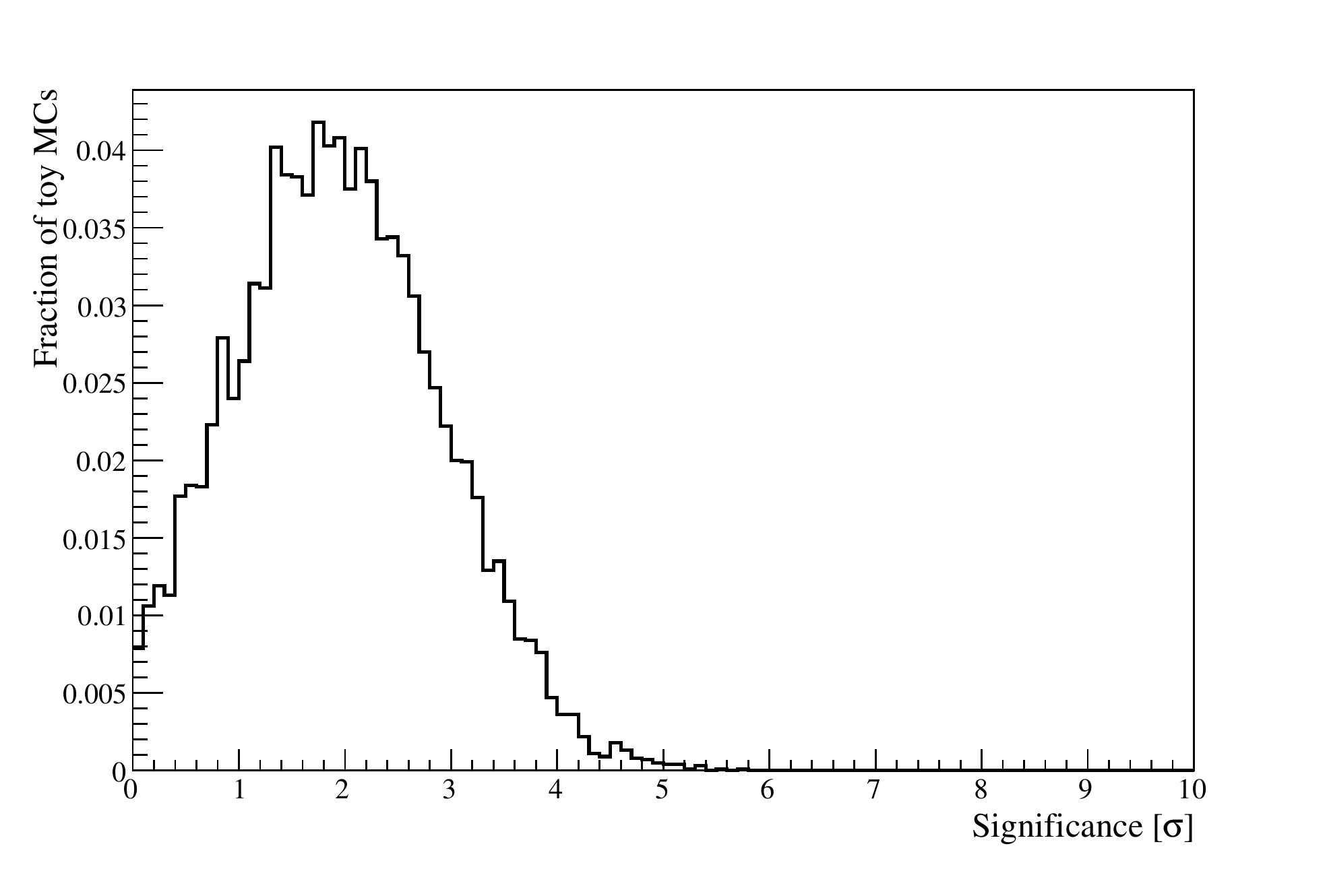}
\end{center}
\caption{ \small
The significance $\sigma$ of rejecting a non-zenith-dependent
$B$ parameter in favour of the MC expectation,
estimated by the result of 10,000 toy MC simulations.
}
\label{azi:fig:sigmas}
\end{figure}

To test the significance of the existence of such zenith-dependent effects,
we perform a $\chi^2$ test of the measured $B$ parameters across the zenith bins,
first against a constant (zenith-independent) $B$ parameter,
then secondly against the nominal MC shape.
For the first test, the normalization of the constant $B$ parameter 
is fitted to the data, for both $\nu_e$ and $\nu_\mu$ events separately,
then the $\chi^2$ statistic is taken and denoted as $\chi^2_{\rm flat}$.
For the second test, the normalizaton of the MC distributions for the $\nu_e$ and $\nu_\mu$ events
is again fitted, and furthermore the amplitude of the effect
(relative difference of $B$ in each zenith bin from the average in all bins)
is freely scaled for $\nu_e$ and $\nu_\mu$ events, which allows for a stronger or weaker
zenith dependency of the dipole alignment than predicted by MC.
The result of the MC shape fit is denoted $\chi^2_{\rm MC}$.
We then define
\begin{equation}
\Delta\chi^2 \equiv \chi^2_{\rm flat} - \chi^2_{\rm MC}
\end{equation}
which by Wilks' theorem~\cite{wilks} should be distributed as a
$\chi^2_{k=2}$ (where $k$ denotes the number of degrees of freedom),
since the two hypotheses have a difference in parameter space dimension of 2.
The significance of rejecting a constant $B$ parameter in favour of the MC expectation
is then defined in units of $\sigma$ by comparing the obtained $p$-value
\begin{equation}
p = \int_{\Delta\chi^2}^{\infty} \chi^2_{k=2}(x) dx
\end{equation}
with the normal distribution.
By toy MC, we expect a $2.0\pm0.9$~$\sigma$ indication of
a zenith-dependent $B$ parameter, as shown in Fig.~\ref{azi:fig:sigmas}.

\subsection{\label{sec:ew_syst}Systematic Uncertainties}
Since the shape of SK detector is azimuthally symmetric, 
any azimuthal asymmetry in the detector's reconstruction abilities is expected to be small.
Nonetheless, the possible existence of reconstruction errors directly dependent on the azimuthal angle
is investigated. 
The dependence of reconstructed momentum on the azimuth angle is checked 
using the momentum of the Michel electrons coming from the decay of
cosmic muons which stopped in the detector, 
and is found to be uniform to better than 0.6\%.
While the energy range of Michel electrons is lower than that of our event samples,
this value should give a reasonable indication of the maximum energy scale asymmetry
and is implemented as a systematic error. 
Other possible azimuthal biases, such as particle identification or ring counting,
are tested by comparing the shape of the likelihood variable distributions between
east and west-going single-ring events.
The maximum shift in events between classifications is found to be less than 0.1\%
in each case, and is considered negligible~\cite{euan_phd}.
Non-neutrino background events in the SK data, such as cosmic ray muons and other
noise events, are estimated to represent 0.2\% of events in the SK-I to SK-IV dataset~\cite{euan_phd},
and any azimuthal asymmetry present in this background is considered negligible.

\begin{figure}[htbp]
\begin{center}
\includegraphics[width=0.48\textwidth]{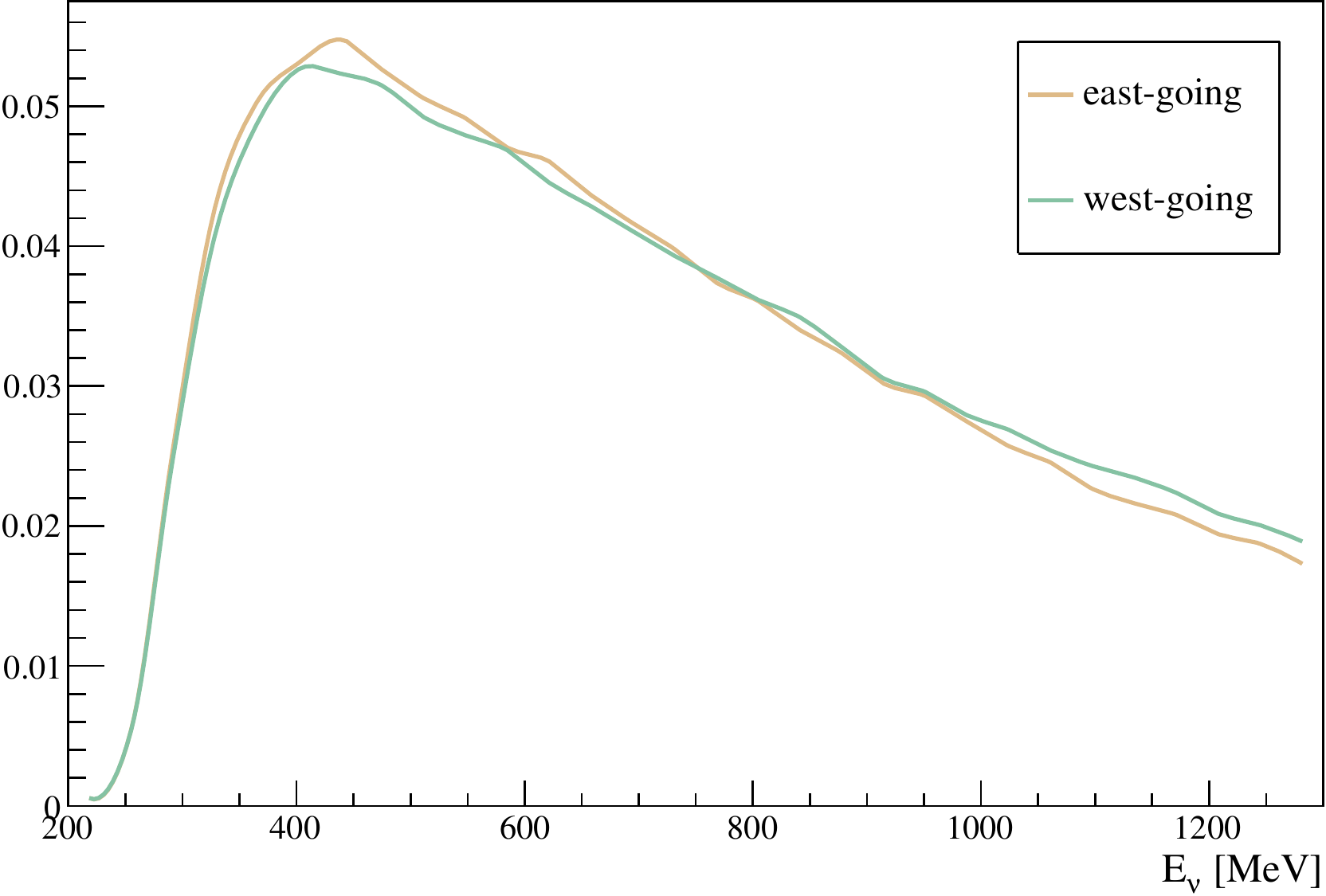}
\end{center}
\caption{ \small
(color online) Energy spectra between east-going
and west-going events, showing the MC true energy spectrum of events reconstructed
as single-ring sub-GeV $\mu$-like. The area under both curves is normalized to unity.
}
\label{fig:systematic_example}
\end{figure}

Considering the neutrino flux at SK, the geomagnetic effects are expected to cause changes in 
the energy spectrum, neutrino path length, and neutrino / antineutrino flux ratio,
depending on the azimuth angle.
Any systematic error that depends on these factors should thus be considered.
To give one clear example of this, we show in
Fig.~\ref{fig:systematic_example}
the MC true energy spectra of $\mu$-like sub-GeV events, separated by east and west-going directions,
where we see that the east-going flux is relatively skewed towards lower energies.
The ``NEUT axial mass parameter'' systematic is at the $\sim10\%$ level when averaging over all events,
but increases sharply for low-energy events in the $O(100)$~MeV range;
the east-going flux bins are thus assigned a systematic due to uncertainty in the neutrino cross section
up to 3 percentage points higher than the west-going flux bins, due to their high proportions of low energy events.

The effects of all applicable errors from the SK systematic error database
(described in Sec.~\ref{sec:flux_errors})
are calculated in the azimuthal binning, including cross section errors (15 errors),
reconstruction-related detector errors (53 errors), neutrino oscillation model errors (6 errors),
and flux model errors (21 errors). 
The detector-related errors are numerous as most of them are modeled
separately for each SK period; however since the selection of single-ring events 
is simpler than the multiple samples used in the
energy spectrum study, fewer reconstruction-related errors are applicable 
than in Section~\ref{sec:flux_errors}.

For the parameter $A$ in Equation~(\ref{eq:a_parameter}), any systematic shift that applies equally to 
$n_{\rm west}$ and $n_{\rm east}$ will cancel out; only systematics that effect the two
unequally, i.e. an azimuthally-dependent component, will contribute to the final systematic error. 
Similarly, when we plot the full azimuthal distributions, we are interested in testing their
shape as a function of azimuth, rather than measuring the absolute normalization of the neutrino flux.
Thus for each systematic error,
we define it's \emph{normalization} component as the average effect on all azimuthal bins,
and it's \emph{azimuthally-dependent} component as the effect on each bin relative to the
average effect. The normalization component is discarded, and the azimuthally-dependent
components of all systematics are combined in quadrature to calculate the final
systematic error in each azimuthal bin.

The systematic errors with the strongest effects on the azimuthal analyses 
are listed in Table~\ref{tab:ew_syst_table}. 
Cross section errors are still dominant, and there are some contributions from relative normalizations,
even though these errors are only weakly azimuthally dependent.
Some detector errors of SK-IV and SK-I are also noticeable, due to their high 
live-times compared with other SK periods.
The total systematic effects are however small, at the 1\% level.

\begin{figure}[htbp]
\begin{center}
\includegraphics[width=0.48\textwidth]{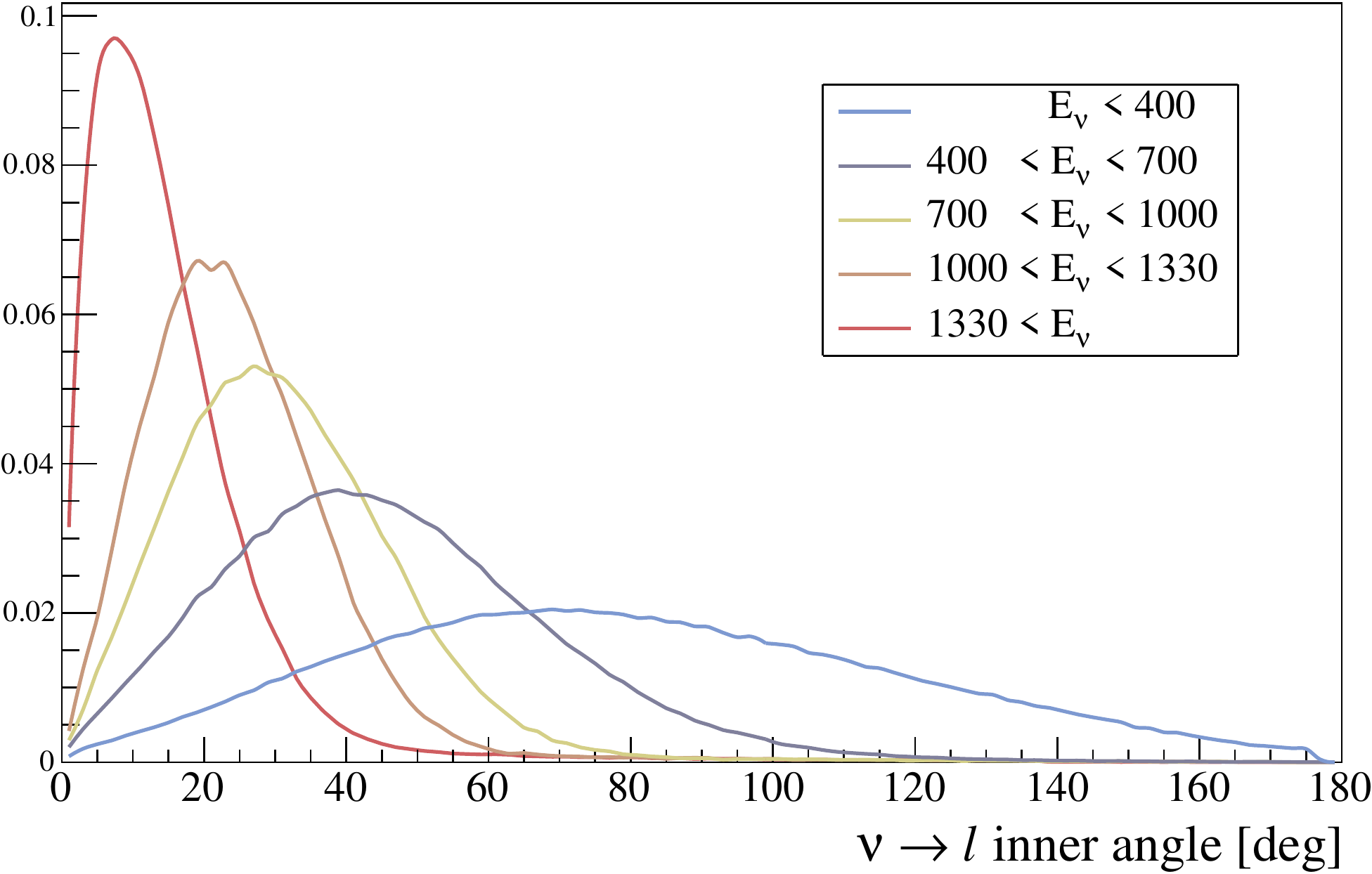}
\end{center}
\caption{ \small
(color online) Directional correlation between the true neutrino and lepton directions
for the single-ring $e$-like events used in the azimuthal analysis,
separated by true neutrino energy $E_{\nu}$ given in MeV.
The correlation for $\mu$-like events is not shown, but is generally similar.
The area under each curve is normalized to unity.
}
\label{fig:dir_correlation}
\end{figure}

Finally, it is important to note that the dominant factor in mis-reconstructing the neutrino direction
does not come from detector inadequacies, but from the fact that in CCQE interactions at lower energies
the neutrino direction and the produced lepton direction are poorly correlated.
This is not modeled as a systematic error, but the effects are accounted for
by the high statistics of the SK MC in the forward-fitting procedure.
The correlation for reconstructed single-ring events is plotted by energy in
Fig.~\ref{fig:dir_correlation},
which shows that for interactions at less than 400~MeV there is only a faint ability to
even discriminate the neutrino's forward direction from its backwards one.
Events in the range $400<E_{\nu}<1330$ are generally correlated at least within 90\degree,
which allows a reasonable separation between east and west-going events.
Finally, multi-GeV events are very well correlated, within $\sim15\degree$.

\begin{table}
\begin{center}
\scalebox{0.8}{
\begin{tabular}{lcl}
\hline
\hline
Systematic error source			& Size of effect [\%] \\ 
\hline
NEUT axial mass parameter		            &	0.59	\\
MC statistical error                        &	0.49	\\
Cross section ratio in CC quasi-elastic 		            &	0.46	\\
Flux relative normalization ($<$1~GeV)       &   0.34    \\
Cross section in single meson production	            &	0.34	\\
Flux relative normalization ($>$1~GeV)       &   0.25    \\
Cross section ratio in neutral / charged pion production		        &	0.22	\\
Coherent pion production cross section		            &	0.17	\\
Deep inelastic scattering $Q^2$ at low $W$	&	0.10	\\
Relative normalization for sub/multi-GeV FC &   0.10    \\
$\nu / \bar{\nu}$ ratio	in single pion production        &	0.08	\\
$\nu / \bar{\nu}$ ratio	in CC quasi-elastic         &	0.08	\\
Deep inelastic scattering model differences	&	0.08	\\
$\Delta m^2_{32}$ error (taken from T2K)	        &	0.07	\\
Azimuthal energy calibration (SK-IV)	    &	0.06	\\
Fiducial volume (SK-IV)			            &	0.06	\\
$\nu / \bar{\nu}$ ratio (1$<$$E_{\nu}$$<$10~GeV)  &   0.05    \\
Overall energy calibration (SK-IV)	        &	0.05	\\
Azimuthal energy calibration (SK-I)	        &	0.05	\\
Others 					                    & 	0.15 	\\
\hline
Total                                       &   1.01    \\
\hline
\hline
\end{tabular}
}
\caption{ \small 
Each systematic error source and the strength 
of their azimuthally-dependant component,
defined as the average percentage shift from the 
nominal MC values in each bin
when binned as in Fig.~\ref{fig:azi_plots},
for a 1$\sigma$ shift in the error source.
The Monte Carlo statistical error does not cancel between east and 
west-going events, but is small, less than 0.5\%.
``Others'' represents the sum of 55 systematics with effects less than 0.05\%,
combined in quadrature, and ``Total'' represents all errors combined in quadrature.
}
\label{tab:ew_syst_table}
\end{center}
\end{table}

\begin{figure*}[htbp]
\begin{center}
\includegraphics[width=0.7\textwidth]{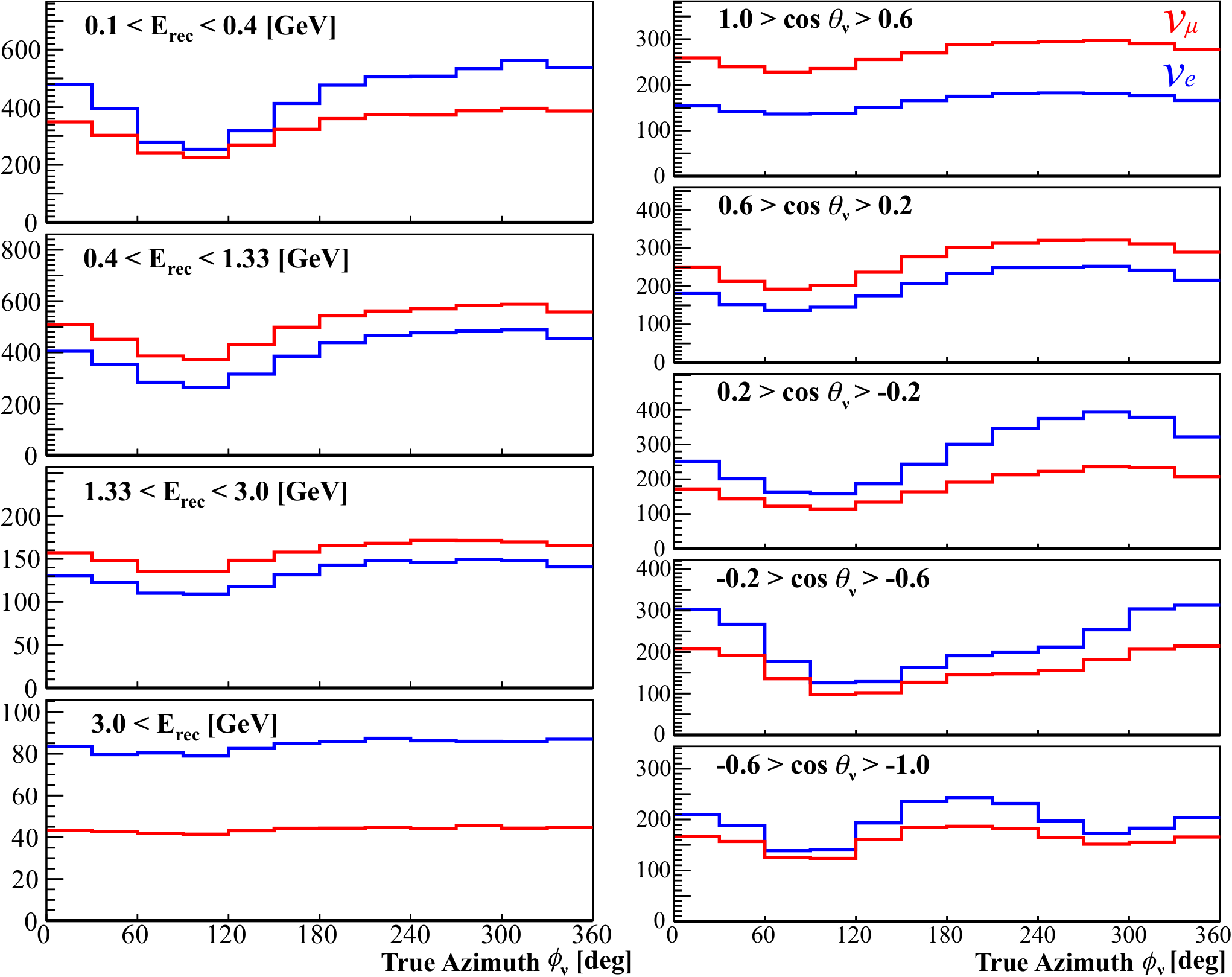}
\end{center}
\caption{ \small
(color online) Azimuthal distributions of the forward-going true neutrino direction, for e-like (blue)
and $\mu$-like (red) MC events. The events were selected as described in Sec.~\ref{sec:ew_data}.
The left plots show reconstructed energy ranges summing over all zenith angles, 
and the right plots show zenith angle ranges summing over all energies. 
The MC truth information is used instead of the reconstructed lepton direction
for both zenith $\theta_\nu$ and azimuthal angle $\phi_\nu$ binning, although the energy binning still uses
the reconstructed energy $E_{\rm rec}$.
}
\label{fig:azi_truth}
\end{figure*}

\begin{figure*}[htbp]
\begin{center}
\includegraphics[width=0.7\textwidth]{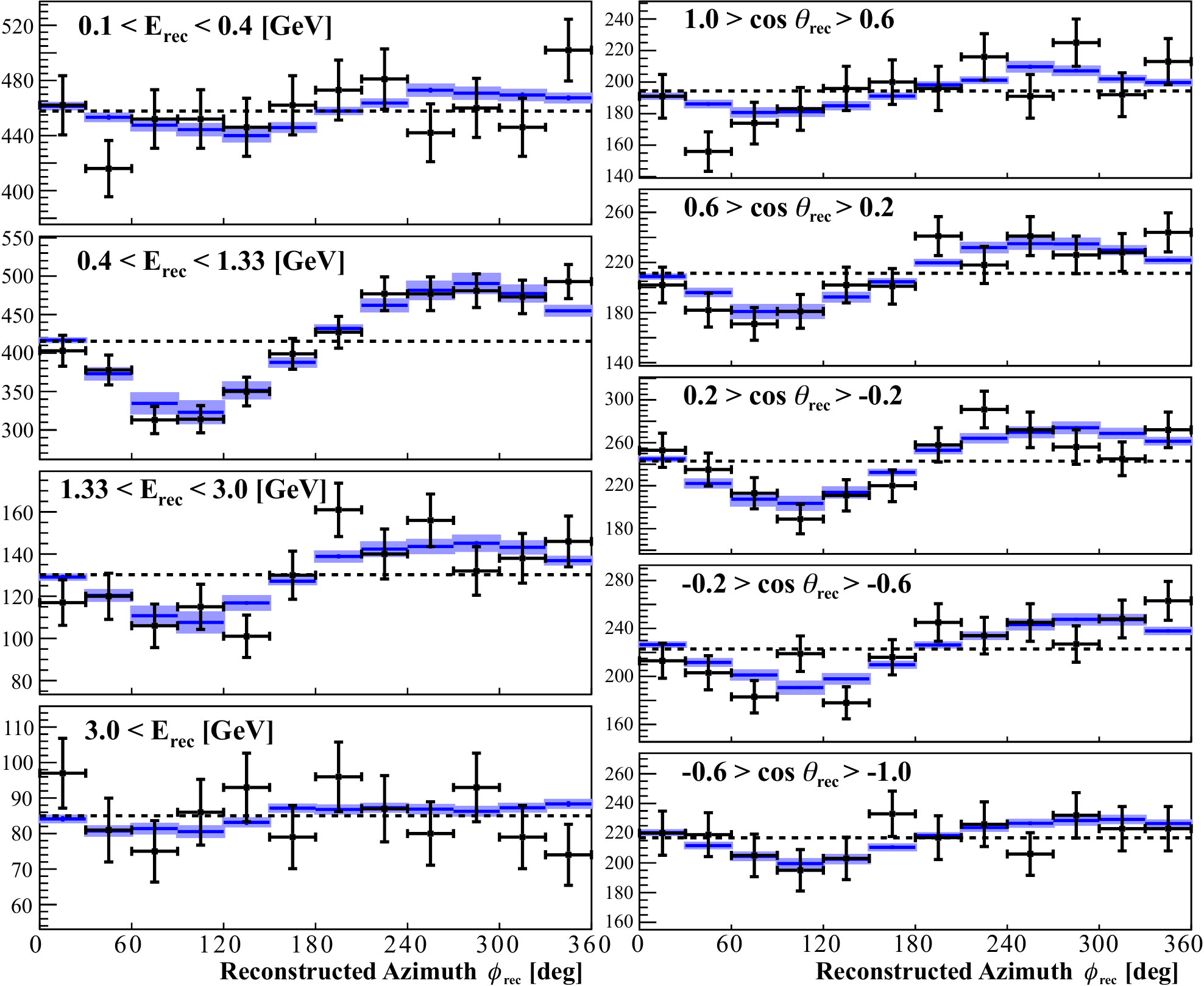}
\includegraphics[width=0.7\textwidth]{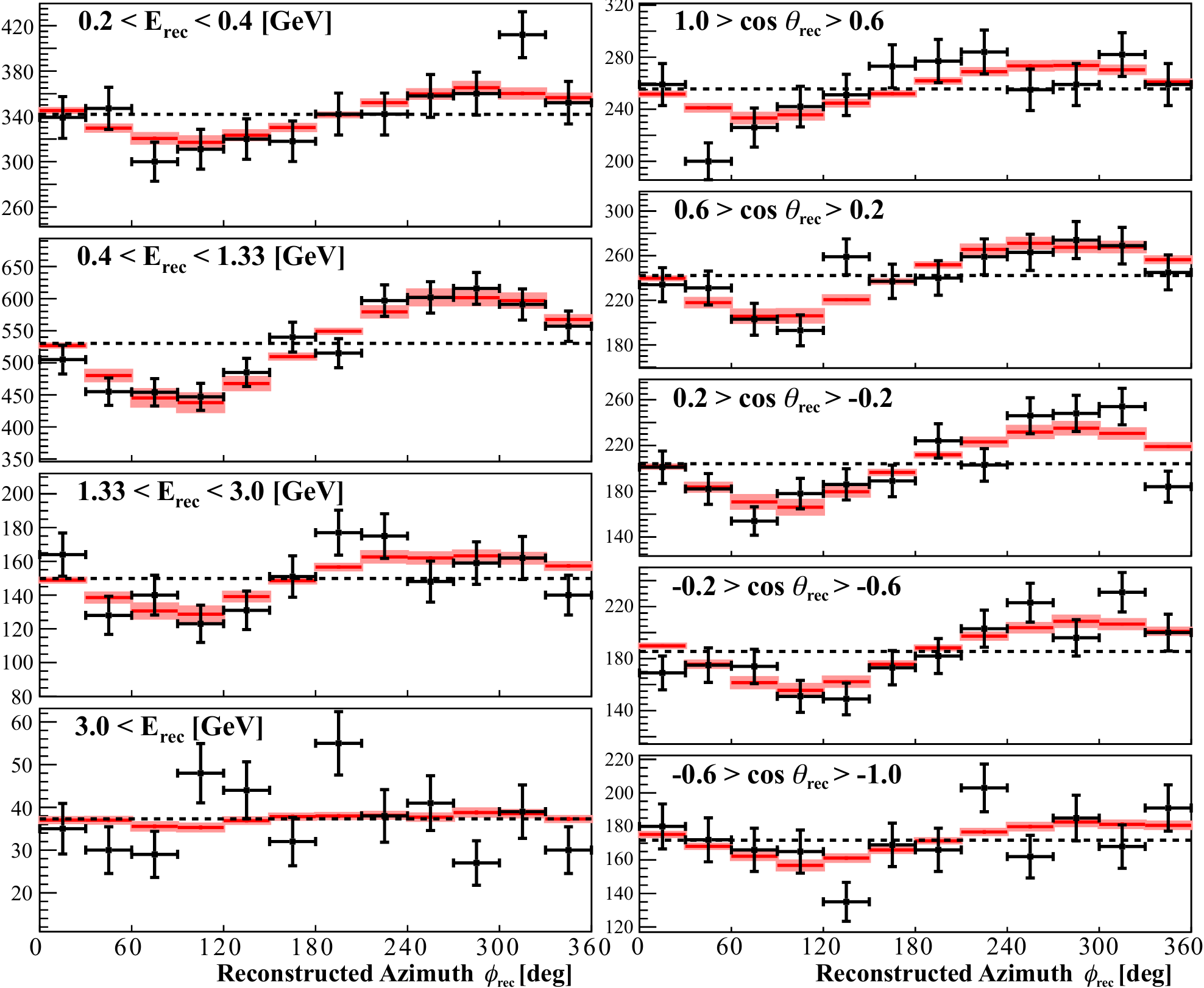}
\end{center}
\caption{ \small
(color online) Azimuthal distributions of the single-ring $e$-like (top) and $\mu$-like (bottom) events for the SK I-IV
data (points) and MC (boxes). The error bars on the data represent the statistical error,
while those on the MC represent the systematic error.
The binning is the same as in Fig.~\ref{fig:azi_truth}, however events are
binned by the reconstructed properties of the lepton.
A dotted line is drawn in each plot at the average value of the MC values
as a visual guide,
and the plots are zero-suppressed to show the data/MC shape comparison in detail.
}
\label{fig:azi_plots}
\end{figure*}

\begin{figure*}[htbp]
\begin{center}
\includegraphics[width=0.75\textwidth]{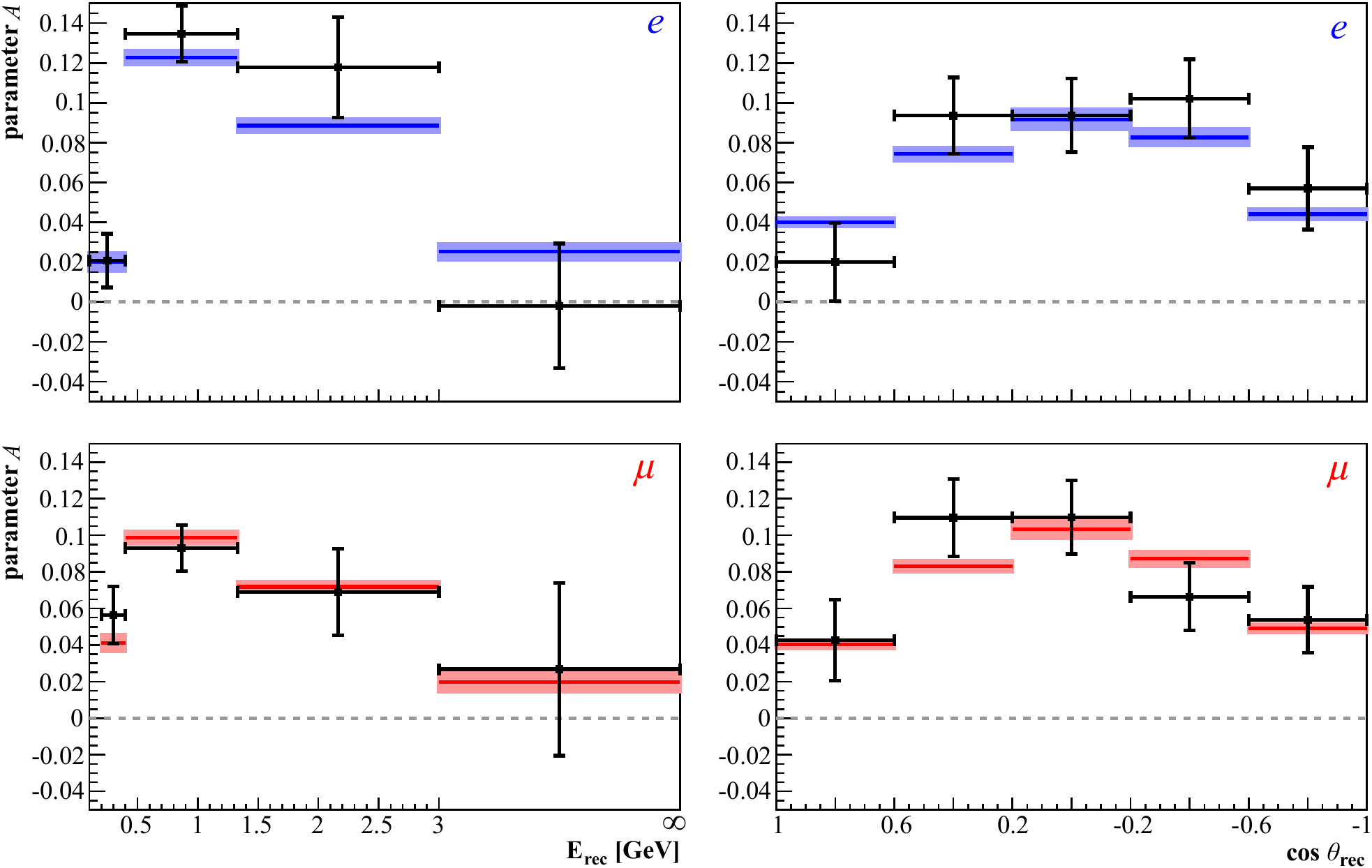}
\end{center}
\caption{ \small
(color online) The parameter $A$ depending on the reconstructed energy (left)
and zenith angle (right), for $e$-like (top) and $\mu$-like (bottom) events, from the
SK I-IV data (points with statistical error) and MC (boxes with systematic error). 
A grey dotted line is drawn at $A=0$, the case of no east-west asymmetry, as a visual guide.
The highest energy bin in the left plots is unbounded from above.
}
\label{fig:a_parameter}
\end{figure*}

\subsection{\label{sec:ew_results}Results and Discussion}

To understand the origin of the reconstructed distributions from the true flux shape,
we first show the azimuthal distributions of MC events binned by true neutrino direction
in Fig.~\ref{fig:azi_truth}, where we denote the true neutrino azimuthal and zenith angles as
$\phi_\nu$ and $\theta_\nu$ respectively.
The azimuthal distributions binned by the lepton's reconstructed
direction $\phi_{\rm rec}$ and $\theta_{\rm rec}$ are shown in Fig.~\ref{fig:azi_plots}, for both data and MC events.
Both of the above figures bin events using the reconstructed particle energy $E_{\rm rec}$.
The values of the dipole asymmetry parameter $A$, corresponding to each plot of Fig.~\ref{fig:azi_plots},
are shown in Fig.~\ref{fig:a_parameter}.

Considering first the MC events, we see that at $E_{\rm rec}<0.4$ GeV,
the dipole asymmetry is high in the true neutrino direction,
but not in the reconstructed lepton direction;
this is due to the poor correlation between the 
incoming neutrino and outgoing lepton directions at these energies.
The asymmetry in the reconstructed lepton distributions
is then expected to peak at around $E_{\rm rec}\sim1$~GeV,
and also for events at around $\cos\theta_{\rm rec}\sim0$.
The prediction of the asymmetry strength is somewhat higher than
in the previous SK analysis~\cite{skeastwest},
at which time the flux simulation considered geomagnetic effects on the primary
cosmic ray particles only;
the additional consideration of the bending of secondary cosmic rays in the atmosphere
is seen to further enhance the dipole asymmetry prediction~\cite{honda2004}.
In the $\cos\theta<-0.6$ bins, the distribution of the true neutrino direction
also shows a different shape compared to the reconstructed lepton directional distribution.
The shape of the true distribution is due to the rigidity cutoff having a suppressive
effect in both east and west directions for up-going ($\cos\theta<0$) neutrinos.
However, as this effect is strongest for sub-GeV events within $\sim10\degree$ of up-going,
it is also washed out after reconstruction using the lepton direction.

Considering next the distributions of the data,
there is in general excellent agreement with the MC expectations
in the azimuthal plots (Fig.~\ref{fig:azi_plots}).
A $\chi^2$ value is calculated based on the weighted-unweighted $\chi^2$ test~\cite{Gagunashvili}
for the energy (zenith) distributions,
and is 87.6 for 96 bins (106.6 for 120 bins)
which gives a p-value of 0.69 (0.79).
The values of the $A$ parameters associated with these plots (Fig.~\ref{fig:a_parameter})
also agree within statistical error.
The HKKM11 simulation thus models the geomagnetic effects well enough
to be consistent with the current data.

Azimuthal distributions were also published by the Bartol group in 2003~\cite{bartol}
for a limited energy and zenith range,
and by reweighting the MC histograms in Fig.~\ref{fig:azi_plots} to the Bartol predictions
we find $\chi^2_{\rm Bartol} - \chi^2_{\rm HKKM11} = 1.0$, which is too low to indicate
any preference between the models from the azimuthal distributions alone. 

\begin{figure}[htbp]
\begin{center}
\includegraphics[width=0.33\textwidth]{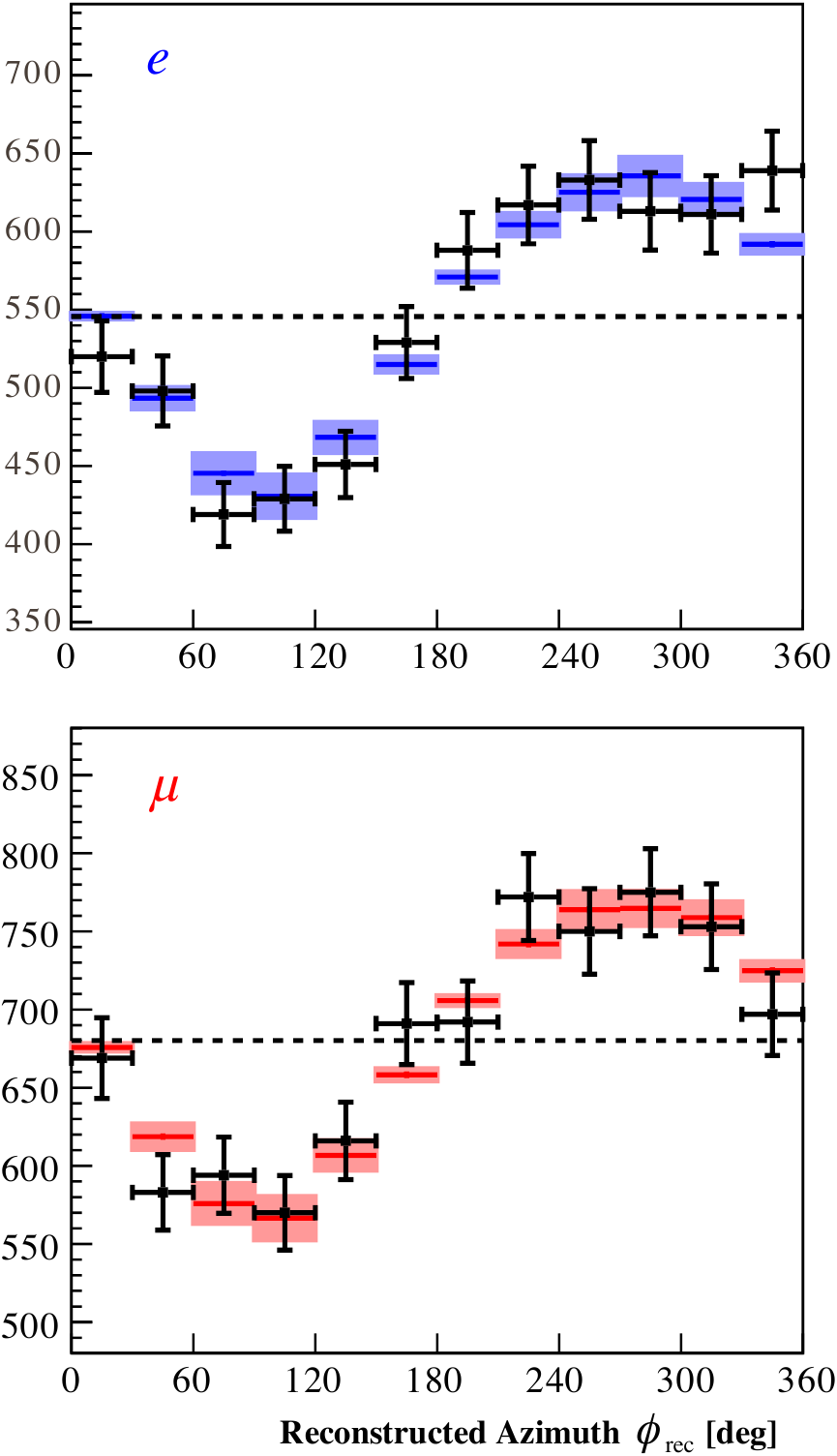}
\end{center}
\caption{ \small
(color online) Azimuthal distributions of a sub-selection of
$e$-like (left) and $\mu$-like (right) events, from the SK I-IV data (points with statistical error)
and MC (boxes with systematic error).
The sub-selection is optimized to obtain the highest significance of the final $A$ parameters,
by using only events with $0.4<E_{\rm rec}<3.0$~[GeV] and $\lvert\cos\theta_{\rm rec}\rvert<0.6$.
}
\label{fig:azi_final}
\end{figure}

The azimuthal distributions of the sub-sample of events selected
to optimize the significance of the east-west dipole asymmetry
are shown in Fig.~\ref{fig:azi_final}.
The result is an asymmetry parameter of
$A_{\mu} = 0.108\pm 0.014 (\text{stat})\pm 0.004 (\text{syst})$ for $\mu$-like, and
$A_{e} = 0.153\pm 0.015 (\text{stat})\pm 0.004 (\text{syst})$ for $e$-like events.
The dipole effect is thus seen at a significance level of
6.0~$\sigma$ (8.0~$\sigma$) for the $\mu$-like ($e$-like) samples,
which represents the discovery of the effect in the $\mu$-like sample.

\begin{figure}[htbp]
\begin{center}
\includegraphics[width=0.35\textwidth]{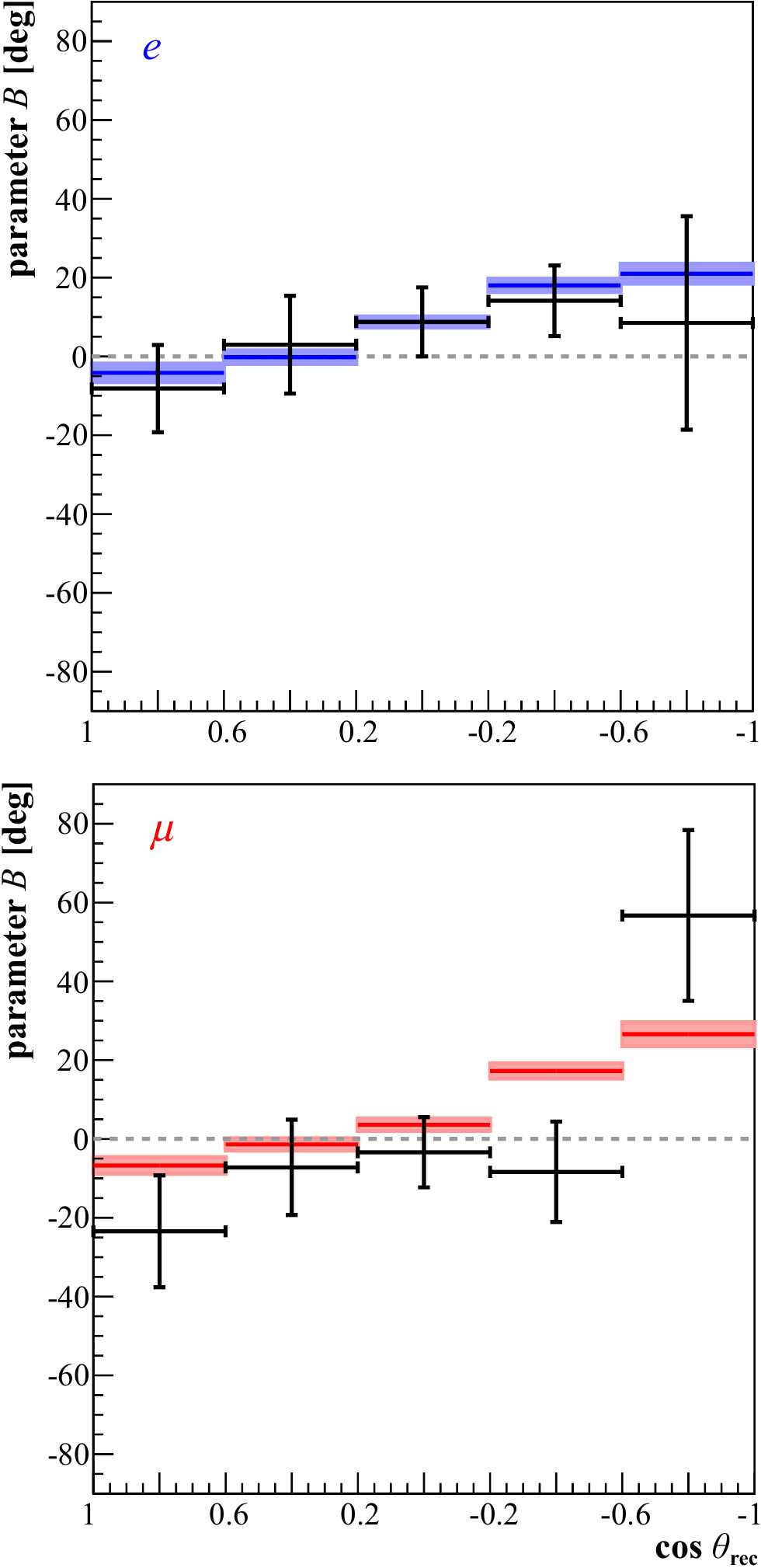}
\end{center}
\caption{ \small
(color online) The parameter $B$ depending on the cosine of the zenith angle, for
$e$-like (left) and $\mu$-like (right) events, from the SK I-IV data (points with statistical error)
and MC (boxes with systematic error).
}
\label{fig:bparam}
\end{figure}

Figure~\ref{fig:bparam} shows the calculated $B$ parameters for data and MC.
The data correspond to a 2.2~$\sigma$ indication for the existence of a
zenith-dependent change in the asymmetry dipole alignment, and
the agreement between data and MC provides further confidence
in the treatment of geomagnetic effects in the HKKM11 model.
While various detailed measurements have been made considering geomagnetic effects on
other cosmic ray particles, such as cosmic ray protons~\cite{AMS2000-3},
this is the first neutrino measurement that
indicates agreement with a more complicated model than a simple
east-west effect based on the geophysical dipole approximation.

%
%
%

\FloatBarrier

\section{Search for a Correlation between the Solar Cycle and the Neutrino
         Flux \label{sec:solarmodulation}}

\FloatBarrier

\subsection{Effect Predictions \label{sec:solmod_prediction}}

Data are provided by the HKKM flux group~\cite{honda_private}
that predict the relative normalization change of the $\nu_{e}$, $\bar{\nu}_{e}$,
$\nu_{\mu}$, and $\bar{\nu}_{\mu}$ fluxes at the SK site depending on the solar cycle.
These data use the count rate of a specific Neutron Monitor, the Climax NM~\cite{climax},
as the parameter corresponding to the degree of solar activity. 
This parameter can be used as the model assumes that, after corrections
for the local air pressure at the monitor, all NM counts are linearly
inversely correlated with the solar flux modulation.
In fact the plasma wind should affect the low-energy primary flux more strongly;
then areas with a low rigidity cutoff, which have relatively more flux at low-energy,
should experience a larger decrease in flux.
However if two NMs are located at different rigidity cutoffs, it is assumed that
while the gradient of the correlation will be different, a linear correlation is still applicable.
The HKKM model calculates the effects at the SK site for a given Climax NM count, 
extrapolating the solar effect on neutrinos coming from all directions,
considering that rigidity cutoff is a function of the direction
as was shown in Fig.~\ref{fig:contour_map}.

\begin{figure}
\begin{centering}
\includegraphics[width=0.49\textwidth]{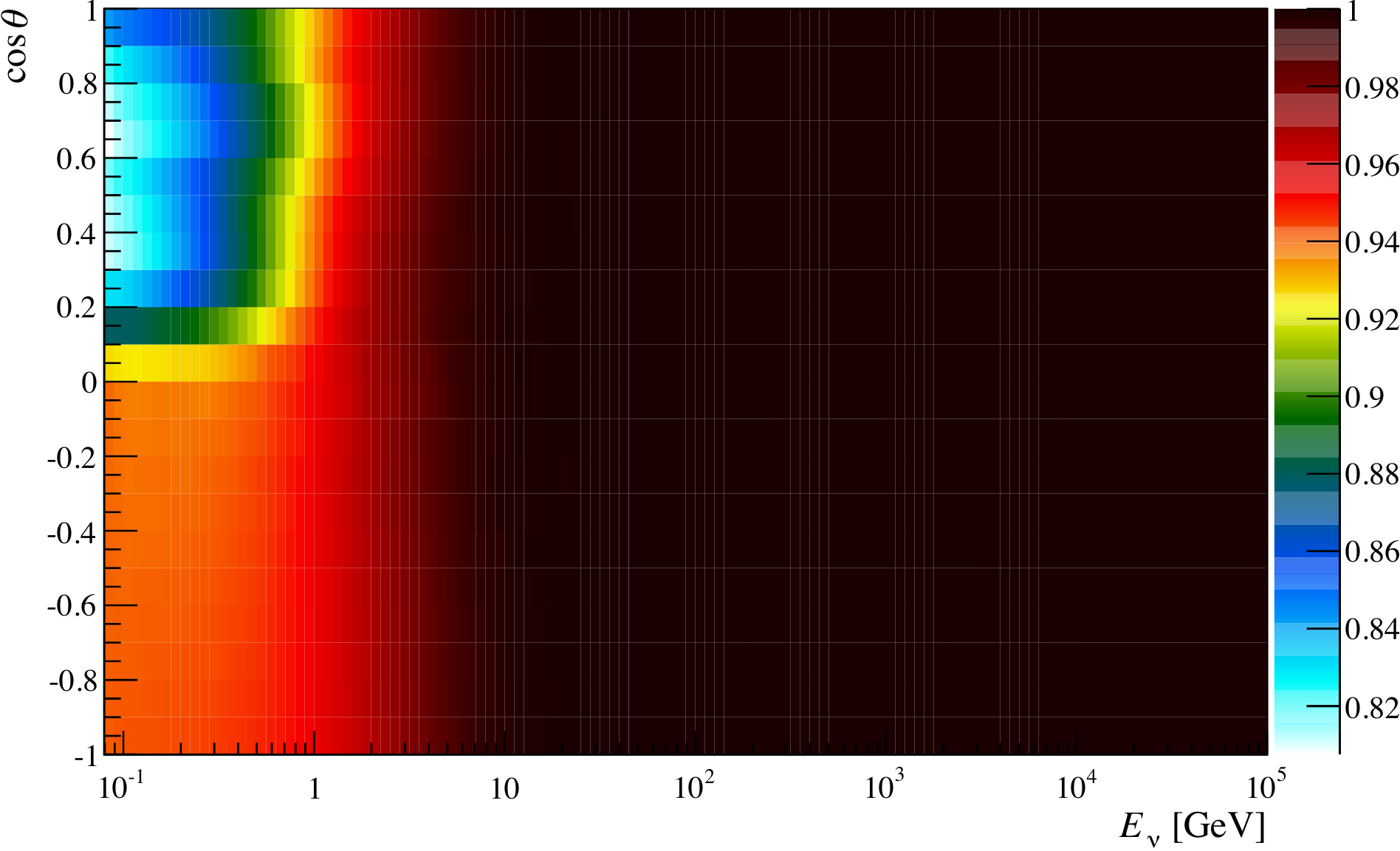}
\par\end{centering}
\caption{\small 
(color online) The predicted normalization change for the $\nu_{e}$ flux due to solar activity,
as a function of the true neutrino energy $E_{\nu}$ and zenith angle $\theta$,
as given by the HKKM model.
The $z$-axis shows the expected relative change in flux
from the solar activity minimum, defined as the activity corresponding to a Climax NM count of
4150~counts~hr$^{-1}\times0.01$, 
to solar maximum, defined as corresponding to a Climax NM count of
3500~counts~hr$^{-1}\times0.01$.
The normalization changes for the $\bar{\nu}_{e}$, $\nu_{\mu}$, and $\bar{\nu}_{\mu}$ fluxes
have the same form, with only small variations within the 2\% level.
}
\label{solmod:fig:reweight}
\end{figure}

As an example, Fig.~\ref{solmod:fig:reweight} shows the expected
normalization change for the $\nu_{e}$ flux at SK site, as a relative decrease from the
\emph{solar minimum} to the \emph{solar maximum}.
Here, solar minimum is defined as a Climax NM count of
4150~counts~hr$^{-1}\times0.01$.
While higher values are occasionally recorded by the Climax NM,
the change in neutrino flux as a function of NM count is expected to be negligible above this value.
The normalization change weights are then provided in intervals of 50 down to
3500~counts~hr$^{-1}\times0.01$, which is defined as the solar maximum
and is the highest solar activity considered by the HKKM model.
The factor of 0.01 is the common notation for NM data.
From here on, we generally refer to the solar modulation effect as a
"relative normalization change", taking the expected flux at solar minimum
as the default normalization.
In the figure, we see that only the neutrino
flux up to around 1~GeV is strongly affected by the effect.
It is also seen that there is a strong zenith-dependence of the normalization change,
which is due to the fact that neutrinos originating at the north and south polar regions
reach the SK detector from below, and the polar regions are the areas of
lowest rigidity cutoff on the Earth.

\subsection{Neutrino and Neutron Monitor Data \label{sec:solmod_data}}

The solar modulation analysis selects the single-ring sub-GeV samples,
which have reconstructed energies $E_{\rm rec}<1330$~MeV 
as described in Section~\ref{sec:det_types}. 
Other samples are not used,
since multi-GeV samples show a negligible solar modulation effect, and 
the sub-GeV multi-ring and PC samples have relative normalization systematic errors
between each SK period that are larger than the expected solar modulation effect.
This analysis uses an additional six months of data in addition to 
the data set described in Section~\ref{sec:det_sets}, up to April~2015, in order to cover more of
the current period of high solar activity.
This gives 10,892 (10,264.3) events in the $e$-like single-ring sub-GeV sample, and
10,763 (10,895.3) events in the $\mu$-like single-ring sub-GeV sample,
where numbers in brackets are the
MC expectations including oscillations. The purity of the samples, defined as the fraction of correct flavour
CC or NC interactions, is estimated by MC as 95.0\% and 98.2\% respectively.
Studies were done to optimize the selected energy ranges of these
samples, in case higher sensitivity could be obtained by using
a sub-selection of the data at lower energy where the solar modulation effect is stronger,
but no significant benefits in sensitivty were found.

In addition to being categorised as either $e$-like or $\mu$-like,
we divide the data into upward-going or downward-going samples
based on the fitted lepton direction,
resulting in a total of four data sub-samples.
For each sample, the SK I-IV periods will be combined
while considering their relative normalization errors,
as explained later in Section~\ref{solmod:sec:analysis}.
The same categorizations are applied for MC events.

For our neutron data, we cannot use directly data from the Climax NM,
which was shut down during the SK II period.
We search for monitors~\cite{nmdb} that have been active
and well-maintained throughout the entire SK I-IV period,
and find four such NMs -- the Thule, McMurdo,
Kerguelen, and Newark monitors~\cite{neutron_data,neutron_data2}. 
These stations all monitor down-going neutrons with relatively low rigidity cutoff
(0.3, 0.3, 1.14, and 2.4~GV respectively), and thus have good
sensitivity to the changes in the solar cycle. 
In particular, the Thule and McMurdo monitors in the north and south polar regions 
are very sensitive to the solar activity.

\begin{figure*}
\begin{centering}
\includegraphics[width=0.8\textwidth]{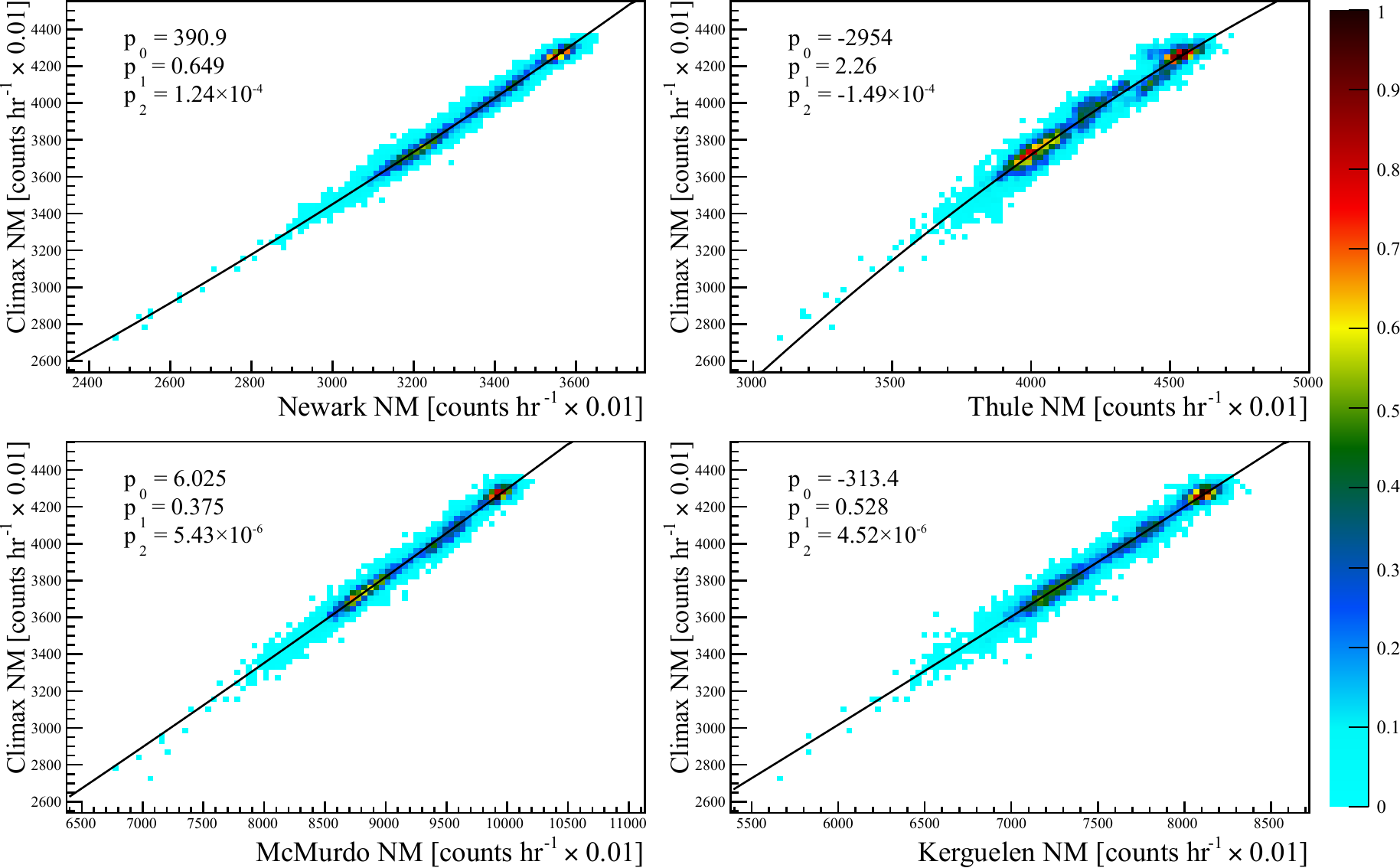}
\par\end{centering}
\caption{\small 
(color online) Correlation of the neutron monitor counts between
the Newark (upper left), Thule (upper right), McMurdo (lower left), and Kerguelen (lower right)
monitors to the Climax NM.
The average pressure-corrected counts per hour were plotted
and fitted with a second-order polynomial,
which are defined as the conversion functions between those NMs.
The figure shows each point binned into a 2D histogram,
where the z-axis represents the arbitrarily normalized number of points,
the conversion functions as black lines, and the polynomial parameters $p_i$.
}
\label{solmod:fig:climax_newark_conversion}
\end{figure*}

\begin{figure*}
\begin{centering}
\includegraphics[width=0.8\textwidth]{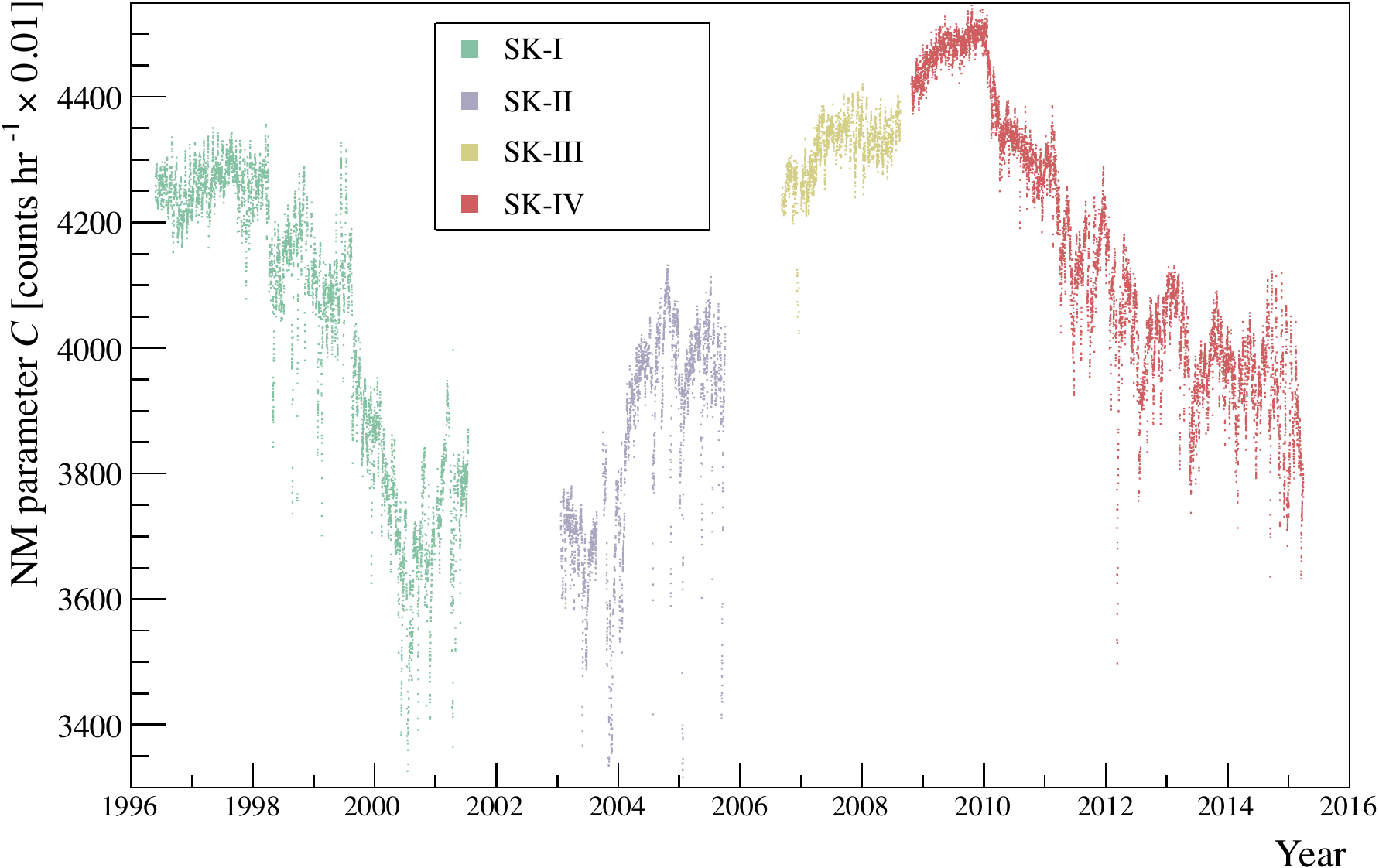}
\end{centering}
\caption{\small 
(color online) The ``NM parameter'' $C$ over the SK operational period,
where low values of $C$ correspond to high solar activity, showing that
almost two solar cycles are covered by the SK data.
\label{solmod:fig:NMcount_over_sk}
}
\end{figure*}

To obtain an ``equivalent Climax NM count'' for each of these NMs,
we compare the counts of the monitors during the times when they were
both operational, as shown in Fig.~\ref{solmod:fig:climax_newark_conversion}.
While the correlation between each monitor is almost linear, we find that
a small improvement in the $\chi^2/$DOF can be obtained using a second-order polynomial fit;
we therefore use second-order polynomials as the conversion functions between monitors.
This non-linearity is mostly due to the outlier events during particularly strong solar activity,
where the assumption of uniformity in the neighborhood of the Earth may be less accurate.
We then define a ``NM parameter'' $C$
as the average of the four NM monitor values after conversion to the
equivalent Climax NM counts.
This parameter is thus directly comparable to the Climax NM count, as used by the HKKM model.
The combination of four monitors also reduces the influence of possible systematic shifts
in a monitor;
the systematic error on $C$ is estimated by taking
the average RMS of the four converted counts, and found to be 15.8~counts~hr$^{-1}\times0.01$
at 1~$\sigma$.
The variance of $C$ over the SK I-IV period
is shown in Fig.~\ref{solmod:fig:NMcount_over_sk}, showing that
with recent data included almost two solar maxima are covered by
the SK data, although there is some down-time between SK periods.

\subsection{Analysis Method\label{solmod:sec:analysis}}

\begin{figure}
\begin{centering}
\includegraphics[width=0.45\textwidth]{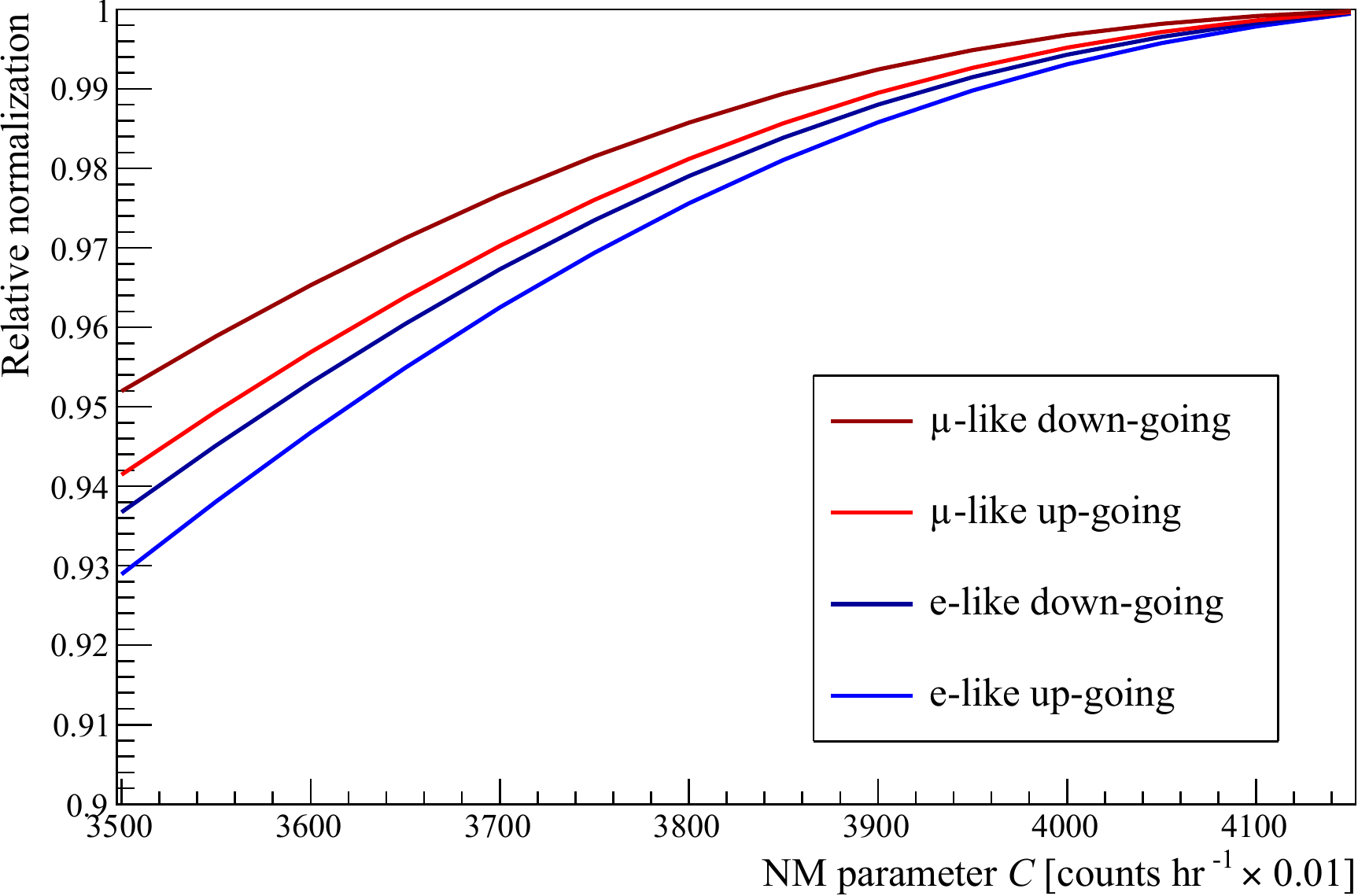}
\end{centering}
\caption{\small
(color online) The relative normalization change on the SK I-IV reconstructed data samples
due to the solar modulation effect, calculated using the SK MC and HKKM model prediction.
The normalization change is defined relative to the expected event rate
corresponding to a NM parameter $C=4150$.
}
\label{solmod:fig:four_models}
\end{figure}

\begin{figure}
\begin{centering}
\includegraphics[width=0.45\textwidth]{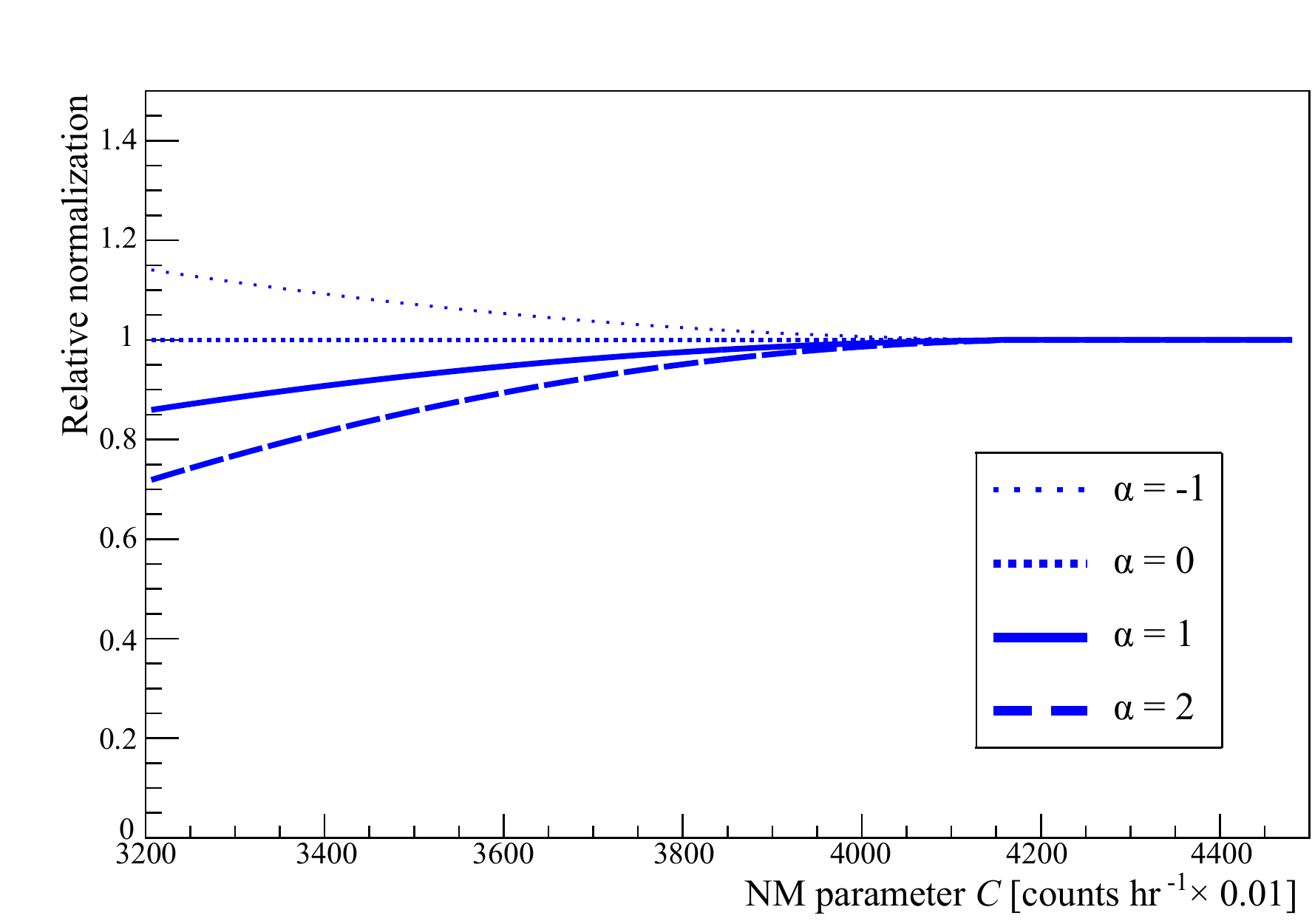}
\end{centering}
\caption{\small
(color online) An example of the relative normalization functions $f_s(C,\alpha)$,
showing the case where sample $s$ is the $e$-like up-going sample,
for various values of $\alpha$.
These functions are based on the $e$-like up-going function shown in Fig.~\ref{solmod:fig:four_models},
but extended below $C=3500$ down to $C=3300$ by a polynomial fitting,
and above $C=4150$ by assuming a relative normalization of unity.
}
\label{solmod:fig:alpha_param}
\end{figure}

The effect of the solar modulation on the four reconstructed data samples
is calculated using the SK MC, by reweighting events based on their true neutrino properties
corresponding to the relative normalization change
predicted by the HKKM model (as described in Section~\ref{sec:solmod_prediction}),
and is shown in Fig.~\ref{solmod:fig:four_models}.
It can be seen that the normalization change in the atmospheric neutrino flux is not linear in the
NM parameter $C$, and only has a significant effect at NM count values
corresponding to particularly high solar activities.
Comparing with Fig.~\ref{solmod:fig:NMcount_over_sk},
we see that only the data obtained during the solar maxima at the
end of SK-I and beginning of SK-II, and some of the recent SK-IV data,
will be strongly affected by the solar modulation. 
Of course, the data in the low-solar-activity periods are also essential,
as we are searching for a relative decrease.

The four functions in Fig.~\ref{solmod:fig:four_models} 
give a single prediction of the relative normalization change
in more accurate way than simple linear fitting would achieve,
however it is possible that in reality the effect is stronger or weaker than the prediction. 
To test for this possibility, we allow these functions to be scaled by
a continuous parameter $\alpha$, where $\alpha=0$ would represent no correlation between
the neutrino flux and the NM count,
and $\alpha=1$ represents the default predicted correlation.
These relative normalization functions are denoted as $f_s(C,\alpha)$, for each data sample $s$.
As an example, Fig.~\ref{solmod:fig:alpha_param} shows $f_s(C,\alpha)$
in the case of the $e$-like up-going sample,
for various fixed values of $\alpha$.
This set of functions will be fitted simultaneously to the four samples,
in order to measure the overall strength of the solar modulation effect
in terms of the single parameter $\alpha$.
Higher values $\alpha>1$ or lower values $\alpha<0$ are allowed in the fit,
although $\alpha<0$ would represent the unexpected case for which
the atmospheric neutrino flux increases during high solar activity.


In terms of systematic errors on the SK data,
as the solar modulation effect is a relative normalization shift
depending on the solar activity, any systematic error that affects
the overall normalization of all SK data,
such as neutrino cross section and oscillation parameters,
may be ignored. 
However, we should consider detector reconstruction errors that are dependent on the detector period,
because the solar activity is a function of detector period,
and thus we must allow for systematic shifts when combining the SK I-IV data.
Such errors can arise from the detector changes such as the difference in configuration
of the PMT coverage or the replacement of the electronics.
Table~\ref{solmod:tab:errors} shows the effect of all SK-period-dependent
systematic errors for our selected data samples.
The dominant contributions, at the $>1\%$ level, 
are the fiducial volume cut and the ring separation uncertainty;
for the $e$-like sample the single-ring $\pi^{0}$ rejection uncertainty also contributes, 
and for the $\mu$-like sample the
Michel electron tagging uncertainty contributes,
although this error becomes smaller in SK-IV due to the improved electronics.
Each error source is combined in quadrature, such that there are a total of 8 systematic errors,
representing the normalization of the $e$-like and $\mu$-like samples in each of
the SK I-IV periods. The same error is assigned for both up-going and down-going samples.

\begin{table*}
\begin{centering}
\begin{tabular}{lcccccccccc}
\hline 
\hline 
\multirow{2}{*}{Systematic error source} && \multicolumn{4}{c}{single-ring sub-GeV $e$-like (\%)} && \multicolumn{4}{c}{single-ring sub-GeV $\mu$-like (\%)}\tabularnewline
 && SK I & SK II & SK III & SK IV && SK I & SK II & SK III  & SK IV\tabularnewline
\hline 
FC reduction && 0.20 & 0.20 & 0.80 & 0.30    && 0.2 & 0.20 & 0.80 & 0.30\tabularnewline
FC / PC separation && -- & -- & -- & --  && 0.01 & $<$0.01 & 0.01 & $<$0.01\tabularnewline
Non-$\nu_{e}$ background && 0.50 & 0.20 & 0.10 & 0.10   && -- & -- & -- & -- \tabularnewline
Non-$\nu_{\mu}$ background && -- & -- & -- & -- &&   0.21 & 0.11 & 0.11 & 0.12\tabularnewline
Ring separation && 1.49 & 1.47 & 1.88 & 0.51 && 0.70 & 1.33 & 1.56 & 0.84\tabularnewline
Single ring particle ID && 0.23 & 0.66 & 0.26 & 0.28 && 0.18 & 0.50 & 0.19 & 0.22\tabularnewline
Absolute energy calibration && 0.12 & 0.18 & 0.30 & 0.26 && 0.40 & 0.61 & 0.99 & 0.85\tabularnewline
Up / down energy calibration && $<$0.01 & $<$0.01 & 0.01 & $<$0.01 && 0.03 & 0.03 & 0.06 & 0.01\tabularnewline
Azimuthal energy calibration && 0.01 & 0.01 & 0.01 & 0.01 && 0.03 & 0.01 & 0.02 & 0.02\tabularnewline
$\pi^{0}$ rejection in $e$-like sample && 1.12 & 1.12 & 0.83 & 1.00 && -- & -- & -- & -- \tabularnewline
$\mu$$\to$$e$ decay tagging && $<$0.01 & $<$0.01 & 0.01 & $<$0.01 && 1.29 & 1.35 & 1.32 & 0.78\tabularnewline
Fiducial volume cut && 2.00 & 2.00 & 2.00 & 2.00 &&  2.00 & 2.00 & 2.00 & 2.00\tabularnewline
\hline 
Total && 2.80 & 2.82 & 3.00 & 2.35 && 2.53 & 2.87 & 3.13 & 2.48\tabularnewline
\hline 
\hline 
\end{tabular}
\end{centering}
\caption{\small 
Contributions from each systematic error source on 
the uncertainty in the number of observed single-ring $e$-like and $\mu$-like events
in the solar modulation analysis,
at the $1\sigma$ level,
separated by detector period.
As described in the text, overall normalization errors which affect
all SK periods in the same way are not considered.
The ``Total'' row represents the combination of all sources in quadrature.
}
\label{solmod:tab:errors}
\end{table*}


\begin{figure}
\begin{centering}
\includegraphics[width=0.45\textwidth]{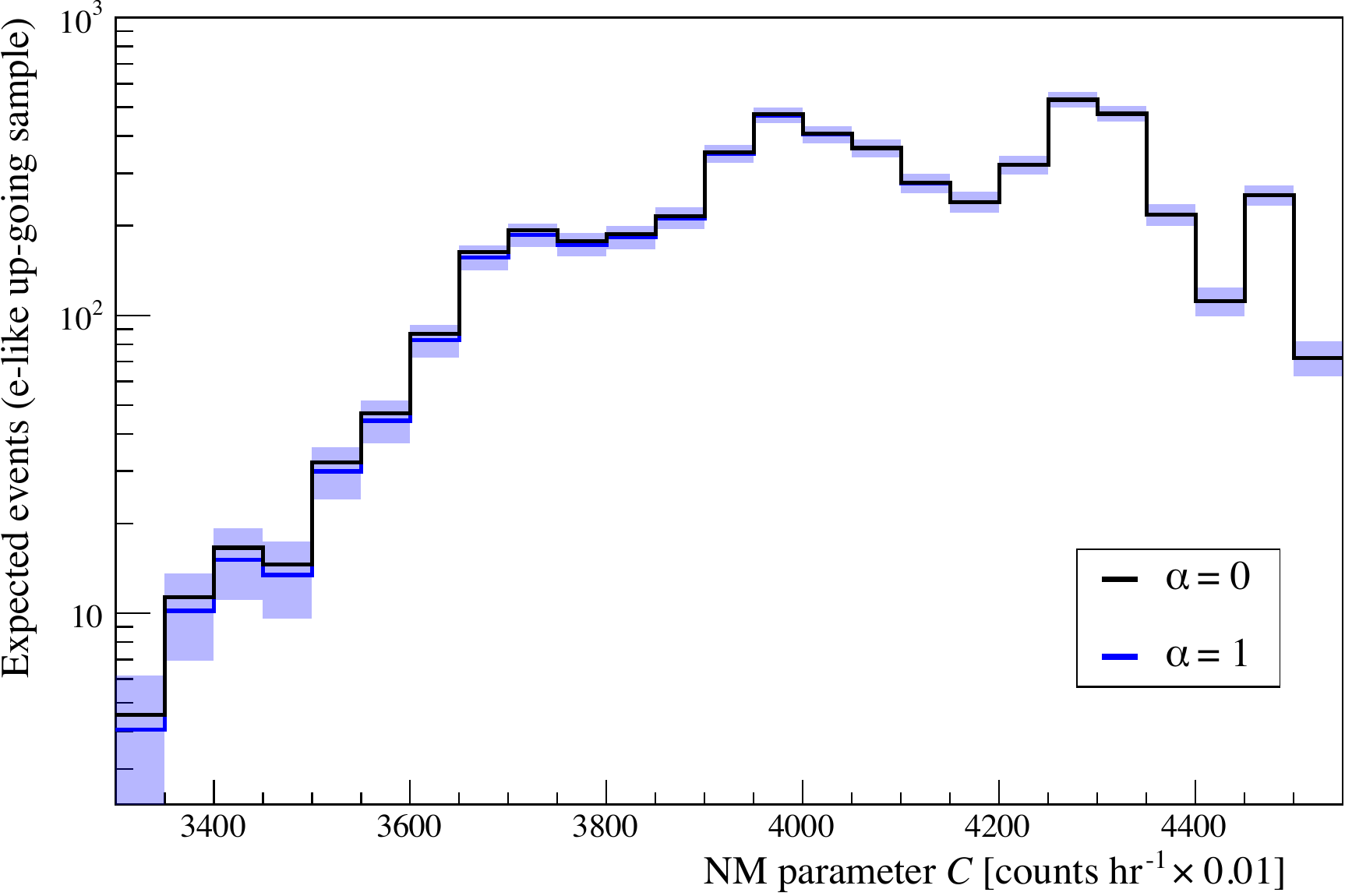}
\end{centering}
\caption{\small
(color online) Monte Carlo predictions for the number of events in the SK I-IV data,
showing two example hypotheses $H_{s,i}(\alpha)$ for the cases
of no solar activity correlation ($\alpha=0$, black), 
and a correlation as predicted by the HKKM model
($\alpha=1$, blue, with statistical error band).
The blue histogram is thus essentially the black histogram
scaled by the function $f_s(C,\alpha=1)$, shown in Fig.~\ref{solmod:fig:alpha_param}.
The sample $s$ is the $e$-like up-going sample.
}
\label{solmod:fig:flat_vs_honda}
\end{figure}

We discuss next the fitting method between the neutrino and neutron data.
Firstly, each hourly period from Jan.~1996 onwards is associated with an NM count,
which is the average NM parameter $C$ during that hour.
Neutrino events are then binned by the parameter $C$ at their observation time,
in 25 equally sized bins from 3300 to 4550~counts~hr$^{-1}\times0.01$.
We define our hypothesis for the expected number of events
in each bin $i$ in each of the four samples $s$ as
\begin{equation}
H_{s,i}(\alpha) = t^{\rm obs}_{i} \times r_{s} \times f_{s,i}(\alpha)
\end{equation}
where $t^{\rm obs}_{i}$ represents the total observation time by the SK detector
at each value of $C$, $r_{s}$ is the nominal event rate of that sample, and $f_{s,i}(\alpha)$ is
the solar modulation function $f_s(C,\alpha)$ for that sample
(an example of which was shown in Fig.~\ref{solmod:fig:alpha_param})
evaluated at the central value of $C$ for the bin $i$.

Figure~\ref{solmod:fig:flat_vs_honda} shows the binning as a function of $C$, and compares the
expected event rate $H$ of the $e$-like up-going sample
in the case of no solar modulation effect 
to the case of the standard solar modulation effect.
The minimum value on the Climax NM parameter in these plots is 3300~counts~hr$^{-1}\times0.01$;
while there is a small amount of exposure time below this value,
these data will be treated separately as discussed later.
We considered the possibility that the small systematic error in the NM parameter
could cause events to migrate between bins and affect the final result.
Using the size of the systematic, estimated in Section~\ref{sec:solmod_data},
to randomly shift the NM parameter for each hour of observation, such migration
was shown to have a negligible impact on the final measurement of $\alpha$.

From Fig.~{\ref{solmod:fig:flat_vs_honda} we see that some bins will have low statistics,
and the Gaussian approximation for statistical errors will be poor.
We thus use a Poisson likelihood-ratio method to compare the $\alpha=0$ and $\alpha\neq0$ hypotheses.
%
%
To take account of the 8 systematic errors as shown in Table~\ref{solmod:tab:errors},
we first consider that we can vary the strength of their effects compared to their default strength
by writing a vector of error pulls $\boldsymbol{{\epsilon}}$, referring to the strength
of each systematic in units of $\sigma$.
We then modify our predictions for the number of events as
\begin{equation}
H^{\prime}_{s,i}(k_s,\boldsymbol{\epsilon},\alpha)
    = H_{s,i}(\alpha) \times S_{s,i}(\boldsymbol{\epsilon}) \times k_s,
\end{equation}
where $S_{s,i}$ is the systematic modification based on the vector
of systematic error pulls, and $k_s$ is a factor
allowing a free renormalization of each of the four data-samples.
While the modification $S_{s,i}$ simply shifts the relative normalization
of each SK period, the shape of the modification on the hypotheses $H_{s,i}$
is somewhat complicated, as the relative contribution of each SK period varies
depending on each bin $i$.
If we denote our data results by $N_{s,i}$, 
the Poisson log-likelihood including systematic error is defined as
\begin{equation}
\ln L(N|H) = \sum_s \sum_i \ln\left( \frac{H_{s,i}^{\prime N_{s,i}} e^{-H^{\prime}_{s,i}}}
             {N_{s,i}!} \right)   + \boldsymbol{\epsilon}^2
\end{equation}
where the second term is a penalty term introduced for the systematic errors,
to avoid unphysically large changes of the relative normalization in each SK period.
The final likelihood ratio statistic is then defined as
\begin{equation}
\Lambda = 2\ln \frac{L_{M} \left( N|H^{\prime} \right) }
                    {L_{M} \left( N|H^{\prime}(\alpha=0) \right)},
\end{equation}
where the subscript $M$ denotes the maximum likelihood estimator,
i.e. the best-fit over all parameters of $H^{\prime}$.
In the denominator however, $\alpha$ is explicitly fixed to be zero.
By Wilks' theorem, as the difference in parameter space is one, the test statistic
$\Lambda$ is expected to be distributed as a $\chi^2$ with one degree of freedom.
The significance of rejecting the $\alpha=0$ hypothesis is then given by $\sqrt{\Lambda}$.


Toy datasets are generated for each of the four samples, for both $\alpha=0$
and $\alpha=1$ hypotheses, in order to test the sensitivity to observe
a long-term solar activity correlation. 
Each toy was created by generating a random set of systematics $\boldsymbol{{\epsilon}}$
according to a Gaussian distribution, calculating the associated hypothesis $H^{\prime}_{s,i}$,
then generating events in each bin based on a Poisson distribution.
One thousand toys are generated and passed through the fitting procedure, and
several tests are performed on the results to ensure that the test statistic
is distributed as expected and the fitting procedure is unbiased.
By taking the significance $\sqrt{\Lambda}$ for each toy,
we expect on average a 1.75~$\sigma$ sensitivity
to observe a non-zero solar activity correlation.

\subsection{Seasonal Correlation Analysis\label{solmod:sec:seasonal}}

The HKKM group has predicted that at the SK site,
the winter decrease in the neutrino flux normalization is much less than 1\%~\cite{honda2015}.
The effect is stronger at higher energies, but even at
the TeV scale the change is only at the 1\% level.
As the SK detector has accumulated limited statistics at such energies,
there is essentially no sensitivity to measure the predicted changes.

Nonetheless, we perform a simple search for a seasonal effect, as a sanity check of
our data and the model predictions.
For this test all of the sub-GeV data samples used in the solar
correlation analysis are combined into one sample,
however as the effect is predicted to exist only at
higher energies, we also combine and test all of the multi-GeV samples as defined
in Table~\ref{tab:sub_sample} in Section~\ref{sec:det_sets}, including PC and UPMU data.
The search is done by plotting the
average daily event rate over each month of the year, then performing
a $\chi^2$ comparison with the best-fit constant event rate.
We also use the unbinned data from this plot and perform a Kuiper test,
which is a type of Kolmogorov-Smirnov test that is invariant under cyclic transformations,
i.e. is not biased by the fact that we start counting in January
and is able to detect a deviation from a constant event rate
with equal significance across the year. The comparison function for the Kuiper
test is taken as the average event rate over all the data, with exact consideration
of the livetime accumulated throughout the year, accounting for the various downtime
in each SK period.

Applying the predicted 11-year solar modulation effect to the data when
binned by month is calculated to cause less than a 0.3\% difference between bins,
so is considered negligible in this test; similarly the effects of SK period-dependent
systematics are disregarded, as each period contributes roughly equally across the year.
Conversely, assuming that no unexpectedly strong seasonal correlation is seen,
the effects from the seasonal correlation on the solar correlation study
were also calculated to be negligible and disregarded.


\begin{figure*}
\begin{centering}
\includegraphics[width=0.8\textwidth]{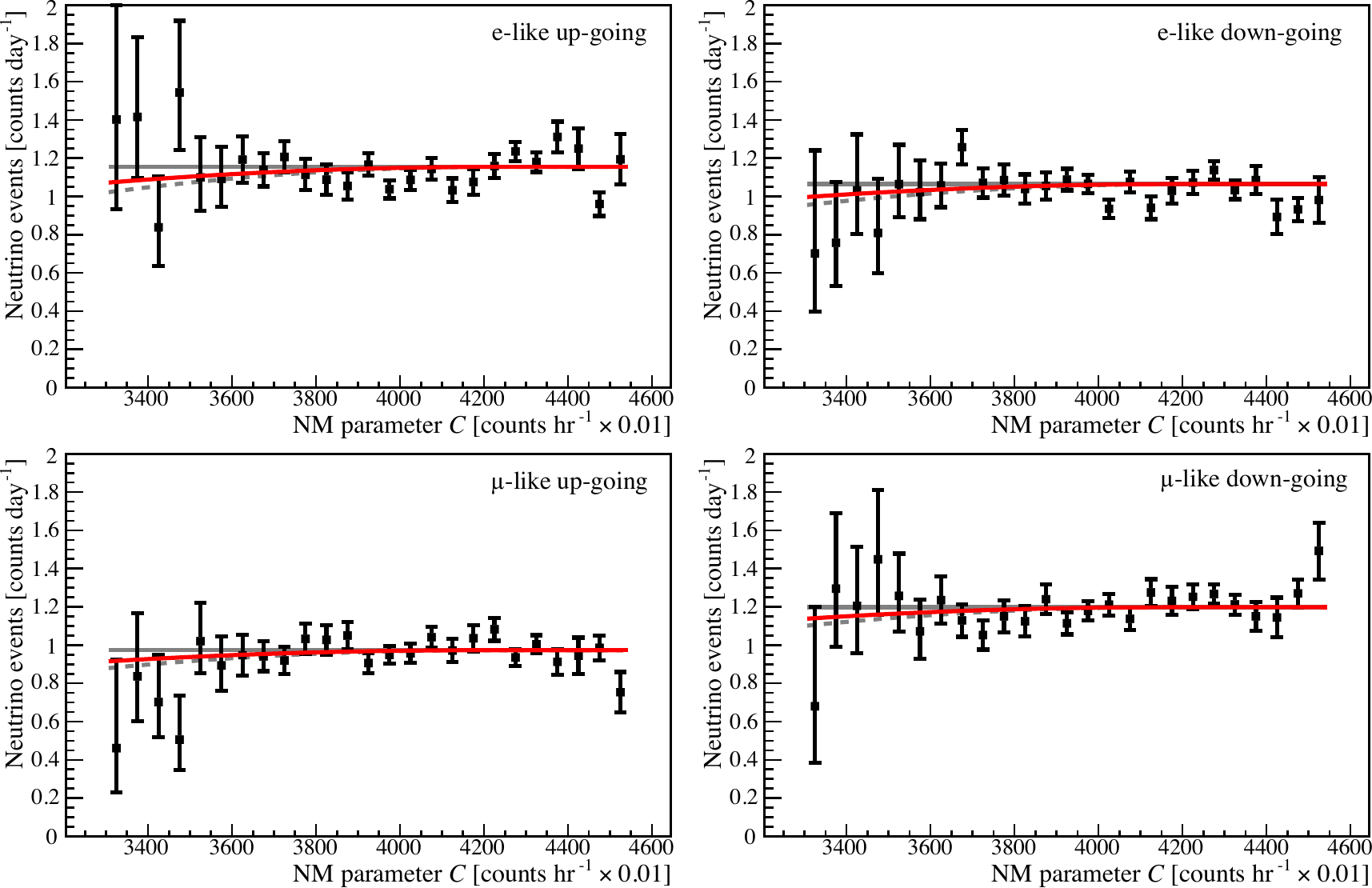}
\par\end{centering}

\caption{\small 
(color online) The test for a solar modulation correlation using the SK I-IV data (points).
The solar correlation hypotheses $H^{\prime}_{s,i}(\alpha)$,
as explained in Sec.~\ref{solmod:sec:analysis},
are shown for no correlation ($\alpha=0$, grey)
best fit ($\alpha=0.62$, red) and the default prediction ($\alpha=1$, grey dotted)
for each of the four data samples.
For clearer visualization, the $H^{\prime}_{s,i}$ are actually shown
without the systematic pulls $S_{s,i}(\boldsymbol{{\epsilon}})$,
but equivalent and opposite shifts are instead applied to the data points.
The statistical error bars are also drawn on the data instead of the model,
using Pearson's-$\chi^{2}$ based Poisson errors as defined in~\cite{poisson_error_bars}.
\label{solmod:fig:open_data}
}
\end{figure*}

\begin{figure}
\begin{centering}
\includegraphics[width=0.45\textwidth]{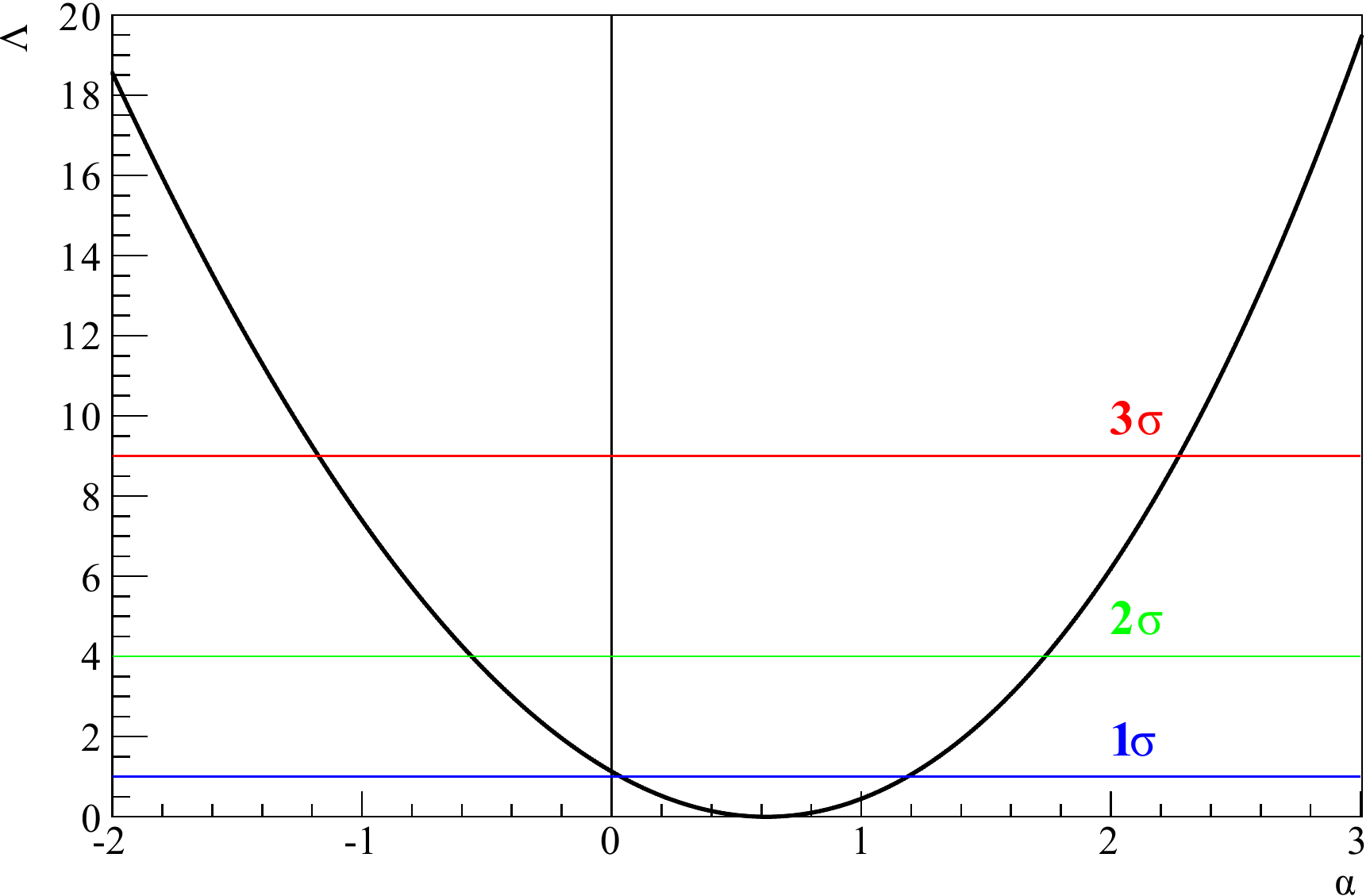}
\par\end{centering}

\caption{\small 
(color online) The test statistic $\Lambda$ as a function of
the solar modulation strength parameter $\alpha$,
both of which are defined in the text.
The significance levels are drawn assuming the validity of Wilks' theorem.
This statistic represents the combined measurement of all four data samples.
\label{solmod:fig:alpha_vs_chi2}
}
\end{figure}

\begin{figure}
\begin{centering}
\includegraphics[width=0.45\textwidth]{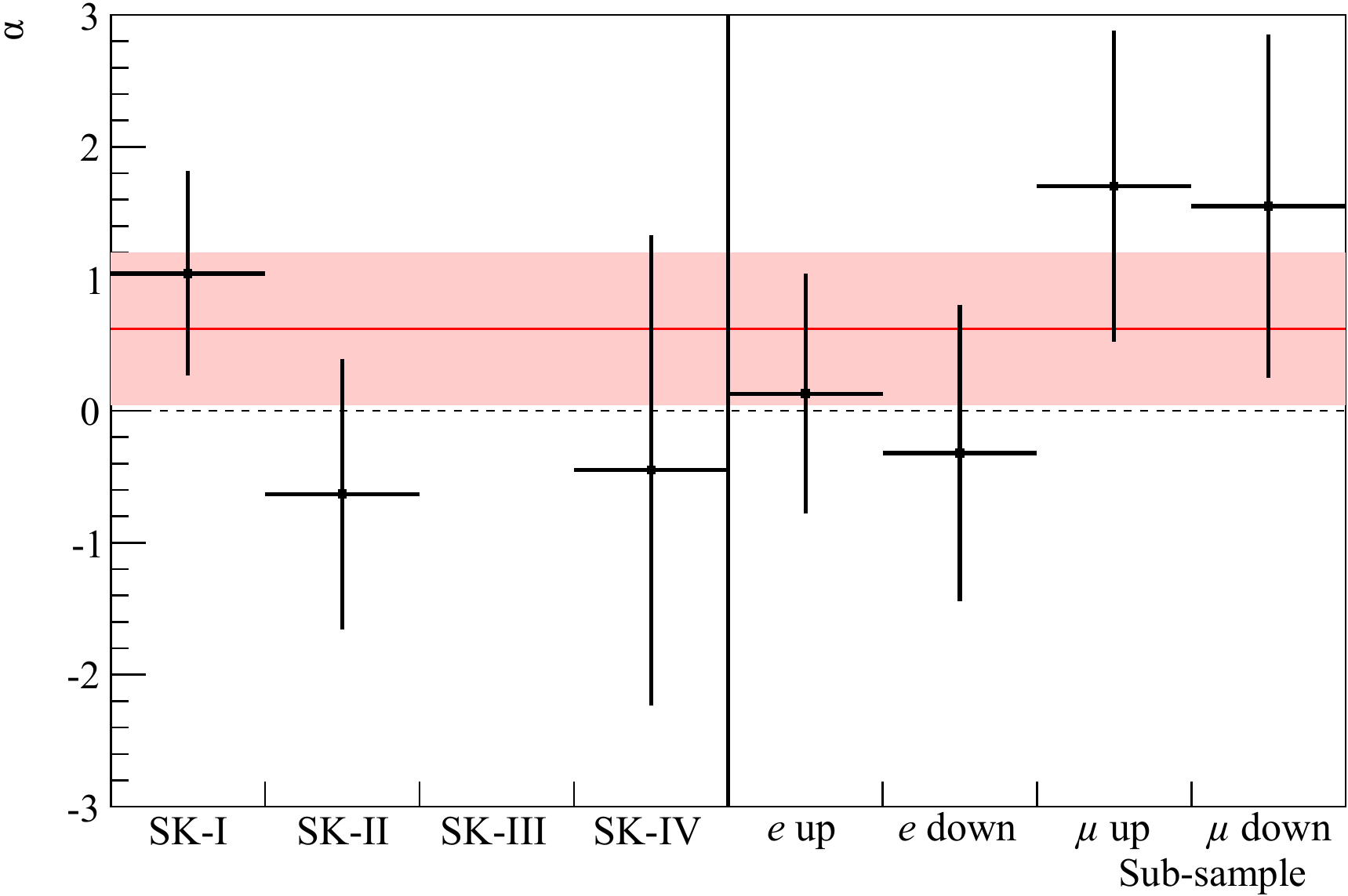}
\par\end{centering}
\protect\caption{(color online) The best-fit $\alpha$ parameter obtained by performing the analysis
on sub-samples of the data, showing SK periods on the left, and data
sub-sample types on the right. The red band shows the combined result using all data.
The SK-III data point is absent, due to a lack of any data in the
sensitive solar-active region, and thus an inability to perform
any fit by itself.\label{solmod:fig:by_sub_sample}}
\end{figure}

\subsection{Extreme Solar Events Analysis\label{solmod:sec:forbrush_search_method}}

Some sharp downwards peaks can be seen in Fig.~\ref{solmod:fig:NMcount_over_sk};
the cause of these events are usually large coronal mass ejections, causing
relatively short and high-intensity increases in the solar plasma
flux~\cite{forbrush_CME}. On Earth these are associated with particularly large decreases
in the cosmic ray flux and are often termed ``Forbush decreases''~\cite{10.2307/89609}.
While there is some ambiguity in the term, a common
definition is any event that causes a decrease greater than 10\% in the intensity
of cosmic rays on the Earth's surface.

So far, we have discussed a search for a long-term correlation
between the neutrino flux and the solar activity, using events down to a 
NM parameter $C$ of 3300~counts~hr$^{-1}\times0.01$. In fact,
$C$ sometimes drops significantly below this value,
although such observations are rare and spread over a range of values of $C$.
These periods are listed in Table~\ref{solmod:tab:forbrush}, and correspond to
Forbush decrease events occuring during solar maxima periods (Forbush decreases occuring
during solar minima, on the other hand, did not generally cause
$C$ to fall below 3300~counts~hr$^{-1}\times0.01$).

\begin{table}
\begin{centering}
\begin{tabular}{lclccc}
\hline 
\hline 
Start time  && End time  && Hour\tabularnewline
\hline 
15 Jul. 2000, 18:00 && 17 Jul. 2000, 21:00 && 50\tabularnewline
11 Apr. 2001, 23:00 && 13 Apr. 2001, 14:00 && 38\tabularnewline
29 Oct. 2003, 11:00 && 01 Nov. 2003, 00:00 && 61\tabularnewline
01 Nov. 2003, 00:00 && 04 Nov. 2003, 13:00 && 67\tabularnewline
19 Jan. 2005, 00:00 && 19 Jan. 2005, 13:00 && 13\tabularnewline
\hline 
Total  &&  && 229\tabularnewline
\hline 
\hline 
\end{tabular}
\par\end{centering}

\caption{\small 
The periods for which the NM parameter $C$ drops below 3300 counts hr$^{-1}$$\times$0.01, 
corresponding to strong Forbush decrease events.
Start and end times are given in UTC.
}
\label{solmod:tab:forbrush}
\end{table}

These time periods were not used in the long-term correlation analysis
as the HKKM simulation of the solar modulation effect
only provides a prediction down to a Climax NM count of
3500~counts~hr$^{-1}\times0.01$.
While we extended the correlation down to a NM parameter
$C=3300$~counts~hr$^{-1}\times0.01$ by a polynomial fitting,
we suspect that extending the correlation further would be an unrealistic model.
In particular, a simple correlation between the neutron and neutrino fluxes at various
locations on Earth may not hold during such extreme events.

We thus analyse separately these time periods where
$C$ falls below 3300~counts~hr$^{-1}\times0.01$.
While no theoretical prediction of the fractional neutrino flux
decrease is available here, these may be the most sensitive
times to measure a solar effect on the atmospheric neutrino flux.
We define a second search method by making a simple test against the null hypothesis,
by counting the number of events observed during all such periods over all four
data samples, and comparing with the nominal event rate,
taken from the fitted normalization constants $k_s\times r_s$ of the best-fit hypotheses 
in the long-term analysis.

\subsection{Results and Discussions}

The data for the long-term solar cycle correlation search,
summing over SK I-IV and dividing into the four sample types
($e$-like or $\mu$-like, and up-going or down-going) are shown in
Fig.~\ref{solmod:fig:open_data}. The test statistic $\Lambda$ depending
on $\alpha$ is shown in Fig.~\ref{solmod:fig:alpha_vs_chi2}. The
best-fit value of alpha is $\alpha=0.62_{-0.58}^{+0.57}$,
with errors given at 1~$\sigma$. 
Although the sensitivity to measure $\alpha$ is fairly low,
the data in Fig.~\ref{solmod:fig:open_data} show no unexpected disagreements
with the model; the four plots together have a $\chi^{2}$ statistic
of 88.0 for 100 bins.
The rejection power of the null hypothesis ($\alpha=0$) is 1.06~$\sigma$.
This is lower than the mean value predicted by toy MC with $\alpha=1$, but still a reasonably
likely result according to that MC, with a p-value of $\mathrm{p}=0.26$.

We also performed the same analysis on sub-samples of the data,
as shown in Fig.~\ref{solmod:fig:by_sub_sample};
we test each SK period I to IV separately, then
each data sub-sample individually (while including all SK periods).
Data from the SK-III period alone cannot give any result, as it has
no observation time above the minimum solar activity required to cause
any effect according to the HKKM model. Although SK-II and SK-IV prefer
a low value of $\alpha$, the statistical power is lower and not inconsistent
with the overall result.
Somewhat interestingly, the $e$-like samples prefer no correlation,
while the $\mu$-like samples prefer more the expected $\alpha=1$
correlation, however the significance is not high enough to draw any
strong conclusions.


\begin{figure}
\begin{centering}
\includegraphics[width=0.45\textwidth]{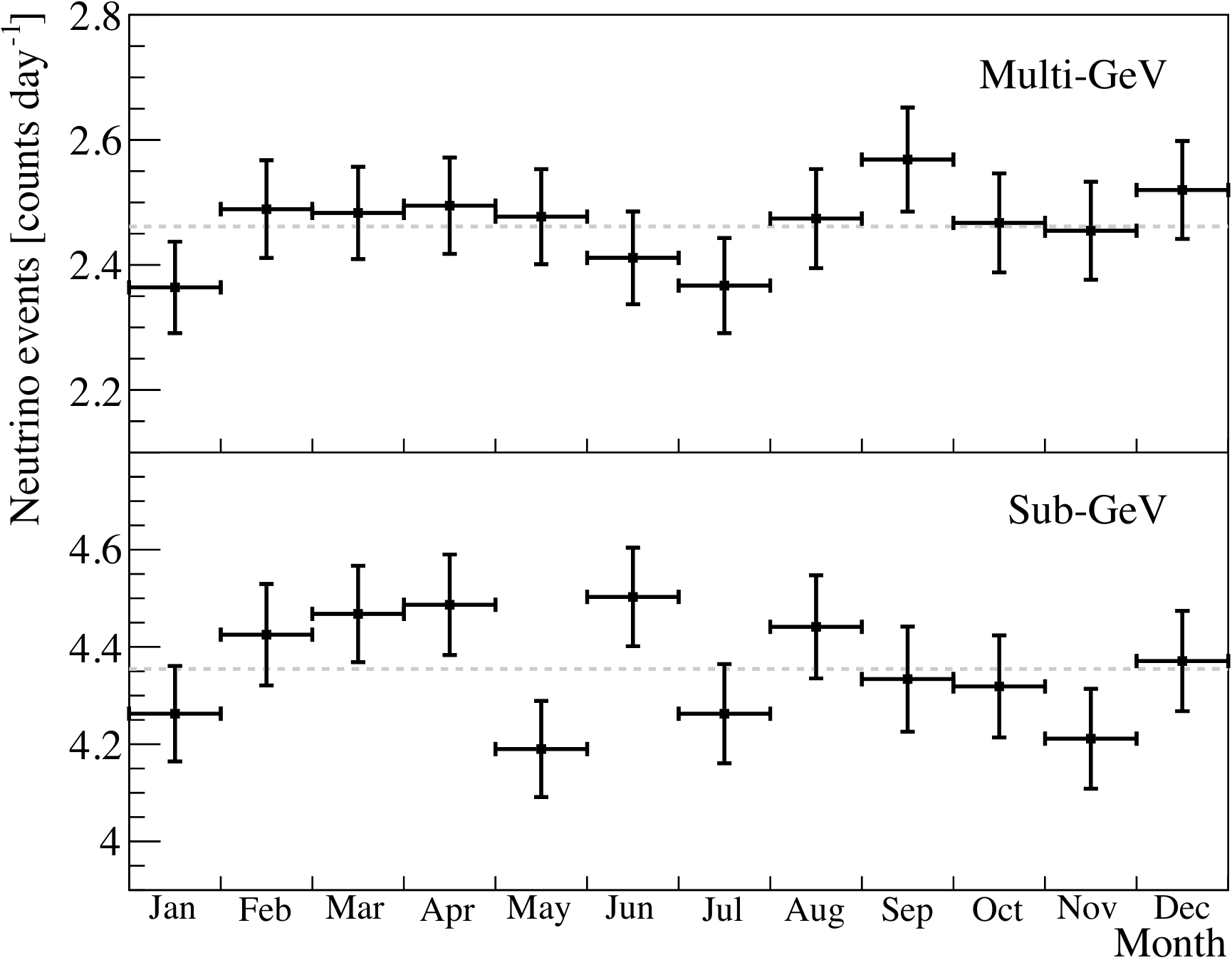}
\par\end{centering}
\protect\caption{
The average daily event rates for each month, using the SK I-IV data
(points with statistical error only).
The dotted lines are the best-fit constant functions.
\label{solmod:fig:monthly}}
\end{figure}

The search for a seasonal correlation is shown in Fig.~\ref{solmod:fig:monthly}.
The $\chi^2$ test statistic, comparing against the best-fit constant functions,
is 12.8 for the sub-GeV data and 6.5 for the multi-GeV data. The Kuiper test
on the unbinned data gives p-values of 0.76 and 0.62 respectively.
As expected, no strong indication is seen for any seasonal correlation
at the SK site.


During the coronal mass ejection periods described in Table~\ref{solmod:tab:forbrush},
the SK detector was operational for a total of 7.21 days. Using the fitted
normalization constants $k_s \times r_s$ from the long-term analysis,
we can calculate the number of events expected in the case of no solar modulation effect;
we find that $31.80\pm0.17$ sub-GeV single-ring events are expected.
The actual number of events recorded
by the detector was $n_{F}=20$, which by consulting the Poisson distribution
gives $\mathrm{P}(n_{F}\leq20)=0.017$, corresponding to rejection
of the null hypothesis of no solar cycle effect at the 98.3\%
(2.38~$\sigma$) significance level.

This significance level is higher than the long-term correlation search,
and while not high enough to claim direct evidence of a solar activity effect,
is nonetheless an indication that it may be possible to measure
such effects with higher accuracy in the next generation neutrino
detectors.

%
%
%
\FloatBarrier

\section{\label{sec:summary}Summary}
The energy spectra of the directionally-integrated atmospheric $\nu_{e}$ and $\nu_{\mu}$
particle plus antiparticle fluxes
at Super-Kamiokande were measured using a Bayesian iterative unfolding method.
A wide energy range was covered from 100~MeV up to 10~TeV,
with higher accuracy than previous measurements. 
The measured $\nu_\mu$ energy spectrum is consistent with the IceCube and AMANDA results,
which cover the energy range above 100~GeV.

The results were also compared with Monte Carlo predictions,
including the HKKM11, FLUKA, and Bartol models,
by performing a $\chi^2$ test incorporating both the $\nu_e$ and $\nu_\mu$ data.
The models were found to be consistent with the data.
The flux normalizations and spectral inclinations of each model were also tested,
and found to be consistent.


The azimuthal spectra of the fluxes were also compared with
the Monte Carlo predictions, which is the first detailed check of the azimuthal
spectra including a full systematic error analysis.
The data and MC agreed well, which provides confidence in the careful treatment
of geomagnetic effects in the recent flux simulations.
The existence of an east-west dipole asymmetry in the flux was also measured with greatly increased
statistics compared to previous measurements.
The effect was seen at a significance of
6~$\sigma$ for $\nu_{\mu}$ and 8~$\sigma$ for $\nu_e$,
which represents discovery of the effect in the $\nu_{\mu}$ flux.
An indication that the angle of the dipole asymmetry shifts depending on the zenith angle
was found at the 2.2~$\sigma$ level, which is the first measurement that explores
the geomagnetic field effects beyond the simple east-west asymmetry.

The expected correlation between the atmospheric neutrino flux and the solar magnetic activity
was studied, using the SK sub-GeV data sample and the ground level neutron flux
measured at various neutron monitors. 
The predicted effects on the neutrino flux
based on the HKKM model were found to be relatively small in the SK dataset, but by
using data spanning 20 years a weak preference for some correlation was seen at the
1.1~$\sigma$ level. 
We also examined separately several short periods of particularly intense
solar activity, for which no theoretical prediction is available, corresponding to 
7.1~days of detector uptime. During this period, an indication for a decrease in the neutrino
event rate below the normal level was seen at the 2.4~$\sigma$ level.

The seasonal change in the neutrino flux was also examined.
While some seasonal correlation is expected due to changes in the atmospheric density profile over the year,
the effect is predicted to be negligible at the SK site, and as expected
no such correlation was seen.

\begin{acknowledgments}
The authors would like to thank M.~Honda for many suggestions and discussions,
and providing the data on the correlation between the atmospheric flux and neutron count. 
The authors would like to thank S.~Yoshida, K.~Mase, A.~Ishihara, T.~Kuwabara, M.~Schmitz 
for the useful discussions on the atmospheric neutrino flux measurement.
We acknowledge the NMDB database~\cite{nmdb},
founded under the European Union's FP7 programme (contract no. 213007) for providing data.
The Newark, Thule, and McMurdo neutron monitors of the Bartol Research Institute are supported
by the National Science Foundation.
Kerguelen neutron monitor data were kindly provided by the
French Polar Institute (IPEV, Brest) and by Paris Observatory.
The authors gratefully acknowledge the cooperation of the Kamioka Mining and Smelting Company. 
Super-Kamiokande has been built and operated from funds provided by the Japanese Ministry of Education, 
Culture, Sports, Science and Technology, 
the U.S. Department of Energy, and the U.S. National Science Foundation. 
This work was supported by Grant-in-Aid for Scientific Research on Innovative Areas Number 25105004 and 25105005.
This work was partially supported by the joint research program of the Institute for Cosmic Ray Research,
the University of Tokyo.
This work was partially supported by the Research Foundation of Korea (BK21 and KNRC), 
the Korean Ministry of Science and Technology, 
the National Science Foundation of China, 
the European Union H2020 RISE-GA641540-SKPLUS,
the National Science and Engineering Research Council (NSERC) of Canada, 
the Scinet and Westgrid consortia of Compute Canada,
and the National Science Centre, Poland (2015/17/N/ST2/04064, 2015/18/E/
ST200758).
\end{acknowledgments}

%
%
%
\appendix

\section{Correlation matrix of the $\nu_e$ and $\nu_\mu$ energy spectrum \label{sec:corr_matrix}}

\begin{widetext}

\begin{table}
{\small
\begin{tabular}{lc|ccccccccccc}
\hline \hline
       &               & \multicolumn{11}{c}{$\nu_e$}  \\
       &  $\log_{10}(E_\nu{\rm [GeV]})$   &  -0.8 - -0.6 & -0.6 - -0.4 & -0.4 - -0.2 & -0.2 - 0.0 & 0.0 - 0.2 & 0.2 - 0.4 & 0.4 - 0.6 & 0.6 - 0.8 & 0.8 - 1.0 & 1.0 - 1.5 & 1.5 - 2.0 \\
\hline
       &  -0.8 - -0.6 &         1.000  &  0.948  &  0.720  &  0.512  &  0.396  &  0.396  &  0.267  &  0.156  &  0.120  & -0.079  & -0.130  \\
       &  -0.6 - -0.4 &         0.948  &  1.000  &  0.880  &  0.678  &  0.580  &  0.557  &  0.400  &  0.288  &  0.240  & -0.034  & -0.124  \\
       & -0.4 - -0.2 &          0.720  &  0.880  &  1.000  &  0.914  &  0.836  &  0.802  &  0.631  &  0.505  &  0.443  &  0.064  & -0.084  \\
       & -0.2 - 0.0 &           0.512  &  0.678  &  0.914  &  1.000  &  0.931  &  0.915  &  0.776  &  0.633  &  0.566  &  0.131  & -0.048  \\
       & 0.0 - 0.2 &            0.396  &  0.580  &  0.836  &  0.931  &  1.000  &  0.903  &  0.779  &  0.698  &  0.625  &  0.187  & -0.005  \\
$\nu_e$ & 0.2 - 0.4 &           0.396  &  0.557  &  0.802  &  0.915  &  0.903  &  1.000  &  0.909  &  0.753  &  0.700  &  0.195  & -0.024  \\
       & 0.4 - 0.6 &            0.267  &  0.400  &  0.631  &  0.776  &  0.779  &  0.909  &  1.000  &  0.917  &  0.820  &  0.293  &  0.044  \\
       & 0.6 - 0.8 &            0.156  &  0.288  &  0.505  &  0.633  &  0.698  &  0.753  &  0.917  &  1.000  &  0.880  &  0.359  &  0.101  \\
       & 0.8 - 1.0 &            0.120  &  0.240  &  0.443  &  0.566  &  0.625  &  0.700  &  0.820  &  0.880  &  1.000  &  0.462  &  0.130  \\
       & 1.0 - 1.5 &           -0.079  & -0.034  &  0.064  &  0.131  &  0.187  &  0.195  &  0.293  &  0.359  &  0.462  &  1.000  &  0.927  \\
       & 1.5 - 2.0 &           -0.130  & -0.124  & -0.084  & -0.048  & -0.005  & -0.024  &  0.044  &  0.101  &  0.130  &  0.927  &  1.000  \\
\hline
      &  -0.6 - -0.4 &           0.843  &  0.888  &  0.809  &  0.682  &  0.537  &  0.592  &  0.458  &  0.307  &  0.260  & -0.019  & -0.110  \\
      &  -0.4 - -0.2 &           0.692  &  0.832  &  0.940  &  0.899  &  0.842  &  0.812  &  0.655  &  0.533  &  0.467  &  0.085  & -0.069  \\
      &  -0.2 - 0.0 &            0.520  &  0.686  &  0.896  &  0.944  &  0.900  &  0.886  &  0.732  &  0.594  &  0.527  &  0.100  & -0.072  \\
      &  0.0 - 0.2 &             0.423  &  0.597  &  0.839  &  0.915  &  0.941  &  0.900  &  0.790  &  0.698  &  0.624  &  0.169  & -0.030  \\
      &  0.2 - 0.4 &             0.400  &  0.544  &  0.771  &  0.889  &  0.874  &  0.946  &  0.903  &  0.776  &  0.700  &  0.172  & -0.052  \\
      &  0.4 - 0.6 &             0.291  &  0.426  &  0.659  &  0.798  &  0.824  &  0.907  &  0.945  &  0.873  &  0.795  &  0.227  & -0.024  \\
$\nu_\mu$ &  0.6 - 0.8 &         0.252  &  0.380  &  0.603  &  0.732  &  0.774  &  0.849  &  0.929  &  0.906  &  0.837  &  0.281  &  0.021  \\
      &  0.8 - 1.0 &             0.240  &  0.360  &  0.565  &  0.681  &  0.715  &  0.793  &  0.888  &  0.891  &  0.849  &  0.409  &  0.153  \\
      &  1.0 - 1.5 &             0.181  &  0.282  &  0.445  &  0.531  &  0.555  &  0.589  &  0.615  &  0.615  &  0.660  &  0.687  &  0.509  \\
      &  1.5 - 2.0 &             0.069  &  0.106  &  0.167  &  0.200  &  0.207  &  0.215  &  0.230  &  0.260  &  0.368  &  0.763  &  0.692  \\
     &  2.0 - 3.0 &              0.034  &  0.051  &  0.077  &  0.092  &  0.090  &  0.099  &  0.121  &  0.163  &  0.285  &  0.747  &  0.700  \\
     &  3.0 - 4.0 &              0.027  &  0.026  &  0.031  &  0.040  &  0.037  &  0.048  &  0.084  &  0.132  &  0.254  &  0.730  &  0.705  \\
\hline
\end{tabular}
}
\\
{\small 
\begin{tabular}{lc|cccccccccccc}
\hline
\hline
       &               & \multicolumn{12}{c}{$\nu_\mu$}  \\
       &    $\log_{10}(E_\nu {\rm [GeV]})$           &  -0.6 - -0.4 & -0.4 - -0.2 & -0.2 - 0.0 & 0.0 - 0.2 & 0.2 - 0.4 & 0.4 - 0.6 & 0.6 - 0.8 & 0.8 - 1.0 & 1.0 - 1.5 & 1.5 - 2.0 & 2.0 - 3.0 & 3.0 - 4.0 \\
\hline
       &  -0.8 - -0.6  &          0.843  &  0.692  &  0.520  &  0.423  &  0.400  &  0.291  &  0.252  &  0.240  &  0.181  &  0.069  &  0.034  &  0.027  \\ 
       &  -0.6 - -0.4 &           0.888  &  0.832  &  0.686  &  0.597  &  0.544  &  0.426  &  0.380  &  0.360  &  0.282  &  0.106  &  0.051  &  0.026  \\ 
       & -0.4 - -0.2 &            0.809  &  0.940  &  0.896  &  0.839  &  0.771  &  0.659  &  0.603  &  0.565  &  0.445  &  0.167  &  0.077  &  0.031  \\ 
       & -0.2 - 0.0 &             0.682  &  0.899  &  0.944  &  0.915  &  0.889  &  0.798  &  0.732  &  0.681  &  0.531  &  0.200  &  0.092  &  0.040  \\        
       & 0.0 - 0.2 &              0.537  &  0.842  &  0.900  &  0.941  &  0.874  &  0.824  &  0.774  &  0.715  &  0.555  &  0.207  &  0.090  &  0.037  \\ 
$\nu_e$ & 0.2 - 0.4 &             0.592  &  0.812  &  0.886  &  0.900  &  0.946  &  0.907  &  0.849  &  0.793  &  0.589  &  0.215  &  0.099  &  0.048  \\ 
       & 0.4 - 0.6 &              0.458  &  0.655  &  0.732  &  0.790  &  0.903  &  0.945  &  0.929  &  0.888  &  0.615  &  0.230  &  0.121  &  0.084  \\ 
       & 0.6 - 0.8 &              0.307  &  0.533  &  0.594  &  0.698  &  0.776  &  0.873  &  0.906  &  0.891  &  0.615  &  0.260  &  0.163  &  0.132  \\ 
       & 0.8 - 1.0 &              0.260  &  0.467  &  0.527  &  0.624  &  0.700  &  0.795  &  0.837  &  0.849  &  0.660  &  0.368  &  0.285  &  0.254  \\ 
       & 1.0 - 1.5 &             -0.019  &  0.085  &  0.100  &  0.169  &  0.172  &  0.227  &  0.281  &  0.409  &  0.687  &  0.763  &  0.747  &  0.730  \\ 
       & 1.5 - 2.0 &             -0.110  & -0.069  & -0.072  & -0.030  & -0.052  & -0.024  &  0.021  &  0.153  &  0.509  &  0.692  &  0.700  &  0.705  \\ 
\hline
      &  -0.6 - -0.4 &              1.000  &  0.840  &  0.699  &  0.574  &  0.596  &  0.462  &  0.388  &  0.369  &  0.293  &  0.096  &  0.039  &  0.015  \\ 
      &  -0.4 - -0.2 &              0.840  &  1.000  &  0.934  &  0.872  &  0.816  &  0.702  &  0.637  &  0.599  &  0.484  &  0.174  &  0.074  &  0.026  \\ 
      &  -0.2 - 0.0 &               0.699  &  0.934  &  1.000  &  0.942  &  0.889  &  0.788  &  0.720  &  0.674  &  0.543  &  0.196  &  0.086  &  0.027  \\ 
      &  0.0 - 0.2 &                0.574  &  0.872  &  0.942  &  1.000  &  0.912  &  0.853  &  0.807  &  0.753  &  0.593  &  0.220  &  0.100  &  0.039  \\ 
      &  0.2 - 0.4 &                0.596  &  0.816  &  0.889  &  0.912  &  1.000  &  0.952  &  0.885  &  0.836  &  0.626  &  0.226  &  0.107  &  0.048  \\ 
      &  0.4 - 0.6 &                0.462  &  0.702  &  0.788  &  0.853  &  0.952  &  1.000  &  0.970  &  0.907  &  0.637  &  0.226  &  0.111  &  0.058  \\ 
$\nu_\mu$ &  0.6 - 0.8&             0.388  &  0.637  &  0.720  &  0.807  &  0.885  &  0.970  &  1.000  &  0.956  &  0.658  &  0.258  &  0.151  &  0.111  \\ 
      &  0.8 - 1.0 &                0.369  &  0.599  &  0.674  &  0.753  &  0.836  &  0.907  &  0.956  &  1.000  &  0.787  &  0.415  &  0.310  &  0.278  \\ 
      &  1.0 - 1.5 &                0.293  &  0.484  &  0.543  &  0.593  &  0.626  &  0.637  &  0.658  &  0.787  &  1.000  &  0.845  &  0.756  &  0.694  \\ 
      &  1.5 - 2.0 &                0.096  &  0.174  &  0.196  &  0.220  &  0.226  &  0.226  &  0.258  &  0.415  &  0.845  &  1.000  &  0.983  &  0.927  \\ 
     &  2.0 - 3.0 &                 0.039  &  0.074  &  0.086  &  0.100  &  0.107  &  0.111  &  0.151  &  0.310  &  0.756  &  0.983  &  1.000  &  0.955  \\ 
     &  3.0 - 4.0 &                 0.015  &  0.026  &  0.027  &  0.039  &  0.048  &  0.058  &  0.111  &  0.278  &  0.694  &  0.927  &  0.955  &  1.000  \\ 
\hline
\end{tabular}
}
\caption{ Values of correlation matrix element ($\hat{C_{ij}}$) in the $\nu_e$ and $\nu_\mu$ energy spectrum measurement. Upper and lower tables corresponds to $\nu_e$ and $\nu_\mu$ energy spectrum, respectively.}
\end{table}

\end{widetext}


\bibliography{draft}

\end{document}